\documentclass[12pt]{article}
\pdfoutput=1
\usepackage{amsmath,amssymb,epsfig,amsfonts}
\usepackage{graphicx}
 
\usepackage{subcaption}
\usepackage[usenames, dvipsnames]{color} 
\usepackage{hyperref}
\usepackage[dvipsnames]{xcolor}
\usepackage{cite}
\usepackage{multirow} 
\usepackage{slashed} 
\usepackage{enumitem}
\usepackage{rotating}
\usepackage{multirow}
\usepackage{bm}
\usepackage[normalem]{ulem} 
\usepackage{cleveref}  

\usepackage{enumitem}
\usepackage{mathtools} 
\definecolor{forestForestGreen(web)}{rgb}{0.13, 0.55, 0.13}
\definecolor{blue(munsell)}{HTML}{005567}
\definecolor{oxfordblue}{rgb}{0.0, 0.13, 0.28}
\definecolor{bblue}{rgb}{0.0, 0.58, 0.71}
\hypersetup{
pdfstartview={FitH}, 
pdftitle={}, 
colorlinks=true, 
linkcolor=oxfordblue, 
citecolor=oxfordblue, 
filecolor=oxfordblue, 
urlcolor=oxfordblue
}
\usepackage{pgfplots}
\pgfplotsset{compat=1.18}
\usepackage{amssymb}
\usepackage{amsfonts}
\usepackage{epstopdf}
\usepackage{dcolumn}
\usepackage{latexsym,bm}
\usepackage{amsthm}
\usepackage{float}
\usepackage{color}
\usepackage{url} 
\usepackage{rotating}
\usepackage{tikz-cd} 
\usepackage{array}

\usepackage{tabularx}
\usepackage{makecell} 
\usepackage{changepage}

\newtheorem{theorem}{Theorem}[section]
\newtheorem{proposition}[theorem]{Proposition}
\newtheorem{lemma}[theorem]{Lemma}

\theoremstyle{definition}
\newtheorem{definition}[theorem]{Definition}
\theoremstyle{remark}

\newcommand{\CC} {\mathbb{C}}

\newcommand{\PP} {\mathbb{P}}

\newcommand{\ZZ} {\mathbb{Z}}

\newcommand{\cX} {\mathcal{X}}
\newcommand{\cP} {\mathcal{P}}
\newcommand{\tcP} {\widetilde{\mathcal{P}}}

\usetikzlibrary{shapes,arrows,chains,calc}

\definecolor{darkgreen}{RGB}{0,100,0}

\newtheoremstyle{defstyle}
  {.6em} 
  {.1em} 
  {} 
  {} 
  {\bfseries} 
  {.} 
  {.5em} 
  {} 

\newtheorem{remark}[theorem]{Remark}
\numberwithin{equation}{section}
\numberwithin{figure}{section}

\date{\today}

\definecolor{britishracinggreen}{rgb}{0.0, 0.26, 0.15}

\newcommand{\be}{\begin{equation}}
\newcommand{\ee}{\end{equation}}
\newcommand{\ba}{\begin{aligned}}
\newcommand{\ea}{\end{aligned}}

\usetikzlibrary{patterns.meta}
\tikzset{seven/.style={circle, draw,inner sep=3pt}}
\tikzset{bd/.style={circle, draw=black, inner sep=0pt, fill=black, minimum size=2mm}}
\tikzset{wd/.style={circle, draw=black, inner sep=0pt, fill=white, minimum size=2mm}}
\tikzset{od/.style={circle, draw=orange, inner sep=0pt, fill=orange, minimum size=2mm}}

\begin{document}

\baselineskip=15pt  
\numberwithin{equation}{section}  
\thispagestyle{empty}
\vspace*{1cm} 
\begin{center}
{\huge 5d SCFTs and Intersection Theory of \\

\vspace*{0.3cm}
Generalized Toric Threefolds
}

\vspace*{1.3cm}

H\"ulya Arg\"uz$^\bullet$, 
Antoine Bourget$^\circ$, 
Pierrick Bousseau$^\bullet$, \\ 

\vspace*{0.3cm}

Julius F.\ Grimminger,  and 
Sakura Sch\"afer-Nameki$^\bullet$

\vspace*{0.5cm} 

 {\it $^\bullet$  Mathematical Institute, University of Oxford, \\
 Andrew Wiles Building,  Woodstock Road, Oxford, OX2 6GG, UK}\\

{\it $^\circ$ Institut de Physique Théorique, Université Paris-Saclay, \\ CEA, CNRS, F-91191 Gif-sur-Yvette, France}

 \end{center}
\vspace*{0.2cm}

\noindent

\begin{abstract}
\noindent
Five-dimensional superconformal field theories are intrinsically strongly coupled
fixed points that can be realized as M-theory on local Calabi-Yau threefolds with
canonical singularities. The low-energy effective theory on their extended Coulomb
branch is governed by a cubic prepotential, determined by the triple intersection
numbers of divisors in the resolved geometry. For toric threefolds, dual to 5-brane
webs, these numbers are computed combinatorially from the toric polygon.
However, no such
description has been available for generalized toric polygons (GTPs), dual to
5-brane webs with 7-branes, since the corresponding generalized toric Calabi–Yau threefolds are generally non-toric. We show that the 7-brane data are encoded by
additional lines in the GTP, which correspond to non-toric, non-compact divisors
that we call exuberant divisors. Building on a local analysis of these divisors and
on the associativity of the intersection product, we develop an algorithm that
computes the triple intersection numbers, and hence the prepotential, directly from
the combinatorial data of the GTP. We illustrate it in examples and provide a mathematica code
implementation of the intersection algorithm. 

\end{abstract}
 
\newpage

\tableofcontents

\section{Introduction}
\label{sec:introduction}

Five-dimensional $\mathcal{N}=1$ superconformal field theories (SCFTs)
provide a particularly rich meeting point between strongly-coupled quantum
field theory, string theory, and algebraic geometry. 
The geometric realization in M-theory on singular Calabi--Yau threefolds
\cite{seiberg1996five,Intriligator:1997pq} is particularly insightful as it provides an alternative way, using geometry, to potentially classify such theories. 
Several complementary geometric constructions have been used to explore and map out the vast landscape of such theories, including toric Calabi--Yau threefolds \cite{Xie:2017pfl}, using constraints from the local surface geometry \cite{Jefferson:2018irk, Bhardwaj:2018vuu, Bhardwaj:2018yhy}, using M-/F-theory duality which realizes canonical singularities starting with 
elliptically fibered Calabi--Yau threefolds with non-minimal singular fibers
\cite{Apruzzi:2019opn,Apruzzi:2019enx,Apruzzi:2019kgb}, isolated hypersurface singularities 
\cite{Xie:2017pfl, Closset:2020scj,Closset:2020afy,Closset:2021lwy}, or non-toric orbifolds \cite{Collinucci:2025rrh}.  Complementary to these geometric approaches, webs of $(p,q)$ 5-branes in Type IIB
string theory, possibly ending on 7-branes or in the presence of orientifold planes,
have been used extensively to construct 5d SCFTs and to map out their gauge-theory
deformations, UV dualities, Higgs branches and RG flows
\cite{Bergman:2013aca, O7planes, O5planes, Hayashi:2015zka,
Jefferson:2017ahm, Hayashi:2018lyv, Bourget:2019rtl, Bourget:2020gzi, vanBeest:2020kou, Bourget:2020mez, VanBeest:2020kxw},
with the predicted enhanced global symmetries cross-checked against the
superconformal index \cite{Kim:2012gu, Bergman:2013aca}.

M-theory sets up a dictionary between the SCFT and geometric data as follows: 
M-theory on the canonical singularity corresponds to the SCFT.  
The crepant resolution of the singularity provides a geometric description of the extended Coulomb branch,
with K\"ahler parameters identified with Coulomb-branch parameters and mass
deformations. The low-energy effective theory on the extended Coulomb branch
is encoded by a cubic prepotential, whose coefficients are given geometrically
by triple intersection numbers of divisors in the resolved Calabi--Yau
threefold. 

For toric Calabi--Yau threefolds this relation is particularly concrete. 
The
geometry is encoded by a two-dimensional toric polygon, and, after choosing a
triangulation, divisors, curves, and their intersection numbers can be
extracted directly from the combinatorics of the polygon. This provides an
efficient route from simple discrete data to the five-dimensional cubic
prepotential, see for example 
\cite{Xie:2017pfl,Closset:2018bjz,Eckhard:2020jyr}. 

A large class of five-dimensional SCFTs also admits a complementary, dual
description in terms of webs of $(p,q)$ 5-branes in Type IIB string theory
\cite{Aharony:1997ju,Aharony:1997bh,DeWolfe:1999hj}. For theories engineered
by toric Calabi--Yau threefolds, the two descriptions are directly related:
the brane web is dual to the toric polygon, and the M-theory and Type IIB
constructions engineer the same five-dimensional theory
\cite{Aharony:1997bh,Leung:1997tw}. More general brane webs naturally involve
$[p,q]$ 7-branes and are no longer described by ordinary toric polygons.
Generalized toric polygons (GTPs) provide a combinatorial framework for these
more general configurations \cite{Benini:2009gi, vanBeest:2020kou, VanBeest:2020kxw, AABwebs}. The corresponding Calabi--Yau geometries, referred to as \emph{generalized toric Calabi--Yau threefolds}, are generically non-toric, so that the familiar toric dictionary and the standard combinatorial rules for computing intersection numbers no longer apply directly. 

Brane webs and GTPs provide powerful combinatorial descriptions of
these five-dimensional theories, which allow one to probe indirectly certain aspects of generalized toric Calabi--Yau threefolds, see for instance \cite{Hayashi:2015xla,Bourget:2026hsg} for computations of BPS invariants using this combinatorics via a generalization of the topological vertex, and \cite{Kho:2026zwc} for associated integrable systems. 
Explicit algebraic descriptions of the generalized toric Calabi--Yau threefolds have been developed for particular classes of GTPs through complex-structure deformations of toric geometries \cite{Bourget:2023wlb,Collinucci:2026kom}.
Other approaches explore the relation between Hanany--Witten transitions and polytope mutations \cite{Arias-Tamargo:2024fjt}, the role of primitive $T$-cones as building blocks of GTP tessellations \cite{CarrenoBolla:2024fxy}, and the computation of Hilbert and Ehrhart series from GTP data
\cite{CarrenoBolla:2025rkv,Franco:2026bie}.

For consistent 5-brane webs with 7-branes, a general algebro-geometric construction of generalized toric Calabi--Yau threefolds associated to a GTP was obtained in  \cite{AABwebs}, building on \cite{AAB_KSBA, MR3825608, MR4255071}. This approach is expanded upon in \cite{BCL} to provide an explicit patch by patch description. Once the geometry is explicitly defined, it becomes a
mathematically well-posed problem to compute the triple intersection numbers
of its divisors and, in particular, to determine whether the cubic
prepotential can be recovered directly from the combinatorial data of the
GTP.

The purpose of this paper is to answer this question. We determine the
relevant intersection theory of the
generalized toric Calabi--Yau threefolds constructed in
\cite{AABwebs} and show that their triple intersection numbers can be
reconstructed directly from the combinatorial data of the GTP. The result is
a finite algorithm that extends the computational power of toric methods to
this class of non-toric Calabi--Yau threefolds.

Let us briefly describe the geometric input. Starting from a GTP with a fixed
polyhedral subdivision, the construction of \cite{AABwebs} produces an
associated \emph{Symington polygon} by a sequence of cuts and integral-affine
identifications. The Symington polygon is an integral-affine manifold with singularities, together with an induced unimodular triangulation. The vertices and edges of this 
triangulation encode a collection
of log Calabi--Yau surfaces and the curves along which they are glued. 
Their union forms a normal crossings surface that admits a one-parameter smoothing.
Then, the generalized toric Calabi--Yau threefold associated with the GTP is obtained as the total space of this one-parameter smoothing.
In the toric case, the Symington
polygon is an ordinary toric polygon and the construction reduces to the
familiar toric geometry.

In addition to the components of the central fiber, the geometry contains
further non-compact divisors, which we call \emph{exuberant divisors}. These
are encoded combinatorially by so-called $E$-lines associated with the GTP. An $E$-line
determines an exuberant divisor together with its intersection curve with the
central fiber. The divisor and curve data relevant to the intersection
problem are therefore encoded jointly by the subdivided Symington polygon
and the $E$-lines inherited from the GTP.

Our main result is that this combinatorial information suffices to reconstruct
the intersection data relevant for five-dimensional physics. Intersections
among the components of the central fiber are determined from the
vertex--edge--triangle incidence relations of the unimodular subdivision,
together with the integral-affine lattice directions of adjacent edges and
the $SL(2,\mathbb{Z})$ monodromies across cuts. Intersections involving
exuberant divisors are controlled by the primitive directions and incidence
data of the $E$-lines. The coefficients not fixed by these rules are determined
by compatibility conditions following from associativity of the intersection
product.
We show that the resulting compatibility system has a unique solution and determines all intrinsic triple intersection numbers. These
intersection numbers determine the five-dimensional cubic prepotential. This
gives a combinatorial algorithm for computing the prepotential directly from
the GTP, extending the toric computation to the generalized toric Calabi--Yau
threefolds constructed in \cite{AABwebs}. 

We provide a public implementation of our algorithm in Mathematica, available on GitHub,\footnote{\href{https://github.com/AntoineBrgt/GTP-Intersections}{https://github.com/AntoineBrgt/GTP-Intersections}} and briefly described in Appendix \ref{app:code}. It takes the GTP as input and returns all triple intersection numbers. 

The organization of the paper is as follows.
Section~\ref{sec:CanSing5d} reviews five-dimensional SCFTs, their geometric
engineering by Calabi--Yau threefolds, and toric intersection theory.
Section~\ref{sec:brane-webs-and-GTPs} reviews $(p,q)$ 5-brane webs, 7-branes,
and GTPs. Section~\ref{sec:CYGTP} reviews the
construction of generalized toric Calabi--Yau threefolds following \cite{AABwebs} and the associated divisor and
curve data. Section~\ref{sec:intersectionTheory} develops the intersection
theory, and Section~\ref{sec:intersectionAlgorithm} gives the combinatorial
algorithm for computing the triple intersection numbers. We conclude with a
series of examples in Section~\ref{sec:examples}.

\section{Canonical Singularities and 5d SCFTs}
\label{sec:CanSing5d}

In this section we review some salient properties of five-dimensional $\mathcal N=1$ SCFTs and their geometric engineering that will be needed throughout the paper. We first review the Coulomb and extended Coulomb parameter spaces. 
For each chamber of the extended Coulomb branch, we then discuss the cubic prepotential and its relation to the effective gauge coupling and Chern--Simons terms. 
We then review the geometric engineering of such theories in  M-theory on resolved Calabi--Yau threefold singularities, where Coulomb and mass parameters are identified with K\"ahler parameters and the cubic prepotential is determined by triple intersection numbers of divisors in the Calabi--Yau threefold. 
Finally, we review the toric case, in which these intersection numbers can be computed directly from the combinatorics of a triangulated toric polygon.

\subsection{Coulomb Branches and Prepotentials}
\label{sec:coulomb-branch}

Consider a five-dimensional \(\mathcal N=1\) superconformal field theory (SCFT). Its Coulomb branch is the locus in the moduli space of vacua along which the \(SU(2)_R\) R-symmetry remains unbroken. The rank $r$ of the SCFT is defined as the real dimension of its Coulomb branch.
At a generic point on the Coulomb branch, the low-energy effective theory is an abelian $U(1)^r$ gauge theory comprising $r$ vector multiplets. Locally, the Coulomb branch is parametrized by the vacuum expectation values
$\phi^i$, $i=1,\ldots,r$,
of the real scalar fields in these vector multiplets.

In addition to moving on the Coulomb branch, one may turn on real mass deformations. Let $G_F$ denote the flavor symmetry of the SCFT and let $m^\alpha$, $\alpha=1,\ldots,f$, be real mass parameters associated with its Cartan subalgebra. We refer to the space locally parametrized by
\be
t^A=(\phi^i,m^\alpha)
\ee
as the extended Coulomb parameter space. 
Fixing the masses gives the Coulomb branch of the corresponding mass-deformed theory, while setting them to zero gives the Coulomb branch of the undeformed SCFT.

The extended Coulomb parameter space is divided into finitely many chambers. Inside each chamber, the low-energy description by a $U(1)^r$ gauge theory is smooth and the two-derivative effective action is encoded by a cubic prepotential: 
\be
\mathcal F(t)
=
\frac16\,c_{ABC}t^At^Bt^C\,,
\label{eq:extended-prepotential}
\ee
with coefficients $c_{ABC}$ constant within the chamber. For fixed mass parameters, the two-derivative effective Lagrangian is of the form 
\be
\mathcal L_{\mathrm{eff}}
=
-\frac12 G_{ij}(t)\,d\phi^i\wedge\star d\phi^j
-\frac12 G_{ij}(t)\,F^i\wedge\star F^j
+\frac{1}{24\pi^2}\,c_{ijk}\,
A^i\wedge F^j\wedge F^k
+\cdots\,,
\label{eq:effective-lagrangian}
\ee
where $\phi^i$ are the real scalar fields, $A^i$ are the $U(1)$ gauge fields of the vector multiplets, $F^i=dA^i$,
with the effective inverse squared gauge-coupling matrix
\be
G_{ij}(t)
=
\frac{\partial^2\mathcal F}
{\partial\phi^i\partial\phi^j}= c_{ijk} \phi^k + c_{ij\alpha} m^\alpha\,,
\label{eq:effective-coupling}
\ee
and Chern--Simons coupling constants
\be
c_{ijk}
=
\frac{\partial^3\mathcal F}
{\partial \phi^i\partial \phi^j \partial \phi^k}\,.
\label{eq:prepotential-derivatives}
\ee

For instance, consider a chamber of the extended Coulomb branch in which the 5d theory provides a UV completion of a non-abelian $\mathcal N=1$ gauge theory with  gauge group
$G_{\text{gauge}}$ of rank $r$ and with matter hypermultiplets in irreducible representations $\mathbf R_f$ of $G_{\text{gauge}}$. At a generic point of the chamber, the gauge group is broken to its Cartan subgroup $U(1)^r$. The real mass parameters $m^\alpha$ of the 5d SCFT are the inverse square gauge couplings $\frac{1}{2g_{ij}^2}$ and the matter masses $m_f$ from the viewpoint of the non-abelian gauge theory. Then, the prepotential is 
\cite{Intriligator:1997pq,Closset:2018bjz}
\be
\begin{aligned}
\mathcal F={}&
\frac{1}{2g_{ij}^2}\,\phi^i\phi^j
+\frac{k_{ijk}}{6}\,\phi^i\phi^j\phi^k\\
&+\frac{1}{12}\left(
\sum_{\alpha\in\Delta}|\alpha\cdot\phi|^3
-\sum_f\sum_{\lambda\in\mathbf W_f}
|\lambda\cdot\phi+m_f|^3
\right),
\end{aligned}
\label{eq:IMS-prepotential}
\ee
where $k_{ijk}$ denote classical Chern--Simons coupling constants for the non-abelian gauge field, $\Delta$ is the set of non-zero roots of $G_{\text{gauge}}$, and $\mathbf{W}_f$ denotes the set of weights of the representation $\mathbf{R}_f$.
From the non-abelian gauge-theory perspective, \eqref{eq:IMS-prepotential} is the exact perturbative prepotential: the non-classical terms are the one-loop contributions obtained by integrating out the massive W-bosons and matter fields. Note that a complementary approach is provided by the complete prepotential
proposed in \cite{Hayashi:2019jvx}, which reduces to the perturbative
prepotential in the appropriate gauge-theory chambers and makes
the enhanced global symmetry manifest in suitable Coulomb-branch
variables.

\subsection{Geometric Engineering Dictionary}
\label{sec:geometric-engineering}

Five-dimensional SCFTs can be engineered by considering M-theory on a non-compact Calabi--Yau threefold with a canonical singularity $\cX$. 
An in-depth discussion of the gauge theory phases from geometry can be found in \cite{Morrison:1996xf, Intriligator:1997pq, Hayashi:2014kca, Jefferson:2017ahm, Closset:2018bjz, Apruzzi:2019opn, Apruzzi:2019enx}. Here we focus on some of the salient features relating to the intersection theory. 

Chambers of the extended Coulomb parameter space correspond to K\"ahler cones of the crepant resolutions
\be
\pi:\widetilde{\cX}\longrightarrow\cX
\label{eq:crepant-resolution}
\ee
of $\cX$.
Let $D_A$ denote a basis of divisor classes in 
$\widetilde{\cX}$
and let $\omega_A$ be their dual two-forms. Reduction of the eleven-dimensional three-form gauge field $C_3$ with curvature $G_4=dC_3$ gives
\be
C_3\sim A^A\wedge\omega_A+\cdots\,,
\qquad
G_4\sim F^A\wedge\omega_A+\cdots\,,
\qquad
F^A=dA^A\,.
\label{eq:C3-reduction}
\ee
For compact divisor classes, the corresponding $A^A$ are dynamical gauge fields. In the local non-compact limit, the modes associated with non-compact divisors are interpreted as background flavor fields.

The field-theory parameters $t^A$ are identified with K\"ahler parameters. In the same divisor basis, introduce the divisor-valued K\"ahler parameter
\be
S=t^A D_A\,.
\label{eq:S-expansion}
\ee
The coefficients associated with compact divisor classes are the Coulomb parameters,
\be
t^i=\phi^i\,,
\ee
while the independent coefficients associated with non-compact divisor classes are real mass parameters. 

An M2-brane wrapped on a compact curve $C\subset\widetilde{\cX}$ gives a BPS particle. Its charges and mass are
\be
q_A=D_A\cdot C\,,
\qquad
M_C=\int_CJ\,,
\label{eq:M2-charge-mass}
\ee
where $J$ is the K\"ahler form, with $[J] = -S$. Here and in the following, we will always assume that the intersections $\cdot$ are in the Calabi--Yau threefold, unless otherwise stated. 
A wall of the extended Coulomb parameter space is reached when
\be
\operatorname{Vol}(C)=\int_CJ=0\,.
\ee
When the contraction is floppable, crossing the wall replaces $C$ by the flopped curve and moves to an adjacent K\"ahler chamber \cite{Witten:1996qb}. See \cite{Collinucci:2022rii} for an extension of these gauge-theoretic methods for higher-length flops. 

The five-dimensional Chern--Simons couplings $\int A \wedge F \wedge F$ in
\eqref{eq:effective-lagrangian} arise by dimensional reduction from the eleven-dimensional Chern--Simons coupling
in the eleven-dimensional supergravity describing the low-energy limit of M-theory:
\be
\int C_3\wedge G_4\wedge G_4\,.
\label{eq:11d-CS}
\ee
Therefore, upon reduction on $\widetilde{\cX}$, its coefficients are the triple intersection numbers
\be
\kappa_{ABC}
:=
\int_{\widetilde{\cX}}
\omega_A\wedge\omega_B\wedge\omega_C
=
D_A\cdot D_B\cdot D_C\,.
\label{eq:CS-triple-intersection}
\ee
With the normalization in \eqref{eq:effective-lagrangian}, we obtain
\be
c_{ABC}=\kappa_{ABC}\,,
\label{eq:CS-kappa}
\ee
and so the cubic prepotential is
\be
\mathcal F(t)
=
\frac16\,S^3
=
\frac16\,\kappa_{ABC}t^At^Bt^C\,.
\label{eq:geom-prepotential}
\ee
In particular,
\be
\frac{\partial^3\mathcal F}
{\partial t^A\partial t^B\partial t^C}
=
\kappa_{ABC}
=
D_A\cdot D_B\cdot D_C\,.
\label{eq:prepotential-intersection}
\ee

Only triple intersections containing at least one compact divisor enter the dynamical Coulomb-branch couplings. 
Purely non-compact triple intersections depend on choices at infinity and are not intrinsic to the local Calabi--Yau geometry. The intersection theory developed below will therefore extract the intrinsic triple products containing at least one compact divisor.

\subsection{Toric Calabi--Yau Threefolds and Intersection Numbers}
\label{sec:toric-intersections}

We now recall the toric case, which provides the basic combinatorial model for the intersection rules used later \cite{danilov1978geometry,fulton1993introduction,cox2011toric}, as well as the applications to the M-theory setting \cite{Xie:2017pfl,Closset:2018bjz,Eckhard:2020jyr}.

\subsubsection{Toric Polygon}
\label{sec:TPs}

Let $\Sigma$ be a fan in $N\simeq\mathbb Z^3$ defining a smooth toric Calabi--Yau threefold $X_\Sigma$. Its one-dimensional cones are generated by primitive lattice vectors $\bm v_A\in N$ and correspond to toric divisors $D_A$. A two-dimensional cone $\langle\bm v_A,\bm v_B\rangle$ corresponds to the torus-invariant curve
\be
C_{AB}=D_A\cdot D_B\,,
\ee
and a three-dimensional cone $\langle\bm v_A,\bm v_B,\bm v_C\rangle$ determines the corresponding triple intersection.

The canonical class is
\be
K_{X_\Sigma}=-\sum_A D_A\,.
\ee
The Calabi--Yau condition implies that, after an $SL(3,\mathbb Z)$ transformation, the primitive generators may be written as
\be
\bm v_A=(v_A^1,v_A^2,1)\,.
\ee
The fan can therefore be represented by a lattice polygon in $\mathbb Z^2$. For a smooth crepant resolution the polygon is equipped with a regular unimodular triangulation. Interior lattice points correspond to compact divisors and boundary lattice points to non-compact divisors. The toric divisor classes obey the three linear relations
\be
\sum_A (v_A)^a D_A=0\,,\qquad a=1,2,3\,.
\ee

\subsubsection{Triple Intersection Numbers}

For three distinct divisors one has
\be
D_A\cdot D_B\cdot D_C=
\begin{cases}
1\,,& \langle\bm v_A,\bm v_B,\bm v_C\rangle\text{ is a maximal cone of }\Sigma\,,\\
0\,,& \text{otherwise.}
\end{cases}
\label{eq:toric-distinct-intersections}
\ee

Mixed intersections containing a repeated divisor can be computed on the corresponding toric surface. Suppose that $C_{AB}=D_A\cap D_B$ and that the two triangles adjacent to the edge $[\bm v_A,\bm v_B]$ have third vertices $\bm v_C$ and $\bm v_D$. The primitive rays in the fan of the surface $D_A$ satisfy
\be
(C_{AB}|_{D_A})^2(\bm v_B-\bm v_A)
+(\bm v_C-\bm v_A)+(\bm v_D-\bm v_A)=0\,.
\label{eq:toric-self-int-relation}
\ee
Since
\be
(C_{AB}|_{D_A})^2=D_A\cdot D_B^2\,,
\ee
this determines intersections with two coincident divisor factors. If $C_{AB}\simeq\mathbb P^1$, the Calabi--Yau condition on its normal bundle gives
\be
(C_{AB}|_{D_A})^2+(C_{AB}|_{D_B})^2=-2\,.
\label{eq:normal-bundle-degree}
\ee

Finally, let $D_i$ be a smooth compact toric divisor and let $N_i$ denote the number of vertices adjacent to the corresponding interior vertex of the triangulated polygon. Equivalently, $N_i$ is the number of rays in the two-dimensional fan of the smooth complete toric surface $D_i$. Adjunction in the Calabi--Yau threefold gives
\be
D_i^3=K_{D_i}^2\,.
\ee
For a smooth complete toric rational surface,
\be
\chi(\mathcal O_{D_i})=1\,,\qquad
c_2(D_i)=\chi_{\rm top}(D_i)=N_i\,.
\ee
Noether's formula therefore gives
\be
D_i^3=12-N_i\,.
\label{eq:toric-triple-self-intersection}
\ee
Equivalently, $h^{1,1}(D_i)=N_i-2$ and $D_i^3=10-h^{1,1}(D_i)$.

\subsubsection{Example}

Consider the toric polygon
\be
 \begin{tikzpicture}[scale= 1.2]
    \node[bd] (1) at (0,0) {};
    \node[left, blue] at (-0.1,0.2) {\scriptsize{$1$}};
    \node[bd] (2) at (0,1) {};
    \node[left, blue] at (-0.1,1.2) {\scriptsize{$2$}};
    \node[bd] (3) at (0,2) {};
    \node[left, blue] at (0,2.2) {\scriptsize{$3$}};
    \node[bd] (4) at (-1,-1) {};
    \node[blue,below] at (-1.2,-1) {\scriptsize{$4$}};
    \node[bd] (5) at (0,-1) {};
    \node[blue,below] at (0,-1.2) {\scriptsize{$5$}};
    \node[bd] (6) at (1,-1) {};
    \node[blue,below] at (1.2,-1) {\scriptsize{$6$}};
    \node[bd] (7) at (1,0) {};
    \node[blue] at (1.2,0) {\scriptsize{$7$}};
    \draw (1)--(2)--(3)--(4)--(5)--(6)--(7)--(3) (2)--(4)--(1)--(5) (1)--(6)--(2)--(7);
    \draw[thick,-{Stealth},red] (1)--(2);
    \node[red] at (0.4,0.2) {\scriptsize{$\bm w_{1,2}$}};
    \draw[thick,-{Stealth},red] (1)--(4);
    \node[red] at (-0.35,-0.7) {\scriptsize{$\bm w_{1,4}$}};
    \draw[thick,-{Stealth},red] (1)--(6);
    \node[red] at (0.35,-0.7) {\scriptsize{$\bm w_{1,6}$}};
 \end{tikzpicture}
\ee
There are two compact divisors, $D_1$ and $D_2$. The five boundary vertices correspond to non-compact toric divisors. Because of the toric linear equivalences, only two independent non-compact divisor classes remain after the compact classes are fixed.

For three distinct divisors the non-zero intersections are
\be
D_1D_2D_4=D_1D_4D_5=D_1D_5D_6=D_1D_2D_6
=D_2D_6D_7=D_2D_7D_3=D_2D_3D_4=1\,.
\ee
For the curve $C_{12}=D_1\cap D_2$, the fan of $D_1$ gives
\be
(C_{12}|_{D_1})^2\bm w_{1,2}+\bm w_{1,4}+\bm w_{1,6}=0\,.
\ee
Using
\be
\bm w_{1,2}=(0,1)\,,\qquad
\bm w_{1,4}=(-1,-1)\,,\qquad
\bm w_{1,6}=(1,-1)\,,
\ee
we obtain from \eqref{eq:toric-self-int-relation} that
\be
(C_{12}|_{D_1})^2=D_1D_2^2=2\,.
\ee
Similarly, the remaining repeated intersections are obtained from \eqref{eq:toric-self-int-relation} and \eqref{eq:normal-bundle-degree}.

Finally, $D_1$ has four neighboring vertices and $D_2$ has five. Hence \eqref{eq:toric-triple-self-intersection} gives
\be
D_1^3=12-4=8\,,\qquad
D_2^3=12-5=7\,.
\ee

\section{Brane Webs and Generalized Toric Polygons}
\label{sec:brane-webs-and-GTPs}

We begin by reviewing the realization of five-dimensional SCFTs through 5-brane webs and their correspondence with toric polygons. We then discuss 5-brane webs together with 7-branes, before introducing generalized toric polygons (GTPs), Symington polygons, and $E$-lines.

\subsection{5-Brane Webs and 5d SCFTs}

An alternative realization of a class of five-dimensional SCFTs is provided by webs of $(p,q)$ 5-branes in Type IIB string theory \cite{Aharony:1997ju,Aharony:1997bh,DeWolfe:1999hj}. 
We consider Type IIB string theory on $\mathbb R^{1,9}$, with coordinates $x^0,\ldots,x^9$, where each $(p,q)$ 5-brane extends along $\mathbb R^{1,4}$ and a straight line in the $(x^5,x^6)$ plane, as summarized in Table~\ref{table_web5}. The resulting brane web is represented by a planar graph whose edges have rational slopes and carry positive integral weights. Each edge $e$ has a primitive direction $u_e=(p,q)\in\mathbb Z^2$, specifying the charge of the corresponding 5-brane, and a weight $w_e$ giving its multiplicity. Supersymmetry requires charge conservation at every junction $v$:
\be
\sum_{e\ni v}w_e u_e=0\,,
\ee
where the directions $u_e$ are taken to be outgoing from $v$.

The 5-branes share the five-dimensional worldvolume $\mathbb R^{1,4}$. Contracting all finite edges produces the asymptotic web, whose common junction supports the 5d SCFT. Resolving the web by varying its finite segments describes Coulomb-branch deformations, while changing its asymptotic data describes mass deformations.
We depict brane webs by drawing the $(x^5,x^6)$ plane.

\begin{table}[t]
\centering
\begin{tabular} {|c|c|c|c|c|c|c|c|c|c|c|} 
\hline 
Type IIB & $x^0$ & $x^1$ & $x^2$ & $x^3$ & $x^4$ & $x^5$ & $x^6$ & $x^7$ & $x^8$ & $x^9$\\
\hline
$(p,q)$ 5-brane & $\times$ & $\times$ & $\times$ & $\times$ & $\times$ & \multicolumn{2}{c|}{\footnotesize $(p,q)$-line} & & & \\
\hline
\end{tabular}
\caption{Spacetime orientations of $(p,q)$ 5-branes in Type IIB string theory.}
\label{table_web5}
\end{table}

\subsection{Toric Polygons and Calabi--Yau Threefolds}

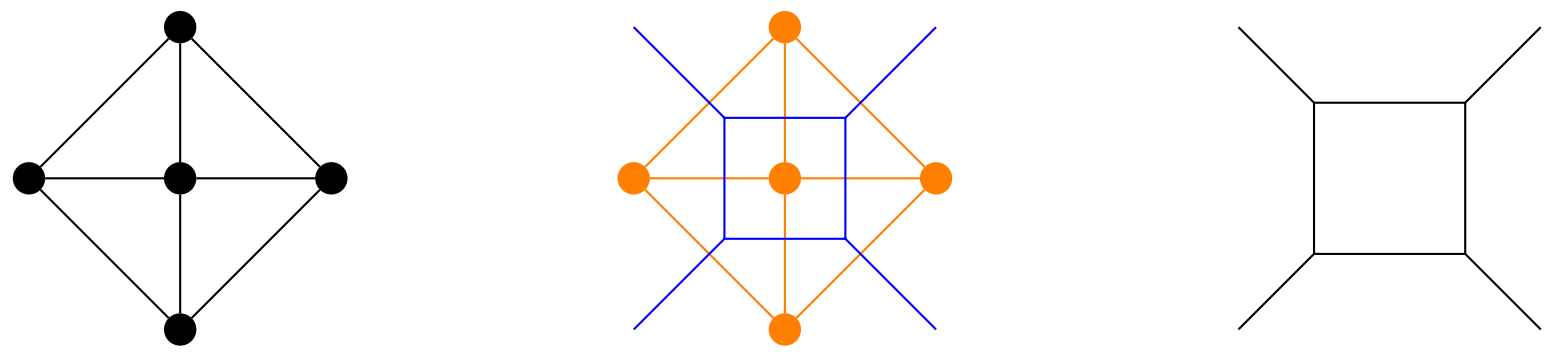
\begin{figure}[htbp]
    \centering
    \begin{tikzpicture}
        \node at (0,0) {\scalebox{1}{\begin{tikzpicture}
            \node[bd] (1) at (0,0) {};
            \node[bd] (2) at (0,1) {};
            \node[bd] (3) at (-1,0) {};
            \node[bd] (4) at (0,-1) {};
            \node[bd] (5) at (1,0) {};
            \draw (1)--(2)--(3)--(4)--(5)--(2)
                  (3)--(1)--(5) (1)--(4);
        \end{tikzpicture}}};
        \node at (0,-2) {(a)};

        \node at (4,0) {\scalebox{1}{\begin{tikzpicture}
            \node[od] (1) at (0,0) {};
            \node[od] (2) at (0,1) {};
            \node[od] (3) at (-1,0) {};
            \node[od] (4) at (0,-1) {};
            \node[od] (5) at (1,0) {};
            \draw[orange] (1)--(2)--(3)--(4)--(5)--(2)
                          (3)--(1)--(5) (1)--(4);
            \draw[blue]
                (-0.4,0.4)--(-0.4,-0.4)--(0.4,-0.4)
                --(0.4,0.4)--(-0.4,0.4)
                (-1,1)--(-0.4,0.4)
                (-1,-1)--(-0.4,-0.4)
                (1,-1)--(0.4,-0.4)
                (1,1)--(0.4,0.4);
        \end{tikzpicture}}};
        \node at (4,-2) {(b)};

        \node at (8,0) {\scalebox{1}{\begin{tikzpicture}
            \draw (-0.5,0.5)--(-0.5,-0.5)
                  --(0.5,-0.5)--(0.5,0.5)--(-0.5,0.5);
            \draw (-1,1)--(-0.5,0.5);
            \draw (-1,-1)--(-0.5,-0.5);
            \draw (1,-1)--(0.5,-0.5);
            \draw (1,1)--(0.5,0.5);
        \end{tikzpicture}}};
        \node at (8,-2) {(c)};
    \end{tikzpicture}

    \caption{(a) A toric polygon.
    (b) The corresponding dual planar graph, with the toric
    polygon shown in orange and the dual graph in blue.
    (c) The corresponding 5-brane web.}
    \label{fig:TPdualBW}
\end{figure}

5-brane webs admit a dual combinatorial description in terms of toric
polygons. A toric polygon is a convex lattice polygon whose edges
are perpendicular to the external 5-branes, with lattice lengths encoding
their multiplicities. More generally, a polyhedral decomposition of the
polygon determines a dual planar graph, in which vertices correspond to
two-dimensional cells, edges to edges, and faces to vertices of the
decomposition. This dual graph describes the associated 5-brane web,
with internal edges corresponding to finite 5-brane segments.
In particular, unimodular triangulations correspond to fully resolved
webs, while coarser subdivisions describe partial resolutions.
The correspondence between a toric polygon and its dual brane web
is illustrated in Figure~\ref{fig:TPdualBW}.

\begin{figure}[htbp]
    \centering
    \scalebox{0.7}{\begin{tikzpicture}
        \node at (0,0) {(a)};
        \node at (2,0) {\scalebox{1}{\begin{tikzpicture}
            \node[bd] (1) at (0,0) {};
            \node[bd] (2) at (0,1) {};
            \node[bd] (3) at (0,2) {};
            \node[bd] (4) at (-1,-1) {};
            \node[bd] (5) at (0,-1) {};
            \node[bd] (6) at (1,-1) {};
            \node[bd] (7) at (1,0) {};
            \draw (1)--(2)--(3)--(4)--(5)--(6)--(7)--(3)
                  (2)--(4)--(1)--(5) (1)--(6) (2)--(7);
            \draw[darkgreen,thick] (2)--(6);
        \end{tikzpicture}}};
        \node at (10,0) {\scalebox{0.7}{\begin{tikzpicture}
            \coordinate (1) at (0,0);
            \coordinate (2) at (8,-5);
            \coordinate (3) at (9,-5);
            \coordinate (4) at (13,-2);
            \coordinate (5) at (15,0);
            \draw (1)--(3,-1)--(7,-3)--(8,-4)--(2)
                  (8,-4)--(9,-4)--(3)
                  (9,-4)--(10,-3)
                  (12,-2)--(4)
                  (12,-2)--(13,-1)--(5)
                  (13,-1)--(3,-1)
                  (7,-3)--(10,-3);
            \draw[darkgreen,thick] (10,-3)--(12,-2);
        \end{tikzpicture}}};

        \node at (0,-5) {(b)};
        \node at (2,-5) {\scalebox{1}{\begin{tikzpicture}
            \node[bd] (1) at (0,0) {};
            \node[bd] (2) at (0,1) {};
            \node[bd] (3) at (0,2) {};
            \node[bd] (4) at (-1,-1) {};
            \node[bd] (5) at (0,-1) {};
            \node[bd] (6) at (1,-1) {};
            \node[bd] (7) at (1,0) {};
            \draw (1)--(2)--(3)--(4)--(5)--(6)--(7)--(3)
                  (2)--(4)--(1)--(5) (1)--(6) (2)--(7);
        \end{tikzpicture}}};
        \node at (10,-5) {\scalebox{0.7}{\begin{tikzpicture}
            \coordinate (1) at (0,0);
            \coordinate (2) at (8,-5);
            \coordinate (3) at (9,-5);
            \coordinate (4) at (11,-3);
            \coordinate (5) at (14,0);
            \draw (1)--(3,-1)--(7,-3)--(8,-4)--(2)
                  (8,-4)--(9,-4)--(3)
                  (9,-4)--(10,-3)--(4)
                  (10,-3)--(12,-1)--(5)
                  (12,-1)--(3,-1)
                  (7,-3)--(10,-3);
        \end{tikzpicture}}};

        \node at (0,-10) {(c)};
        \node at (2,-10) {\scalebox{1}{\begin{tikzpicture}
            \node[bd] (1) at (0,0) {};
            \node[bd] (2) at (0,1) {};
            \node[bd] (3) at (0,2) {};
            \node[bd] (4) at (-1,-1) {};
            \node[bd] (5) at (0,-1) {};
            \node[bd] (6) at (1,-1) {};
            \node[bd] (7) at (1,0) {};
            \draw (1)--(2)--(3)--(4)--(5)--(6)--(7)--(3)
                  (2)--(4)--(1)--(5) (1)--(6) (2)--(7);
            \draw[red,thick] (1)--(7);
        \end{tikzpicture}}};
        \node at (10,-10) {\scalebox{0.7}{\begin{tikzpicture}
            \coordinate (1) at (0,0);
            \coordinate (2) at (7,-5);
            \coordinate (3) at (8,-5);
            \coordinate (4) at (10,-3);
            \coordinate (5) at (12,0);
            \draw (1)--(3,-1)--(5,-2)--(7,-4)--(2)
                  (7,-4)--(8,-4)--(3)
                  (8,-4)--(9,-3)--(4)
                  (9,-3) (9,-2)--(10,-1)--(5)
                  (10,-1)--(3,-1)
                  (5,-2)--(9,-2);
            \draw[red,thick] (9,-3)--(9,-2);
        \end{tikzpicture}}};
    \end{tikzpicture}}
    \caption{A flop transition in the toric polygon and its dual
    5-brane web. (a) A fully triangulated toric polygon and its
    dual web, with the curve to be contracted highlighted in green.
    (b) The configuration after contraction of the green curve.
    (c) The resulting resolution after the flop, with the new
    curve highlighted in red.}
    \label{fig:TPdualBWflop}
\end{figure}
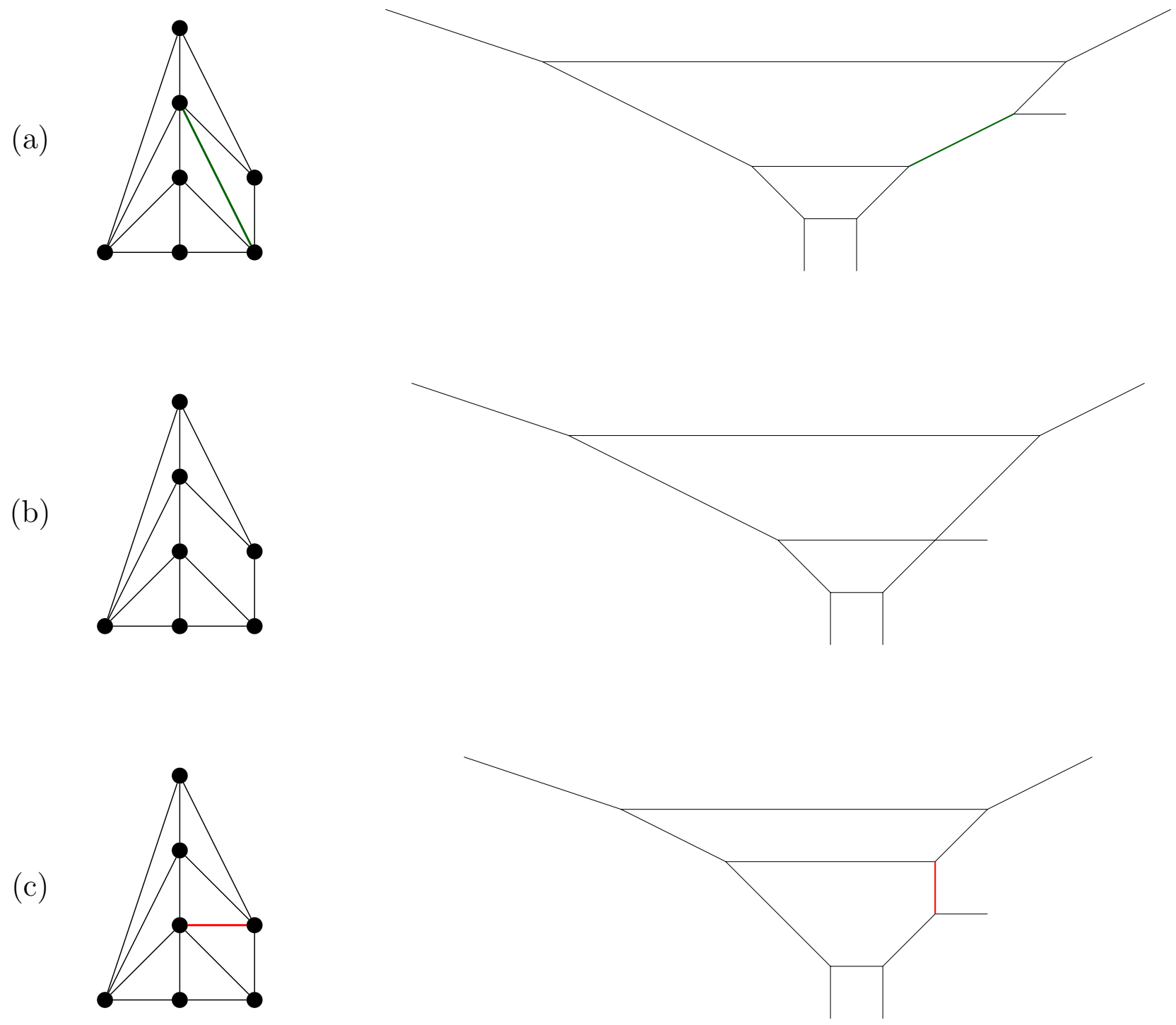

As reviewed in Section~\ref{sec:TPs}, toric polygons also encode toric
Calabi--Yau threefolds, with their polyhedral decompositions describing
partial or complete crepant resolutions. Combining the two descriptions
gives a direct correspondence between the brane web and the geometry:
bounded faces of the web correspond to compact divisors, while unbounded
faces correspond to non-compact divisors. Finite 5-brane segments
correspond to compact curves, whose contraction describes the shrinking
of the corresponding curves in the Calabi--Yau threefold.
Different unimodular triangulations are related by flops, realized in
the web by contracting and re-expanding finite 5-brane segments,
as illustrated in Figure~\ref{fig:TPdualBWflop}.
The five-dimensional SCFT engineered by M-theory on the toric
Calabi--Yau threefold coincides with the theory supported on the dual
5-brane web in Type IIB string theory
\cite{Aharony:1997bh,Leung:1997tw}.

\subsection{5-Brane Webs with 7-Branes}
\label{sec_webs57}

More general Type IIB configurations allow the external $(p,q)$ 5-branes
to end on $[p,q]$ 7-branes. A $[p,q]$ 7-brane occupies the directions
$x^0,\ldots,x^4,x^7,x^8,x^9$,
and is localized at a point in the $(x^5, x^6)$ plane, as summarized in Table \ref{tab:brane-orientations}.
Supersymmetry requires a $(p,q)$ 5-brane ending on a $[p,q]$ 7-brane to have identical $(p,q)$ charge. Several 5-branes of the same charge may
end on the same 7-brane
\cite{Hanany:1996ie,Benini:2009gi,vanBeest:2020kou,Bergman:2020myx}.

\begin{table}[htbp]
\centering
\begin{tabular}{|c|c|c|c|c|c|c|c|c|c|c|}
\hline
Type IIB & $x^0$ & $x^1$ & $x^2$ & $x^3$ & $x^4$ & $x^5$ & $x^6$ & $x^7$ & $x^8$ & $x^9$ \\
\hline
$(p,q)$ 5-brane & $\times$ & $\times$ & $\times$ & $\times$ & $\times$ & \multicolumn{2}{c|}{\footnotesize $(p,q)$-line} & & & \\
\hline
$[p,q]$ 7-brane & $\times$ & $\times$ & $\times$ & $\times$ & $\times$ & & & $\times$ & $\times$ & $\times$ \\
\hline
\end{tabular}
\caption{Spacetime orientations of $(p,q)$ 5-branes and $[p,q]$ 7-branes in Type IIB string theory.}
\label{tab:brane-orientations}
\end{table}

A $[p,q]$ 7-brane carries an $SL(2,\mathbb Z)$ monodromy, represented by a cut with matrix
\begin{equation}
    M_{[p,q]}=\begin{pmatrix}
                1-pq & p^2\\
                -q^2 & 1+pq
            \end{pmatrix}\;, 
\end{equation}
crossing the cut counterclockwise. 
Across the cut, the charge of a 5-brane is transformed by the monodromy:
\begin{equation}
    \begin{tikzpicture}
            \draw (0,0)--(2,0)--(2,-2);
            \node[seven,label=below:{\scriptsize$[p,q]$}] (1) at (1,-1) {};
            \draw[dash dot] (1)--(3,1);
            \node at (1,0.5) {\scriptsize$(r,s)$};
            \node at (3,-1) {\scriptsize$M_{[p,q]}.(r,s)$};
            \node at (4,1.3) {\scriptsize$[p,q]$ monodromy cut};
            \draw[dash dot] (3.6,1.6)--(4,2);
        \end{tikzpicture}\;.
\end{equation}
In figures, we generally suppress cuts that extend to infinity without crossing the 5-brane web.

\begin{figure}[htbp]
\centering
\scalebox{0.8}{\begin{tikzpicture}[scale=1.5]
        \node[bd] (0) at (0,0) {};
        \node[bd] (1) at (0,1) {};
        \node[bd] (2) at (0,2) {};
        \node[bd] (5) at (0,3) {};
        \node[bd] (3p) at (-1,-1) {};
        \node[bd] (w) at (0,-1) {};
        \node[bd] (3pp) at (1,-1) {};
        \node[bd] (4) at (1,0) {};
        \draw (0)--(1)--(2)--(3p)--(w)--(3pp)--(4)--(2) (1)--(3p)--(0)--(w) (0)--(3pp)--(1) (2)--(3pp) (3p)--(5)--(4) (2)--(5);
    \end{tikzpicture}}
\qquad 
    \scalebox{0.4}{\begin{tikzpicture}[scale=1]
        \node[seven] (1) at (0,0) {};
        \node[seven] (2) at (13,-6) {};
        \node[seven] (3) at (15,-7) {};
        \node[seven] (4) at (23,-2) {};
        \node[seven] (5) at (26,0) {};
        \draw (1)--(4,-1)--(10,-3)--(12,-4)--(13,-5)--(2) (13,-5)--(15,-5)--(3) (15,-5)--(16,-4)--(18,-3)--(21,-2)--(4) (21,-2)--(23,-1)--(5) (4,-1)--(23,-1) (10,-3)--(18,-3) (12,-4)--(16,-4);
    \end{tikzpicture}}
\caption{A toric polygon and its dual 5-brane web, with all external 5-branes ending on 7-branes. This web contains in particular parallel external 5-branes ending on distinct 7-branes.}
\label{fig:BW7initial}
\end{figure}
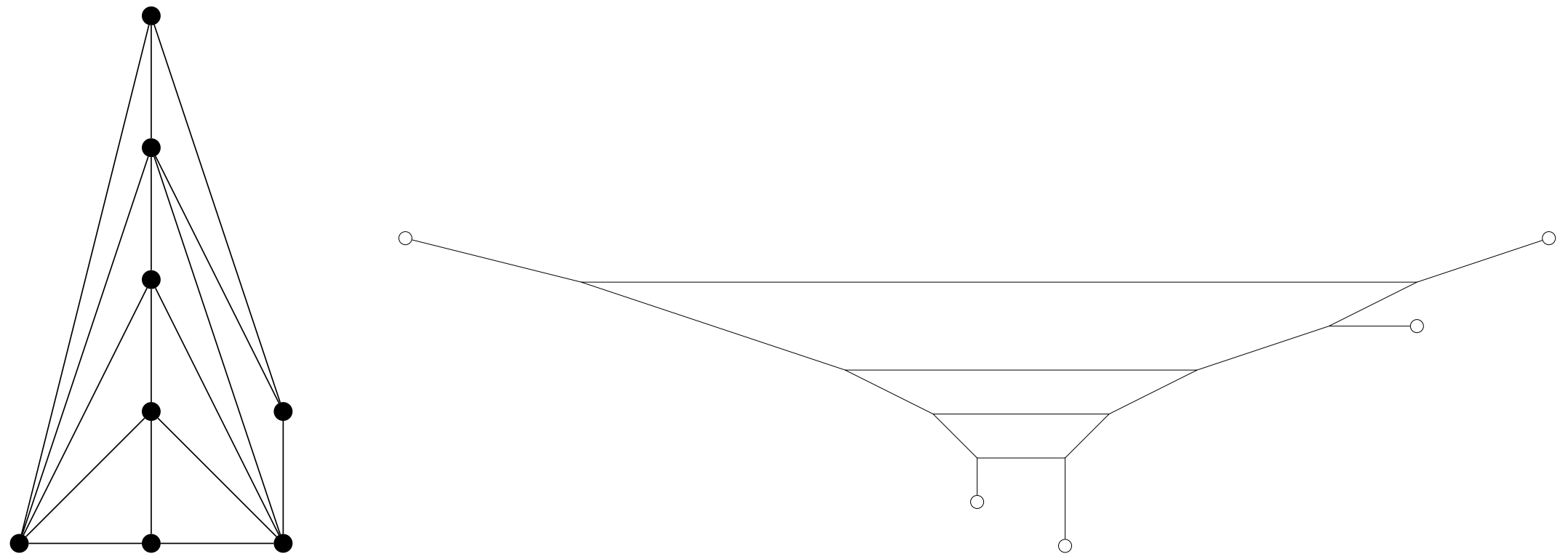

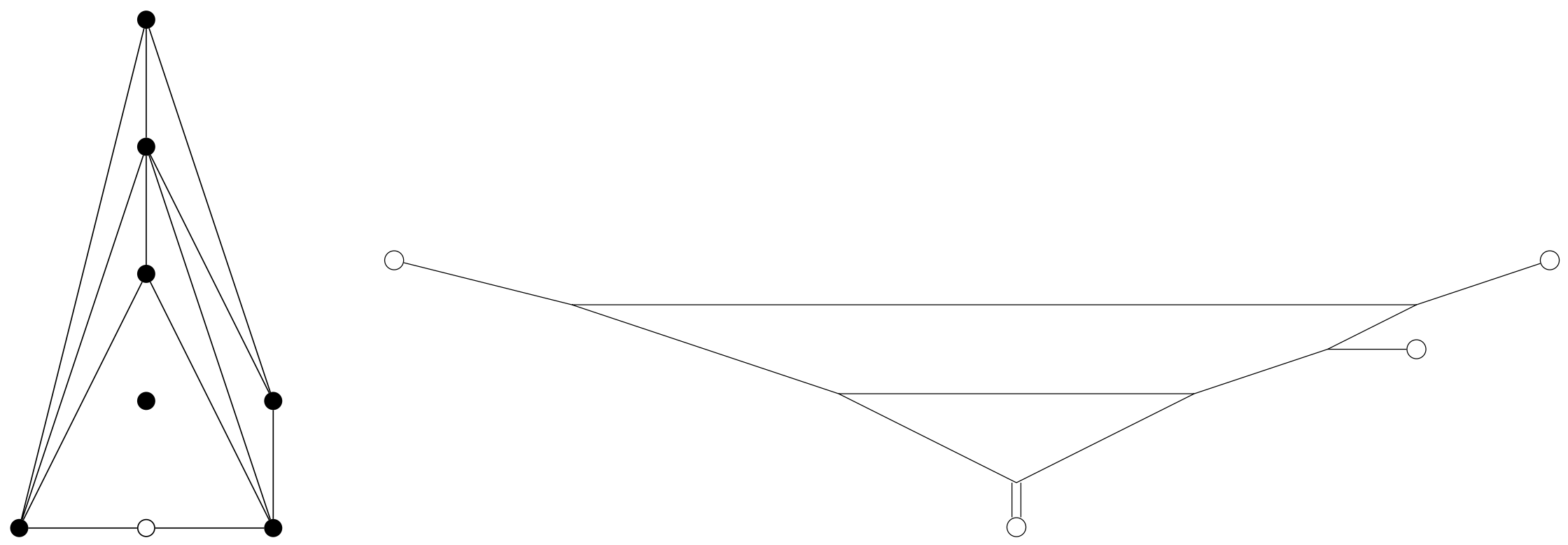
\begin{figure}[htbp]
\centering
\scalebox{0.8}{\begin{tikzpicture}[scale=1.5]
    \node[bd] (0) at (0,0) {};
        \node[bd] (1) at (0,1) {};
        \node[bd] (2) at (0,2) {};
        \node[bd] (5) at (0,3) {};
        \node[bd] (3p) at (-1,-1) {};
        \node[wd] (w) at (0,-1) {};
        \node[bd] (3pp) at (1,-1) {};
        \node[bd] (4) at (1,0) {};
        \draw (1)--(2)--(3p)--(w)--(3pp)--(4)--(2) (1)--(3p) (3pp)--(1) (2)--(3pp) (3p)--(5)--(4) (2)--(5);
    \end{tikzpicture}}
\qquad 
    \scalebox{0.6}{\begin{tikzpicture}[scale=0.7]
        \node[seven] (1) at (0,0) {};
        \node[seven] (2) at (14,-6) {};
        \node[seven] (4) at (23,-2) {};
       \node[seven] (5) at (26,0) {};
        \draw (1)--(4,-1)--(10,-3)--(14,-5)  (14,-5)--(18,-3)--(21,-2)--(4) (21,-2)--(23,-1)--(5) (4,-1)--(23,-1) (10,-3)--(18,-3);
        \draw[transform canvas={xshift=-2pt, yshift=0pt}] (14,-5)--(2);
       \draw[transform canvas={xshift=2pt, yshift=0pt}] (14,-5)--(2);
    \end{tikzpicture}}
\caption{A toric polygon and its dual 5-brane web, with all external 5-branes ending on 7-branes. The white lattice point on the boundary of the polygon indicates that the dual web contains parallel external 5-branes ending on the same 7-brane. }
\label{fig:BW7partial}
\end{figure}

In what follows, we consider 5-brane webs whose external legs can end on 7-branes, and assume that they are maximally resolved subject to these endpoint conditions. When several 5-branes end on the same 7-brane, such a configuration need not be maximally resolved as a web of 5-branes alone, since the endpoint constraints may obstruct certain resolutions. This distinction is illustrated in Figures~\ref{fig:BW7initial}--\ref{fig:BW7partial}.

\begin{figure}[htbp]
\centering
\begin{tikzpicture}
        \node at (0,0) {\scalebox{1}{\begin{tikzpicture}[scale=1]
            \node[seven] (1) at (0,0) {};
            \draw (-1,2)--(0,1)--(1,2);
            \node at (-1.1,1.5) {\scriptsize$(1,-1)$};
            \node at (1,1.5) {\scriptsize$(1,1)$};
            \draw[transform canvas={xshift=-2pt, yshift=0pt}] (0,1)--(1);
            \draw[transform canvas={xshift=2pt, yshift=0pt}] (0,1)--(1);
        \end{tikzpicture}}};
        \node at (0,-1.5) {\color{red}breaks s-rule};
        \node at (6,0) {\scalebox{1}{\begin{tikzpicture}[scale=1]
            \node[seven] (1) at (0,0) {};
            \draw (-2,2)--(0,1)--(2,2);
            \node at (-1.7,1.4) {\scriptsize$(2,-1)$};
            \node at (1.7,1.4) {\scriptsize$(2,1)$};
            \draw[transform canvas={xshift=-2pt, yshift=0pt}] (0,1)--(1);
            \draw[transform canvas={xshift=2pt, yshift=0pt}] (0,1)--(1);
        \end{tikzpicture}}};
        \node at (6,-1.5) {\color{darkgreen}obeys s-rule};
    \end{tikzpicture}
\caption{Two configurations of parallel 5-branes ending on a common 7-brane, violating and satisfying the s-rule, respectively. In the configuration on the left, pulling the 7-brane through the junction would require a negative number of 5-branes, whereas in the configuration on the right, the 7-brane can be pulled through the junction and becomes detached from the 5-branes.}
\label{fig:BW7srule}
\end{figure}
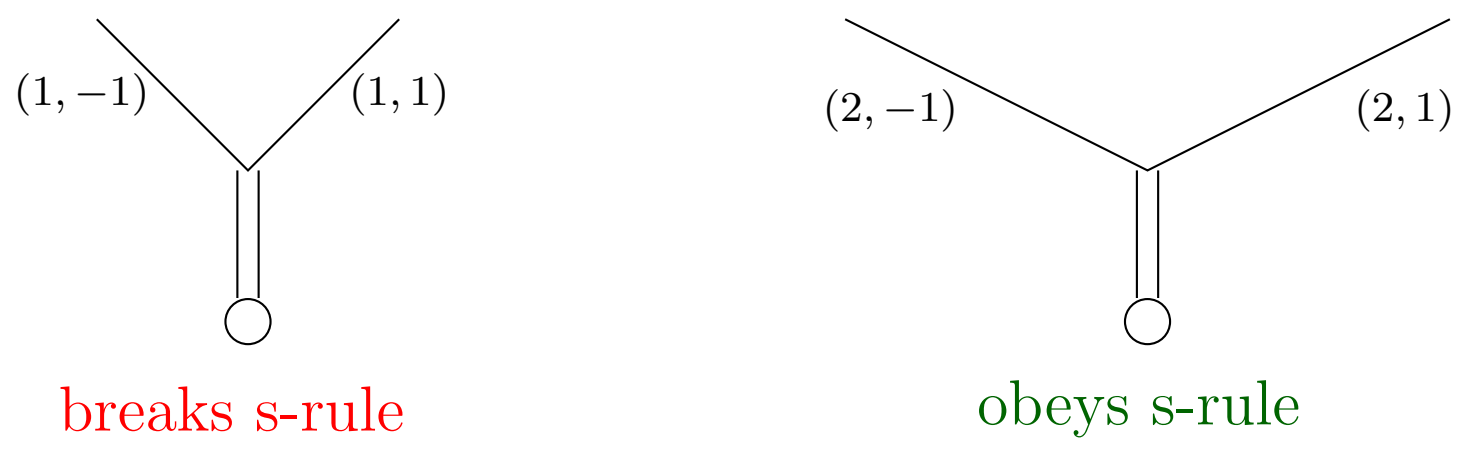

In general, configurations of 5-branes ending on 7-branes need not preserve supersymmetry and therefore need not engineer 5d SCFTs. In addition to charge conservation at junctions, supersymmetry imposes further constraints, notably the s-rule \cite{Hanany:1996ie,Benini:2009gi,vanBeest:2020kou,Bergman:2020myx}, illustrated in Figure~\ref{fig:BW7srule}. Following \cite{AABwebs}, we consider a stronger notion of \emph{consistency}, formulated in terms of the successive pulling of 7-branes into the web, as discussed below. A local version of this has recently been discussed in \cite{Franco:2026gnm}. 

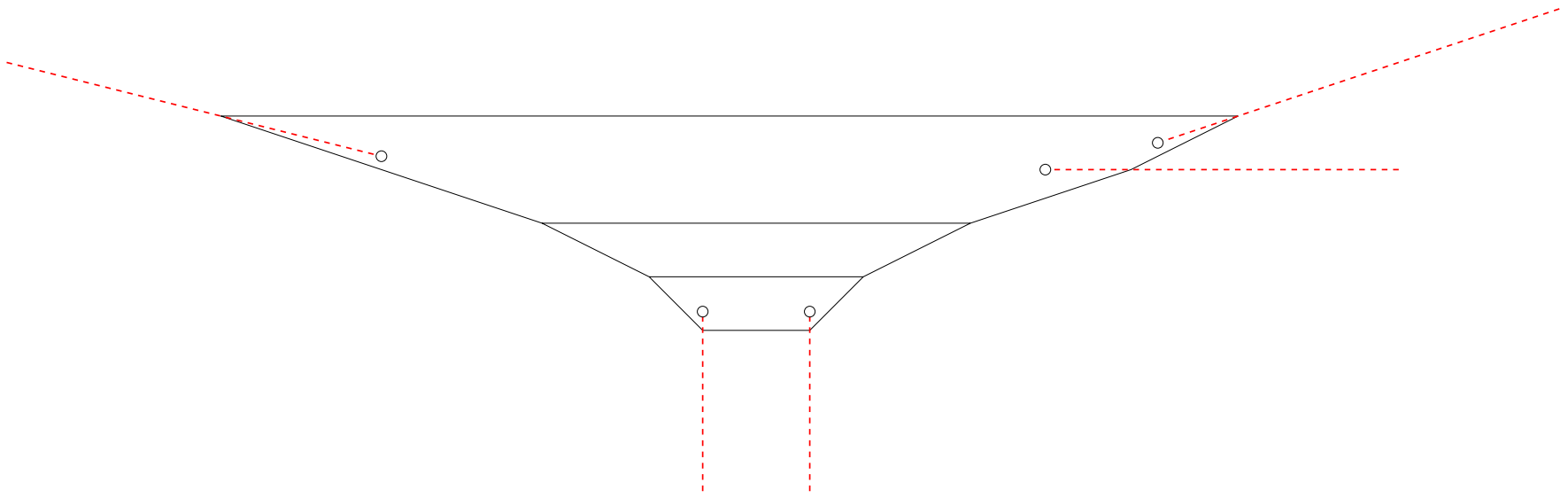
\begin{figure}[htbp]
\centering
\scalebox{0.4}{\begin{tikzpicture}[scale=1]
    \draw[black]
        (4,-1)--(10,-3)--(12,-4)--(13,-5)
        (13,-5)--(15,-5)--(16,-4)--(18,-3)--(21,-2)
        (21,-2)--(23,-1)
        (4,-1)--(23,-1)
        (10,-3)--(18,-3)
        (12,-4)--(16,-4);

    \draw[red,dashed,thick] (0,0)--(7,-1.75);
    \node[wd] at (7,-1.75) {};

    \draw[red,dashed,thick] (13,-8)--(13,-4.65);
    \node[wd] at (13,-4.65) {};
    \draw[red,dashed,thick] (15,-8)--(15,-4.65);
    \node[wd] at (15,-4.65) {};

    \draw[red,dashed,thick] (26,-2)--(19.4,-2);
    \node[wd] at (19.4,-2) {};

    \draw[red,dashed, thick] (29,1)--(21.5,-1.5);
    \node[wd] at (21.5,-1.5) {};
\end{tikzpicture}}
\caption{The 5-brane web of Figure \ref{fig:BW7initial} after all the 7-branes have been pulled into bounded faces.}
\label{fig:pushingin7branes}
\end{figure}

\begin{figure}[htbp]
\centering
\scalebox{0.6}{\begin{tikzpicture}[scale=0.7]
    \draw[black]
        (4,-1)--(10,-3)--(14,-5)
        (14,-5)--(18,-3)--(21,-2)--(23,-1)
        (4,-1)--(23,-1)
        (10,-3)--(18,-3);

    \draw[red,dashed,thick] (0,0)--(7,-1.75);
    \node[wd] at (7,-1.75) {};

    \draw[red,dashed,thick] (14,-7.5)--(14,-4.25);
    \node[wd] at (14,-4.25) {};

    \draw[red,dashed,thick] (23,-2)--(19.4,-2);
    \node[wd] at (19.4,-2) {};

    \draw[red,dashed,thick] (26,0)--(21.2,-1.6);
    \node[wd] at (21.2,-1.6) {};
\end{tikzpicture}}
\caption{The 5-brane web of Figure \ref{fig:BW7partial} after the 7-branes have been pulled into bounded faces.}
\label{fig:pushingin7branes2}
\end{figure}
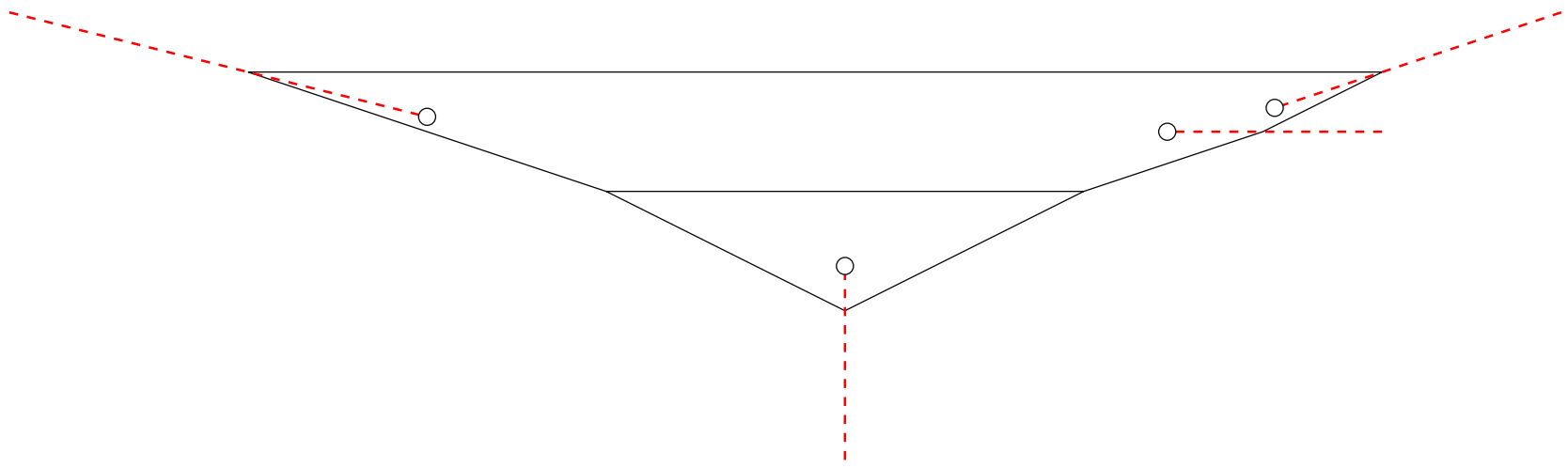

Starting from a configuration in which the external 5-branes end on 7-branes, we move a given $[p,q]$ 7-brane along the direction of the 5-branes attached to it, towards the interior of the web. When the 7-brane reaches a junction, it can be pulled through it by applying a local Hanany--Witten transition, as detailed in \cite[\S 4.5]{AABwebs}. This operation changes the local configuration of 5-branes, while preserving charge conservation. In this way, the 7-brane can pass through
successive junctions, changing the combinatorial structure of the web. By repeating such operations, a 7-brane can be pushed
through several junctions. Eventually, it may become detached from all 5-branes and lie in a bounded face of the web. Although it no longer appears as an endpoint of a 5-brane, the 7-brane still carries its $SL(2,\mathbb Z)$ monodromy, represented by a cut emanating from its position, as illustrated in Figures \ref{fig:pushingin7branes} and \ref{fig:pushingin7branes2}.

Following \cite{AABwebs}, we say that a 5-brane web with 7-branes is
\emph{consistent} if there exists a sequence of pulling operations, such that the resulting
configuration satisfies the following properties:
\begin{enumerate}
    \item All 7-branes lie inside bounded faces of the web, with no
    5-branes ending on them.
    \item Every vertex of the resulting 5-brane web is trivalent and
    has multiplicity one, equivalently, it is dual to a lattice
    triangle of normalized area one.
\end{enumerate}

For example, Figures \ref{fig:pushingin7branes} and \ref{fig:pushingin7branes2} show consistent 5-brane webs with pulled 7-branes satisfying the above properties.

Not every 5-brane web is consistent. Indeed, the possibility
of pulling a 7-brane through a junction without creating negative numbers of 5-branes depends on the charges
and multiplicities of the adjacent 5-branes, and pulling one
7-brane can affect the subsequent motion of the others.
Consistency is therefore a global condition on the web,
stronger than charge conservation at individual junctions
or the local s-rule. In what follows, we restrict to
consistent 5-brane webs.

\subsection{Generalized Toric Polygons}
\label{sec:GTP}

We now explain how the correspondence between 5-brane webs and toric polygons extends to the case of 5-brane webs with 7-branes. 

Consider a consistent 5-brane web with 7-branes, as reviewed above in Section \ref{sec_webs57}. If we temporarily disregard the 7-brane endpoints and extend
the attached external 5-branes to infinity, the resulting web is dual to a toric lattice
polygon $P$ with a regular polyhedral decomposition $\cP$. A vertex
of the web is dual to a two-dimensional cell of $\cP$, and a 5-brane
segment is dual to an edge of $\cP$. In particular, the external
5-brane segments are dual to edges of $\cP$ contained in the boundary
of $P$. The lattice length of each such edge is the number of parallel
5-branes represented by the corresponding segment of the web.

The endpoints on 7-branes supply additional data on these boundary
edges. For each 7-brane $\alpha$, denote by $L_\alpha$ the boundary edge of
$\cP$ dual to the 5-branes ending on it. If $L_\alpha$ has lattice length $m_\alpha$, it
represents $m_\alpha$ parallel 5-branes ending on the same 7-brane. We draw
the $m_\alpha-1$ interior lattice points of $L_\alpha$ in white, with its two
endpoints in black. The other lattice points of the polygon are drawn
in black -- see Figure \ref{fig:BW7partial} for an example. This is the familiar white-dot convention for generalized toric polygons \cite{Benini:2009gi, Bourget:2023wlb}. 
An edge $L_\alpha$ of lattice length one has no interior lattice point to color.

Now fix a sequence of pullings of the 7-branes as in the consistency
condition of Section~\ref{sec_webs57}.
Label the 7-branes by $\alpha$ and color by $\alpha$
each cell of $\cP$ dual to a vertex of the original web through
which the 7-brane $\alpha$ passes during the pulling.  The possible
colored cells are elementary lattice trapezoids and elementary
lattice triangles, as defined in \cite[\S 4.5]{AABwebs}. 
An elementary trapezoid has two parallel edges of lattice lengths $a$ and $b$, with $a>b$, separated by a lattice distance of $a-b$, and two remaining edges of lattice length one.
An elementary triangle has a distinguished base of lattice
length $a$, two remaining edges of lattice length one, and lattice
height $a$ above the base. These are the cells dual to the vertices
across which a 7-brane can be pulled. We refer to
\cite[\S\S 4.5--4.6]{AABwebs} for a detailed description of the corresponding local operations on
the web.

For each $\alpha$, the $\alpha$-colored cells form a consecutive strip
beginning at $L_\alpha$. The strip consists of zero or more elementary
trapezoids, followed by a single elementary triangle
\cite[\S 5.1]{AABwebs}. The edge of the first colored cell facing
the boundary is $L_\alpha$. 
In each trapezoid the edge opposite this
incoming edge is shorter and serves as the incoming edge of the
next cell. The final triangle has no opposite edge: the strip ends
at its third vertex, which we denote by $V_\alpha$. Under duality with
the web, $V_\alpha$ corresponds to the bounded face in which the
7-brane $\alpha$ lies after being fully pulled in. The entire $\alpha$-colored
strip records the vertices of the initial web that have disappeared
during this pulling. Different colored strips contain no common
two-dimensional cells, although their tips $V_\alpha$ may coincide.

All cells of $\cP$ that are not colored are lattice triangles of
normalized area one, or equivalently Euclidean area $1/2$ in a
lattice basis. These cells correspond to the trivalent vertices of
multiplicity one that remain in the web with pulled-in 7-branes. 
In particular,
the distinction is between the uncolored unit triangles, whose
dual web vertices remain, and the elementary triangles and
trapezoids colored by the 7-branes, whose dual vertices have been
removed. Existence of a coloring with precisely these properties
is the polygon counterpart of the consistency condition of
Section~\ref{sec_webs57}. 
The precise polygon
criterion is given in \cite[\S 5.1]{AABwebs}.

\begin{figure}[htbp]
\centering
\begin{tikzpicture}
    \node at (3,-3.5) {\scalebox{1}{\begin{tikzpicture}
        \node[seven] (3) at (0.1,-0.1) {};

        \draw (-0.5,0.5)--(-0.5,-0.5)
              --(0.5,-0.5)--(0.5,0.5)--(-0.5,0.5)
              (-1,1)--(-0.5,0.5)
              (-1,-1)--(-0.5,-0.5)
              (1,1)--(0.5,0.5);

        \draw[red,dashed,thick]
              (3)--(1,-1);
    \end{tikzpicture}}};

    \node at (8,-3.5) {\scalebox{1}{\begin{tikzpicture}
        \fill[red]
             (0,0)--(0,-1)--(1,0)--cycle;
        \node[bd] (1) at (0,0) {};
        \node[bd] (2) at (0,1) {};
        \node[bd] (3) at (-1,0) {};
        \node[bd] (4) at (0,-1) {};
        \node[bd] (5) at (1,0) {};
        \draw (1)--(2)--(3)--(4)--(5)--(2)
              (3)--(1)--(5) (1)--(4);
    \end{tikzpicture}}};
\end{tikzpicture}

\caption{A 7-brane pulled into a bounded face of the
web (left), with its monodromy cut indicated.
The dual GTP (right) shows the colored triangular
region dual to the vertex of the original web that disappears after the 7-brane is pulled.}
\label{fig:BW7pull}
\end{figure}
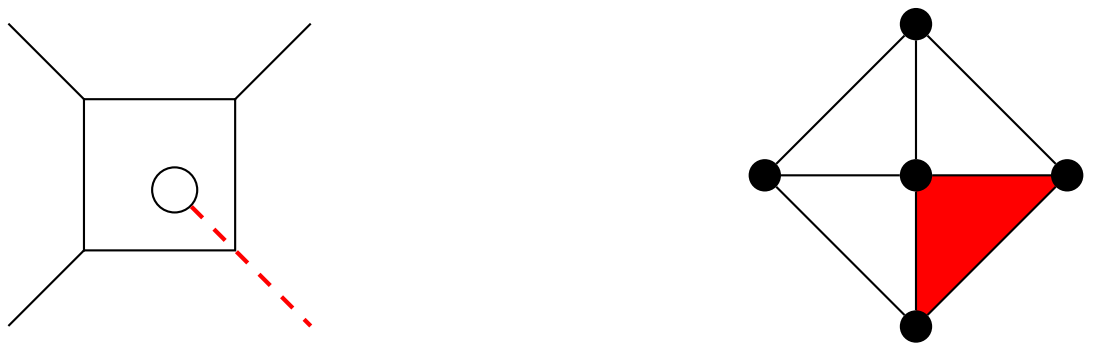

We call the resulting data---the toric polygon $P$, its chosen
regular polyhedral decomposition $\cP$, the boundary edges
$L_\alpha$, and the coloring of its cells determined by a complete
pulling sequence---a \emph{generalized toric polygon}, or \emph{GTP}.
In \cite{AABwebs}, this structure is 
called a decorated toric polygon equipped with a consistent regular polyhedral decomposition. 
In this paper, we use the shorter name GTP
for this data. A simple example is given in Figure \ref{fig:BW7pull}.

\subsection{Symington Polygons and $E$-lines}
\label{sec:symington-E-lines}

Starting from a GTP $(P,\cP)$ as defined in Section~\ref{sec:GTP}, we now
construct an integral-affine manifold with singularities, called the
\emph{Symington polygon} $\widetilde P$ associated with  $(P,\cP)$ \cite[\S 5.2]{AABwebs}.

Recall that, for each 7-brane $\alpha$, the $\alpha$-colored cells form a
strip beginning at the marked boundary edge $L_\alpha$: a possibly empty
sequence of elementary trapezoids followed by a single elementary
triangle with tip $V_\alpha$. 
We remove the interior of this strip,
together with $L_\alpha$, and identify the two lateral edges of each
colored cell, starting at $L_\alpha$ and proceeding towards $V_\alpha$.
The lateral edges are lattice segments of length one, so each
pair can be identified while preserving its lattice subdivision.
At the final triangle the two sides meeting at $V_\alpha$ are identified,
leaving the image of $V_\alpha$ as a point of the resulting polygon.
We carry out this operation for every colored strip. In the dual
brane web, the colored cells correspond to the junctions through
which the 7-branes passed during the pulling procedure. Removing
them eliminates precisely those junctions, while the edge
identifications reproduce the adjacencies of the fully pushed web.

The identifications also specify the integral-affine structure of
the resulting polygon. On each remaining triangle we retain the
usual lattice-affine coordinates. Across a pair of identified
edges, these coordinates are related by an affine transformation
whose linear part belongs to $SL(2,\mathbb Z)$. For each colored
cell this transformation preserves the lattice direction parallel
to its incoming edge and identifies the primitive directions of
its two lateral edges. Along a strip the transformations are
composed in the order of the successive gluings. Thus the
remaining triangles fit together by integral-affine coordinate
changes, except possibly at the images of the tips $V_\alpha$.
Their union $\widetilde P$ with these identifications is an integral-affine manifold with singularities at the vertices $V_\alpha$, and is called a Symington polygon.

The uncolored cells of $\cP$ are triangles of normalized lattice
area one. They survive
the cutting and gluing, and so their images give a
unimodular triangulation $\tcP$ of $\widetilde P$. This
triangulation is dual to the web after all 7-branes have been
pulled in: faces of the web correspond to vertices of $\tcP$,
5-brane segments to edges, and web junctions to triangles.
By the consistency condition of Section~\ref{sec_webs57}, every
junction of the fully pushed web is trivalent and has multiplicity
one, that is, its dual cell is a triangle of normalized area
one. We summarize the cut-and-glue construction by
\begin{equation}
(P,\cP)\longmapsto(\widetilde P,\tcP)\,.
\label{eq:section3-GTP-Symington}
\end{equation} 
If there are no 7-branes, nothing
is removed or glued, and $\widetilde P=P$ with its ordinary
affine structure and triangulation.

\begin{figure}[!htbp]
\centering
    \begin{tikzpicture}
        \node (a) at (0,0) {\scalebox{1}{\begin{tikzpicture}[scale=1.5]
            \fill[red] (0,1)--(-1,-1)--(1,-1)--(0,1);
            \node[bd] (1) at (0,1) {};
            \node[blue] at (-0.1,1.2) {\scriptsize{$1$}};
            \node[bd] (2) at (0,2) {};
            \node[blue] at (-0.1,2.2) {\scriptsize{$2$}};
            \node[bd] (5) at (0,3) {};
            \node[blue] at (0,3.2) {\scriptsize{$5$}};
            \node[bd] (3p) at (-1,-1) {};
            \node[blue] at (-1.2,-1) {\scriptsize{$3'$}};
            \node[wd] (w) at (0,-1) {};
            \node[bd] (3pp) at (1,-1) {};
            \node[blue] at (1.3,-1) {\scriptsize{$3''$}};
            \node[bd] (4) at (1,0) {};
            \node[blue] at (1.2,0) {\scriptsize{$4$}};
            \draw (1)--(2)--(3p)--(w)--(3pp)--(4)--(2) (1)--(3p) (3pp)--(1) (2)--(3pp) (3p)--(5)--(4) (2)--(5);
            \draw[orange,->,line width=0.45mm] ($(3p)!0.2!(5)$)--($(3p)!0.5!(5)$);
            \node[orange] at (-1,0.2) {\scriptsize$\begin{pmatrix}
                1\\
                4
            \end{pmatrix}$};
            \draw[darkgreen,->,line width=0.45mm] ($(3p)!0.3!(2)$)--($(3p)!0.8!(2)$);
            \node[darkgreen] (gv1) at (-0.8,1.4) {\scriptsize$\begin{pmatrix}
                1\\
                3
            \end{pmatrix}$};
            \draw[darkgreen,transform canvas={xshift=-4pt, yshift=0pt}] (gv1)--(-0.2,1.2);
        \end{tikzpicture}}};
        \node (b) at (8,0) {\scalebox{1}{\begin{tikzpicture}[scale=1.5]
            \node[bd] (1) at (0,1) {};
            \node[blue] at (-0.1,1.2) {\scriptsize{$1$}};
            \node[bd] (2) at (0,2) {};
            \node[blue] at (-0.1,2.2) {\scriptsize{$2$}};
            \node[bd] (5) at (0,3) {};
            \node[blue] at (0,3.2) {\scriptsize{$5$}};
            \node[bd] (3pp) at (1,-1) {};
            \node[blue] at (1.2,-1) {\scriptsize{$3$}};
            \node[bd] (4) at (1,0) {};
            \node[blue] at (1.2,0) {\scriptsize{$4$}};
            \coordinate (35) at (-0.5,1);
            \coordinate (32) at (-1/3,1);
            \draw (1)--(2)--(32)--(3pp)--(4)--(2) (1)--(3pp) (2)--(3pp)--(35)--(5)--(4) (2)--(5);
            \draw[red,dashed,thick] (1)--(-1,1);
            \node[red] at (-1.5,1.3) {\scriptsize{monodromy cut}};
            \node[red] at (-1.5,1) {\scriptsize{$M_{[0,1]}$}};
            \draw[orange,->,line width=0.45mm] ($(3pp)!0.5!(35)$)--($(3pp)!0.9!(35)$);
            \node[orange] at (-2,0.2) {\scriptsize{$M_{[0,1]}^T\begin{pmatrix}
                1\\
                4
            \end{pmatrix}=\begin{pmatrix}
                1 & -1\\
                0 & 1
            \end{pmatrix}\begin{pmatrix}
                1\\
                4
            \end{pmatrix}=\begin{pmatrix}
                -3\\
                4
            \end{pmatrix}$}};
            \draw[orange,->,line width=0.45mm] ($(35)!0.2!(5)$)--($(35)!0.9!(5)$);
            \node[orange] at (-0.5,2.5) {\scriptsize{$\begin{pmatrix}
                1\\
                4
            \end{pmatrix}$}};
            \draw[darkgreen,->,line width=0.45mm] ($(3pp)!0.2!(32)$)--($(3pp)!0.9!(32)$);
            \node[darkgreen] at (-1.5,-0.5) {\scriptsize{$M_{[0,1]}^T\begin{pmatrix}
                1\\
                3
            \end{pmatrix}=\begin{pmatrix}
                1 & -1\\
                0 & 1
            \end{pmatrix}\begin{pmatrix}
                1\\
                3
            \end{pmatrix}=\begin{pmatrix}
                -2\\
                3
            \end{pmatrix}$}};
            \draw[darkgreen,->,line width=0.45mm] ($(32)!0.2!(2)$)--($(32)!0.9!(2)$);
            \node[darkgreen] (gv2) at (-0.9,1.8) {\scriptsize{$\begin{pmatrix}
                1\\
                3
            \end{pmatrix}$}};
            \draw[darkgreen,transform canvas={xshift=-4pt, yshift=0pt}] (gv2)--(-0.1,1.5);
        \end{tikzpicture}}};
        \draw[->] (a)--(b);
    \end{tikzpicture}
\caption{Cutting and gluing a colored triangle. The shaded triangle is removed and its two lateral edges are identified. On the right, the resulting Symington polygon is displayed using a monodromy cut and the indicated lattice vectors are related across the cut by $M_{[0,1]}^T$. The dashed cut is not a boundary of the polygon.}
\label{fig:GTP-affine-cut}
\end{figure}
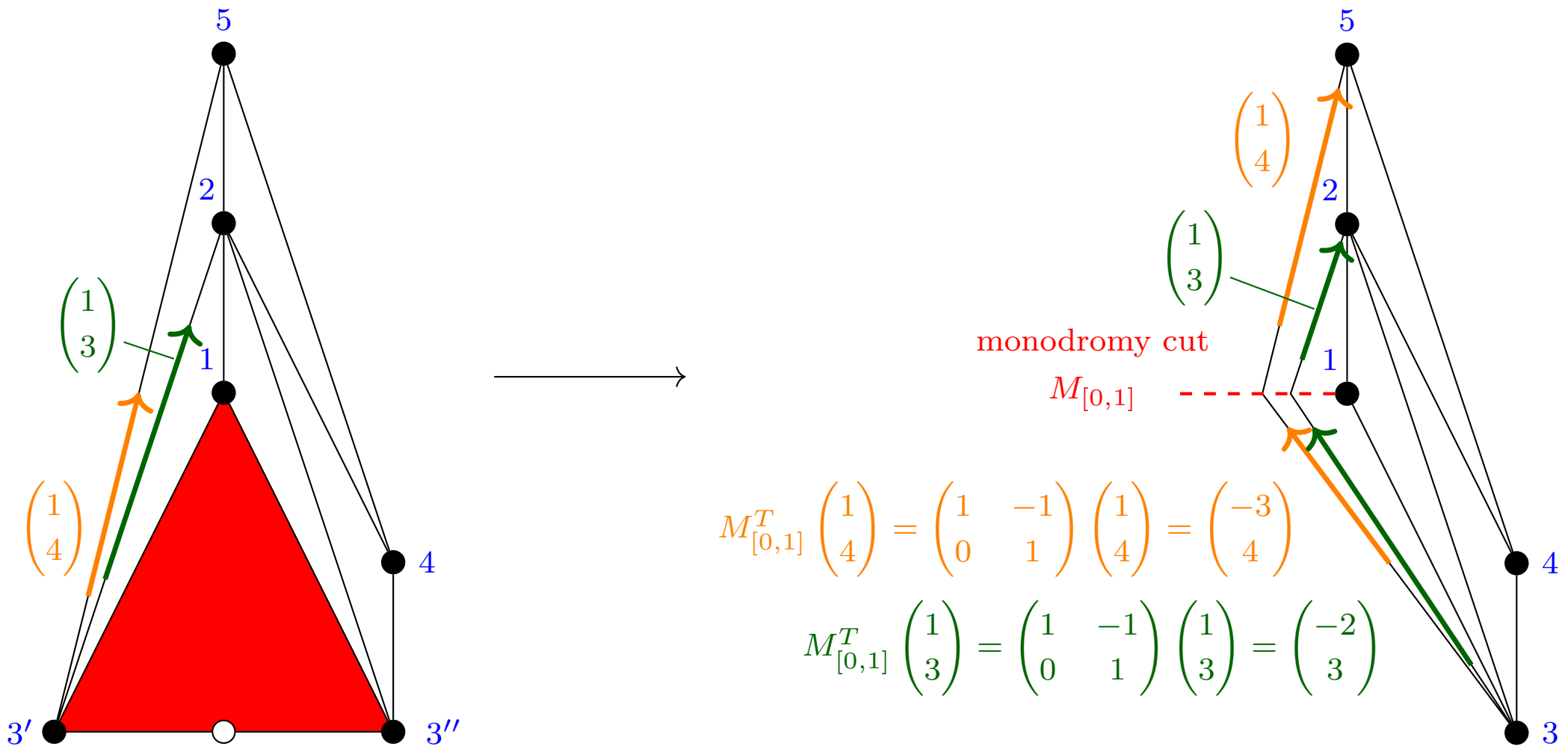

The simplest nontrivial cut-and-glue operation, involving a
single colored triangle, is shown in
Figure~\ref{fig:GTP-affine-cut}. Its two lateral edges are glued
after the shaded triangle is removed. To draw the result in the
plane, one may introduce a \emph{monodromy cut} starting at the
image of the tip. This cut records the change of affine coordinates
between the two sides. With the crossing
orientation used in the figure, the lattice directions of the
polygon transform according to
$M_{[0,1]}^T=\left(\begin{smallmatrix}1&-1\\0&1\end{smallmatrix}\right)$.
For example, the vectors $(1,4)$ and $(1,3)$ are represented on
the other side by $(-3,4)$ and $(-2,3)$, respectively. Crossing
the cut in the opposite direction uses the inverse matrix.
The transpose here follows the convention for polygon lattice
directions dual to the 5-brane charges, whose monodromy
$M_{[p,q]}$ was introduced in Section~\ref{sec_webs57}.

The only points at which the integral-affine structure is singular are the images of the tips $V_\alpha$. 
If a vertex is the tip
of a single colored strip, its image is a \emph{focus--focus
singularity}, as familiar from almost-toric geometry \cite{symington, leungsymington} and the theory of integral-affine manifolds with singularities \cite{gross_book, Gross:2006gyk, GS_real_affine, MR3825608, MR4255071}. Transporting a lattice direction around a small loop
encircling this point gives a nontrivial integral shear. If two or
more strips have the same tip, their images
produce one singular point whose monodromy combines the
contributions of the corresponding 7-branes. Such a point need not
be a single focus--focus singularity. Away from the images of the
tips, the coordinate changes constructed above provide an ordinary
integral-affine structure.

It will often be convenient to separate any coincident singularities.
This can be achieved by an arbitrarily small perturbation of the
integral-affine structure near a common tip, replacing it by
separate focus--focus singularities, one for each 7-brane. The
perturbation is not unique: in particular, it involves a choice
of the relative positions of the singularities. We fix such a
choice and continue to write $V_\alpha$ for the singularity associated
with the 7-brane $\alpha$. 

The monodromy around each $V_\alpha$ fixes a primitive lattice
direction. In local affine coordinates this direction is parallel
to the marked boundary edge $L_\alpha$ of the original GTP, transported
to a neighborhood of $V_\alpha$ through the gluings. Thus it is
determined by the external 5-branes that originally ended on the
7-brane $\alpha$, to which $L_\alpha$ is dual. The invariant direction is
unoriented and determines two opposite rays issuing from $V_\alpha$.
Either ray may be continued in the Symington polygon.

An \emph{$E$-line} is the continuation of one of these rays as a
straight line in the integral-affine structure of $\widetilde P$.
Away from the singularities, its direction is constant when
expressed in any local affine coordinates. If a planar drawing
uses a monodromy cut, the coordinates on its two sides differ by
the corresponding $SL(2,\mathbb Z)$ transformation. An $E$-line may
therefore appear to bend upon crossing the cut, although it is
straight in the integral-affine sense. The change in its drawn
direction is the same change of coordinates illustrated in
Figure~\ref{fig:GTP-affine-cut}. Moreover, for a generic perturbation of all the singularities $V_\alpha$, an $E$-line will never cross a singularity, except possibly when the direction of the $E$-line coincides with the monodromy-invariant direction of the singularity. In this latter case, the $E$-line is extended across the singularity without ambiguity  by following the monodromy-invariant direction.

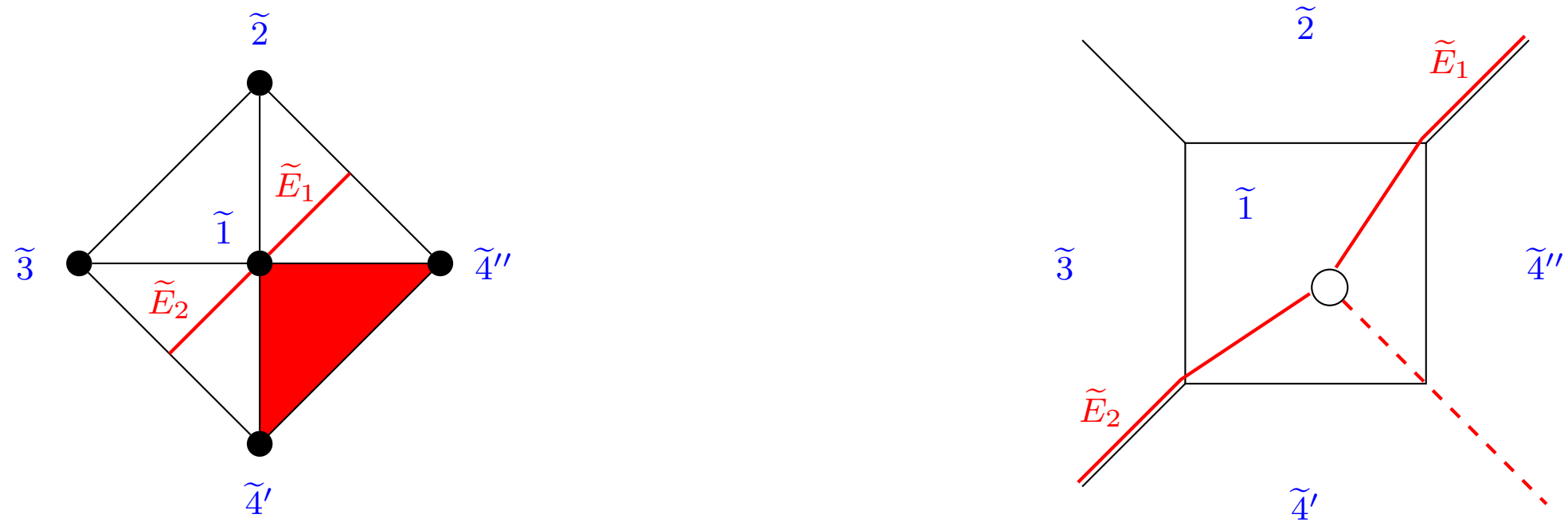
\begin{figure}[!htbp]
\centering
\begin{minipage}[c]{0.48\textwidth}
    \centering
    \begin{tikzpicture}[scale=1.5]
        \fill[red] (0,0)--(0,-1)--(1,0)--(0,0);
        \node[bd] (1) at (0,0) {};
        \node[blue] at (-0.2,0.2) {\scriptsize{$\widetilde{1}$}};
        \node[bd] (2) at (0,1) {};
        \node[blue] at (0,1.3) {\scriptsize{$\widetilde{2}$}};
        \node[bd] (3) at (-1,0) {};
        \node[blue] at (-1.3,0) {\scriptsize{$\widetilde{3}$}};
        \node[bd] (4) at (0,-1) {};
        \node[blue] at (0,-1.3) {\scriptsize{$\widetilde{4}'$}};
        \node[bd] (5) at (1,0) {};
        \node[blue] at (1.3,0) {\scriptsize{$\widetilde{4}''$}};
        \draw (1)--(2)--(3)--(4)--(5)--(2)
              (3)--(1)--(5) (1)--(4);
        \draw[red,thick] (1)--(-0.5,-0.5);
        \node[red] at (-0.5,-0.2)
            {\scriptsize{$\widetilde{E}_2$}};
        \draw[red,thick] (1)--(0.5,0.5);
        \node[red] at (0.2,0.45)
            {\scriptsize{$\widetilde{E}_1$}};
    \end{tikzpicture}
\end{minipage}
\hfill
\begin{minipage}[c]{0.48\textwidth}
    \centering
    \begin{tikzpicture}[scale=2]
        \node (1) at (-1,1) {};
        \node (2) at (-1,-1) {};
        \node[seven] (3) at (0.1,-0.1) {};
        \node (4) at (1,1) {};
        \draw (-0.5,0.5)--(-0.5,-0.5)--
              (0.5,-0.5)--(0.5,0.5)--(-0.5,0.5)
              (1)--(-0.5,0.5)
              (2)--(-0.5,-0.5)
              (4)--(0.5,0.5);
        \draw[red,dashed,thick] (3)--(1,-1);
        \node[blue] at (-0.25,0.25)
            {\scriptsize{$\widetilde{1}$}};
        \node[blue] at (0,1)
            {\scriptsize{$\widetilde{2}$}};
        \node[blue] at (-1,0)
            {\scriptsize{$\widetilde{3}$}};
        \node[blue] at (0,-1)
            {\scriptsize{$\widetilde{4}'$}};
        \node[blue] at (1,0)
            {\scriptsize{$\widetilde{4}''$}};
        \draw[red,thick,
              transform canvas={xshift=-1pt,yshift=1pt}]
              (3)--(-0.5,-0.5)--(2);
        \node[red] at (-0.85,-0.6)
            {\scriptsize{$\widetilde{E}_2$}};
        \draw[red,thick,
              transform canvas={xshift=-1pt,yshift=1pt}]
              (3)--(0.5,0.5)--(4);
        \node[red] at (0.6,0.85)
            {\scriptsize{$\widetilde{E}_1$}};
    \end{tikzpicture}
\end{minipage}

\caption{On the left: the $E$-lines $\widetilde E_1$ and
$\widetilde E_2$ emerging from a singularity in a
Symington polygon. The shaded triangle marks the
region removed in the cut-and-glue construction;
its lateral edges are identified. On the right: the dual paths starting from the 7-branes in the brane web.}
\label{fig:E-lines-first-example}
\end{figure}

For the Symington polygons arising from the consistent GTPs
considered here, every $E$-line reaches the boundary of
$\widetilde P$ rather than continuing indefinitely in its
interior. This can be deduced from 
the existence of an associated Calabi--Yau geometry, as described in Section~\ref{sec:CYGTP}. A simple example of $E$-lines is illustrated in 
Figure~\ref{fig:E-lines-first-example}.

\begin{figure}[!htbp]
\centering
    \begin{tikzpicture}
        \node at (0,0) {\scalebox{1}{\begin{tikzpicture}[scale=1.5]
            \fill[red] (0,0)--(0,-1)--(1,-1)--(0,0);
            \node[bd] (1) at (0,0) {};
            \node[blue] at (-0.1,0.2) {\scriptsize{$\widetilde{1}$}};
            \node[bd] (2) at (0,1) {};
            \node[blue] at (-0.1,1.2) {\scriptsize{$\widetilde{2}$}};
            \node[bd] (3) at (0,2) {};
            \node[blue] at (0,2.2) {\scriptsize{$\widetilde{3}$}};
            \node[bd] (4) at (-1,-1) {};
            \node[blue] at (-1.2,-1) {\scriptsize{$\widetilde{4}$}};
            \node[bd] (5) at (0,-1) {};
            \node[blue] at (0,-1.2) {\scriptsize{$\widetilde{5}'$}};
            \node[bd] (6) at (1,-1) {};
            \node[blue] at (1.2,-1) {\scriptsize{$\widetilde{5}''$}};
            \node[bd] (7) at (1,0) {};
            \node[blue] at (1.2,0) {\scriptsize{$\widetilde{6}$}};
            \draw (1)--(2)--(3)--(4)--(5)--(6)--(7)--(3) (2)--(4)--(1)--(5) (1)--(6)--(2)--(7);
            \draw[red,thick] (1)--(-0.68,0);
            \node[red] at (-0.4,-0.15) {\scriptsize{$\widetilde{E}_2$}};
            \draw[red,thick] (1)--(7);
            \node[red] at (0.4,-0.15) {\scriptsize{$\widetilde{E}_1$}};
        \end{tikzpicture}}};
        \node at (8,0) {\scalebox{0.75}{\begin{tikzpicture}[scale=1]
            \coordinate (1) at (0,0);
            \coordinate (2) at (8,-5);
            \node[seven] (3) at (9,-3.5) {};
            \coordinate (4) at (13,-2);
            \coordinate (5) at (15,0);
            \draw (1)--(3,-1)--(7,-3)--(8,-4)--(2) (8,-4)--(9,-4) (9,-4)--(10,-3)--(12,-2)--(4) (12,-2)--(13,-1)--(5) (13,-1)--(3,-1) (7,-3)--(10,-3);
            \draw[red,dashed,thick] (9,-5)--(3);
            \node[blue] at (8.1,-3.6) {$\widetilde{1}$};
            \node[blue] at (8.5,-2) {$\widetilde{2}$};
            \node[blue] at (8.5,0) {$\widetilde{3}$};
            \node[blue] at (4,-3) {$\widetilde{4}$};
            \node[blue] at (8.5,-4.5) {$\widetilde{5}'$};
            \node[blue] at (11,-3.5) {$\widetilde{5}''$};
            \node[blue] at (13.7,-1) {$\widetilde{6}$};
            \draw[red,thick,transform canvas={xshift=0pt, yshift=-2pt}] (1)--(3,-1)--(7,-3)--(3);
            \node[red] at (5,-2.5) {$\widetilde{E}_2$};
            \draw[red,thick,transform canvas={xshift=0pt, yshift=-2pt}] (3)--(10,-3)--(12,-2);
            \draw[red,thick,transform canvas={xshift=0pt, yshift=-2pt}] (12,-2) .. controls (13,-1.5) .. (15,-1);
            \node[red] at (11.5,-2.7) {$\widetilde{E}_1$};
        \end{tikzpicture}}};
    \end{tikzpicture}
\caption{$E$-lines in the Symington polygon (left) and their dual paths in the 5-brane web with pulled 7-branes (right). The paths follow the dual combinatorial data and need not be straight in the web plane.}
\label{fig:E-lines-rank-two}
\end{figure}
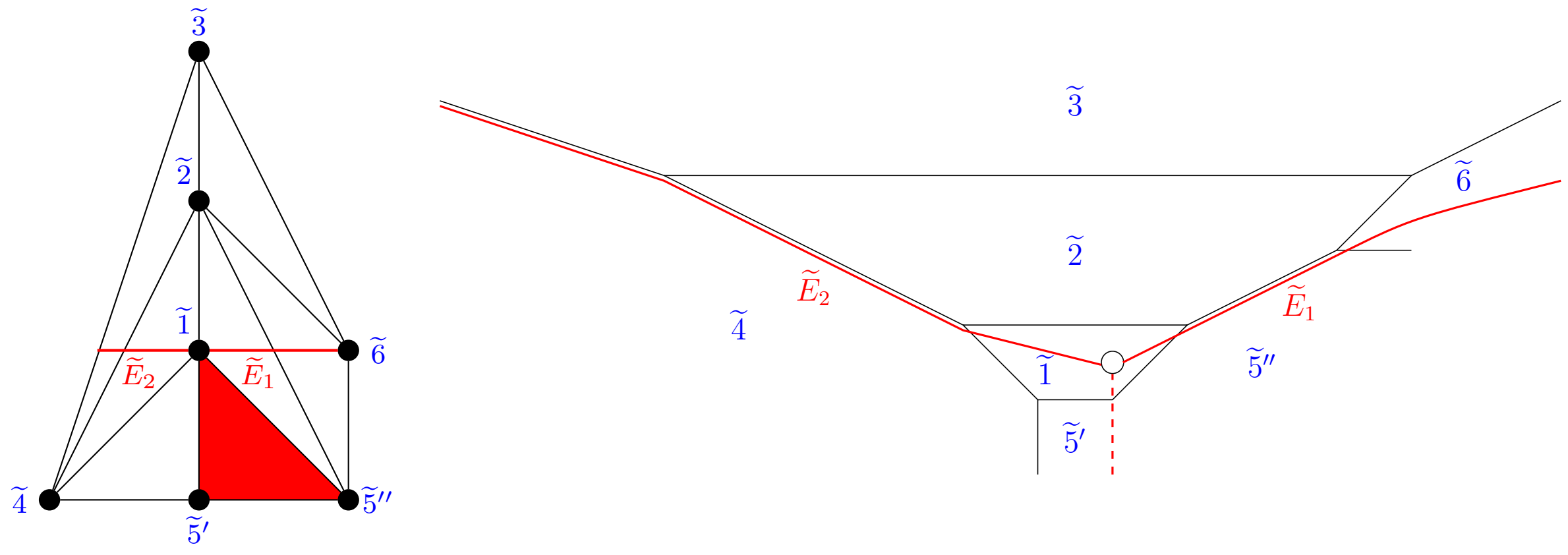

An $E$-line also determines a dual path in the 5-brane web with pulled 7-branes.
The point $V_\alpha$ is dual to the bounded face containing the
7-brane $\alpha$, so the path starts at that 7-brane. When the $E$-line
crosses an edge of $\tcP$ and enters an adjacent triangle, the
dual path follows the corresponding 5-brane segment to the next
web junction. In this way the path records the sequence of edges
and triangles met by the $E$-line. 
However, the exact shape of the path is not fixed beyond these incidence conditions.
In particular, it can be drawn with curved
segments in the $(x^5,x^6)$ plane, whereas the $E$-line is straight
in the integral-affine structure of the Symington polygon. Examples illustrating these dual paths are given in Figures~\ref{fig:E-lines-first-example} and \ref{fig:E-lines-rank-two}.

\section{Generalized Toric Calabi--Yau Threefolds}
\label{sec:CYGTP}

In this section, we review the construction of Calabi--Yau threefolds from GTPs developed in \cite{AABwebs}, and identify the divisors and curves relevant for the intersection theory studied in the following sections. In Section \ref{sec_fromGTP}, we first recall the necessary properties of log Calabi--Yau surfaces, then explain how such surfaces are associated with the vertices of a Symington polygon and glued to form the central fiber of a one-parameter family of surfaces whose total space is a Calabi--Yau threefold. In Section~\ref{sec:DivisorsCurves}, we describe the divisors and curves of the resulting threefold, distinguishing those arising from the central fiber from the additional non-compact divisors associated with $E$-lines, which we call \emph{exuberant divisors}. 

\subsection{From Generalized Toric Polygons to Calabi--Yau Threefolds}  \label{sec_fromGTP}

We fix a GTP $(P,\mathcal P)$ as in Section~\ref{sec:GTP} and consider the associated Symington polygon $\widetilde P$, equipped with its unimodular triangulation $\tcP$, as constructed in Section~\ref{sec:symington-E-lines}. We denote by $\tcP^{[0]}$, $\tcP^{[1]}$, and $\tcP^{[2]}$ the sets of vertices, edges, and triangles of this triangulation, respectively. We first explain how to associate a surface to each vertex of $\tcP$ and then how to glue these surfaces along curves corresponding to its edges. The resulting surface is the central fiber of a one-parameter family of surfaces whose total space will be the Calabi--Yau threefold associated to the GTP.

\subsubsection{Surfaces from Vertices of the Symington Polygon}
\label{sec:LogCYFromSymington}

For each vertex $v\in\tcP^{[0]}$, we explain below how to associate a smooth surface
$X_v$ and a normal crossing divisor $\partial X_v$ on $X_v$, referred to as the \emph{boundary} of $X_v$. 
Every pair $(X_v, \partial X_v)$ will be log Calabi--Yau, that is, satisfying 
\be
K_{X_v} + \partial X_v =0\,,
\ee
where $K_{X_v}$ denotes the canonical divisor of $X_v$.

If $v$ is a smooth point of the integral-affine structure on $\widetilde P$, that is $v \neq V_\alpha$ for every 7-brane $\alpha$, the primitive directions of the edges
emanating from it form a fan $\Sigma_v$. The unimodularity of $\tcP$
ensures that adjacent rays form a basis of the local lattice. We take
$X_v$ to be the corresponding smooth toric surface and $\partial X_v$
its toric boundary. The surface is compact when $v$ is an interior
vertex of $\widetilde P$, and non-compact when $v$ lies on its boundary.
Equivalently, $X_v$ is the toric surface with momentum polytope given by the face of the 5-brane web dual to $v$.

Now suppose that $v$ is a singular point of the integral-affine structure on $\widetilde P$, that is, the image in the Symington polygon of tips $V_\alpha$ of the $\alpha$-colored strips associated to 7-branes $\alpha$. Before the
corresponding cuts producing the Symington polygon, 
the star of the vertex in the polyhedral
decomposition $\mathcal{P}$ determines a toric surface
$(\overline X_v,\partial\overline X_v)$, possibly singular. 
Equivalently, $\overline{X}_v$ is the toric surface with momentum polytope given by the face of the 5-brane web dual to $v$ before pulling the 7-brane. 
Each cone
removed by a cut corresponds to a zero-dimensional stratum of the toric
divisor $\partial \overline{X}_v$.
The removed cone is the cone over an elementary lattice
triangle whose base has the same lattice length $n$ as its lattice
height. 
For $n>1$ the associated affine toric surface has a Wahl
singularity at the corresponding zero-dimensional toric stratum,
\be
\CC^2/\bigl(\ZZ/n^2\ZZ\bigr),
\ee
where the cyclic group acts with weights $(1,an-1)$ for an integer $a$
with $\gcd(a,n)=1$ \cite{KSB}. Equivalently, the singularity is the quotient of
$xy=z^n$ by $\ZZ/n\ZZ$, acting on
$(x,y,z)$ with weights $(1,-1,a)$.
The surface $X_v$ is obtained from $\overline{X}_v$ by smoothing these Wahl singularities.
Locally, these smoothings can be described as the quotient of
\be
xy=z^n+t
\label{eq:WahlSmoothing}
\ee
by the same $\ZZ/n\ZZ$-action. For $t\ne0$ it smooths the surface
singularity and the corresponding corner of $\partial \overline{X}_v$. For $n=1$ there is no surface singularity, and the corresponding
operation replaces the normal-crossing divisor $xy=0$ by $xy=t$
in a smooth surface. These local
deformations can be carried out simultaneously for all cut cones
\cite[\S 6.2, Step I]{AABwebs}.  
The result of this deformation is a smooth log
Calabi--Yau pair $(X_v,\partial X_v)$, where the divisor $\partial X_v$ is obtained by
smoothing precisely the corners of the toric divisor $\partial \overline{X}_v$
selected by the cuts.
The deformation of the toric surface $(\overline{X}_v, \partial \overline{X}_v)$ into the non-toric surface $(X_v, \partial X_v)$ is the geometric counterpart of pulling 7-branes into the face of the 5-brane web dual to $v$. In particular, the face with 7-branes pulled in is a non-toric generalization of the toric momentum polytope, called a Symington polytope for the non-toric surface $(X_v, \partial X_v)$ \cite{symington, leungsymington}. The complement $X_v \setminus \partial X_v$ admits a torus fibration over this polytope, with nodal singular fibers over the 7-branes. In particular, the number of 7-branes in the face dual to $v$ is equal to the \emph{charge}
$Q_{X_v}$ of the log Calabi--Yau surface $(X_v, \partial X_v)$, that is, the topological Euler characteristic of the complement $X_v \setminus \partial X_v$. 

\begin{remark}
    Elementary triangles are also known as \emph{primitive $T$-cones}, and some aspects of their relation with GTPs have also been explored in \cite{Arias-Tamargo:2024fjt, CarrenoBolla:2024fxy, CarrenoBolla:2025rkv, Franco:2026bie, Franco:2026gnm}. For the present paper, the important point is the observation made in \cite{AABwebs} that the smoothing of the corresponding Wahl singularities is a basic operation in the construction of the Calabi--Yau threefold corresponding to a GTP.
\end{remark}

\subsubsection{Gluing, Smoothing, and the Generalized Toric Calabi--Yau Threefold}
\label{sec:GlobalCY}
In Section \ref{sec:LogCYFromSymington}, we constructed smooth surfaces $X_v$ by deformation of possibly singular toric surfaces $\overline{X}_v$. We now explain how to glue these surfaces together to produce a normal crossing surface $\widetilde{\cX}_0$.

For each edge $e=[v,w]\in\tcP^{[1]}$, the boundaries $\partial X_v$ and $\partial X_w$ contain curves corresponding to the primitive directions of $e$ at $v$ and $w$, respectively. We identify these curves according to the gluing data of the Symington triangulation.
Performing
these identifications for every edge produces a
normal crossing surface
\be
\widetilde{\cX}_0=\bigcup_{v\in\tcP^{[0]}}X_v\,.
\label{eq:centralFiber}
\ee
Each triangle of $\tcP$ records a point where three components meet.
Thus the vertices, edges, and triangles of $\tcP$ correspond to the
components, double curves, and triple points of the normal crossing surface $\widetilde{\cX}_0$, as summarized in the table below:
\be
\boxed{\begin{array}{ccccc}
\text{Web with pulled 7-branes}&&\text{Symington triangulation}&&\widetilde{\cX}_0\\[1mm]
\hline\\[-3mm]
\text{face}&\longleftrightarrow&\text{vertex}&\longleftrightarrow&\text{surface   }X_v,\\[2mm]
\text{5-brane segment}&\longleftrightarrow&\text{edge}&\longleftrightarrow&\text{double curve}\\[2mm]
\text{trivalent junction}&\longleftrightarrow&\text{triangle}&\longleftrightarrow&\text{triple point}
\end{array}}
\label{eq:BW-GTP-CY-dictionary}
\ee

The gluing can be chosen so that the normal crossing  surface $\widetilde{\cX}_0$ admits a
one-parameter smoothing \cite[\S 6.2, Step III]{AABwebs}. More precisely, there
is a smooth non-compact Calabi--Yau threefold $\widetilde{\cX}$ and a
holomorphic map
\be
\pi:\widetilde{\cX}\longrightarrow\Delta,\qquad
\pi^{-1}(0)=\widetilde{\cX}_0\,,
\ee
where $\Delta\subset\CC$ is a disk about the origin. 
We refer to the smooth non-compact Calabi–Yau threefold $\widetilde{\cX}$ obtained by this construction as the \emph{generalized toric Calabi–Yau threefold} associated with the given GTP.

Note that the deformation of the normal crossing surface $\widetilde{\cX}_0$ over $\Delta$ to produce the threefold $\widetilde{\cX}$ is distinct from the deformations considered in Section \ref{sec:LogCYFromSymington} to construct the surfaces $X_v$. By construction, the surface components $X_v$ of $\widetilde{\cX}_0$ are divisors in the threefold $\widetilde{\cX}$. Those associated with interior vertices are compact, whereas those associated with boundary vertices are non-compact. Contracting all compact curves and all compact divisors in $\widetilde{\cX}$ yields the associated canonical Calabi--Yau threefold singularity
\be
\widetilde{\cX}\longrightarrow\cX\,.
\ee
According to \cite{AABwebs}, the 5d SCFT defined by a consistent 5-brane web with 7-branes in Type IIB string theory admits a dual description as M-theory on the canonical Calabi--Yau threefold singularity $\cX$. The choice of unimodular triangulation of the Symington polygon specifies the resolved geometry and hence a chamber of the extended Coulomb branch. In the absence of 7-branes, the GTP is just a toric polygon, and the above construction recovers the associated toric Calabi--Yau threefold.

\begin{remark}
   An alternative description of the threefolds $\widetilde{\cX}$ and $\cX$ in terms of scattering diagrams and broken lines is given in \cite[\S 6.4]{AABwebs} and further developed in \cite{BCL}. For our intersection-theoretic purposes, however, it is more convenient to realize the threefold as the total space of a smoothing of an explicit normal crossings surface, as reviewed in this section.
\end{remark}

\subsection{Divisors and Curves}
\label{sec:DivisorsCurves}

We now identify the divisors and curves of the generalized toric Calabi--Yau threefold $\widetilde{\cX}$ relevant for the intersection theory developed in Section~\ref{sec:intersectionTheory}. 
We first describe the divisors arising from the components of the central fiber $\widetilde{\cX}_0$ and the curves along which these components intersect. We then turn to the additional non-compact divisors associated with $E$-lines, which we call \emph{exuberant divisors}, and describe their intersections with the central fiber. Finally, we introduce the notation for these divisors and curves and summarize the incidence data needed to compute their intersection products.

\subsubsection{Divisors and Double Curves of the Central Fiber}

For every vertex $v_i\in\tcP^{[0]}$, consider the divisor
\be
D_i:=X_{v_i}\subset\widetilde{\cX} 
\ee
given by the corresponding irreducible component of the normal crossing surface $\widetilde{\cX}_0$. We denote the index set of these divisors by $I_0$ and write
\be
I_0=I_c\sqcup I_{0,\partial}\,,
\label{eq:centralDivisorIndexSets}
\ee
where $I_c$ indexes the interior vertices and the corresponding
compact divisors, while $I_{0,\partial}$ indexes boundary vertices
and non-compact divisors.

An edge $e_a=[v_i,v_j]\in\tcP^{[1]}$ specifies an irreducible
double curve $C_a\subset D_i\cap D_j$. The full intersection
$D_i\cap D_j$ may contain several such curves if more than one edge
joins $v_i$ and $v_j$ after the affine identifications. An explicit example where this happens is provided in Section \ref{sec:ex1}.
Let $A_0$
index the double curves and set
\be
A_0=A_{0,c}\sqcup A_{0,\partial}\,,
\ee
with $A_{0,c}$ corresponding to interior edges, whose curves are
compact, and $A_{0,\partial}$ to boundary edges, whose curves are
non-compact. The incidence relation is
\be
C_a\subset D_i\quad\Longleftrightarrow\quad
v_i\text{ is an endpoint of }e_a\,.
\ee
In particular, the irreducible curves $C_a$ must be distinguished
from the possibly reducible full intersection of two divisors.

\subsubsection{Exuberant Divisors and $E$-lines}
\label{sec:ExuberantGeometry}

The components $D_i$ are not all the divisors relevant to the
intersection theory of $\widetilde{\cX}$. In the toric presentation,
a divisor is associated with a lattice vertex, or equivalently a face
of the dual 5-brane web. The non-toric deformation changes this
picture: the smoothing of a boundary corner associated with a
pulled-in 7-brane can identify toric boundary pieces, while the
resulting threefold has further non-compact divisors not contained in
$\widetilde{\cX}_0$, that we call \emph{exuberant divisors}.

\begin{figure}[htbp]
\centering
\begin{tikzpicture}[
    x=1cm, y=1cm,
    line cap=round,
    line join=round,
    every path/.style={line width=1.15pt}
]


\coordinate (v) at (1.90,2.86);
\coordinate (Va) at (2.33,2.67);

\draw (v) -- (0.40,5.02);
\draw (v) -- (4.34,3.50);
\draw (v) -- (0.10,1.33);

\draw[red!85!black,
      dash pattern=on 2.2pt off 4.2pt]
      (-0.02,3.79) -- (Va);

\draw[red!85!black] (Va) -- (4.44,1.74);

\fill[red!85!black] (Va) circle[radius=1.8pt];

\node at (2.12,3.12) {$v$};
\node[below, text=red!85!black]
    at (2.10,2.15) {$V_\alpha$};
\node[below, text=red!85!black]
    at (3.65,1.69) {$\ell_\alpha$};


\node at (7.90,5.35)
    {$(\widetilde X_v,\partial\widetilde X_v)$};

\coordinate (A) at (5.76,2.72);
\coordinate (B) at (7.64,4.12);
\coordinate (C) at (8.77,2.39);
\coordinate (D) at (7.25,0.92);

\draw (D) -- (A) -- (B) -- (C);
\draw[red!85!black] (C) -- (D);

\draw[red!85!black] (7.19,2.62)
    .. controls (7.65,2.52) and (8.02,2.15)
    .. (8.12,1.48);


\node at (12.30,5.35)
    {$(X_v,\partial X_v)$};

\coordinate (E) at (10.45,2.80);
\coordinate (F) at (12.20,4.40);
\coordinate (G) at (13.05,1.88);

\draw (E) -- (F) -- (G) -- cycle;

\draw[red!85!black] (11.68,3.17) -- (G);

\node[below, text=red!85!black]
    at (11.95,1.84) {$C_\alpha$};

\end{tikzpicture}
\caption{The exuberant curve $C_\alpha$ in $X_v$, image of an exceptional curve in $\widetilde{X}_v$}
\label{fig_exuberant_curve}
\end{figure}
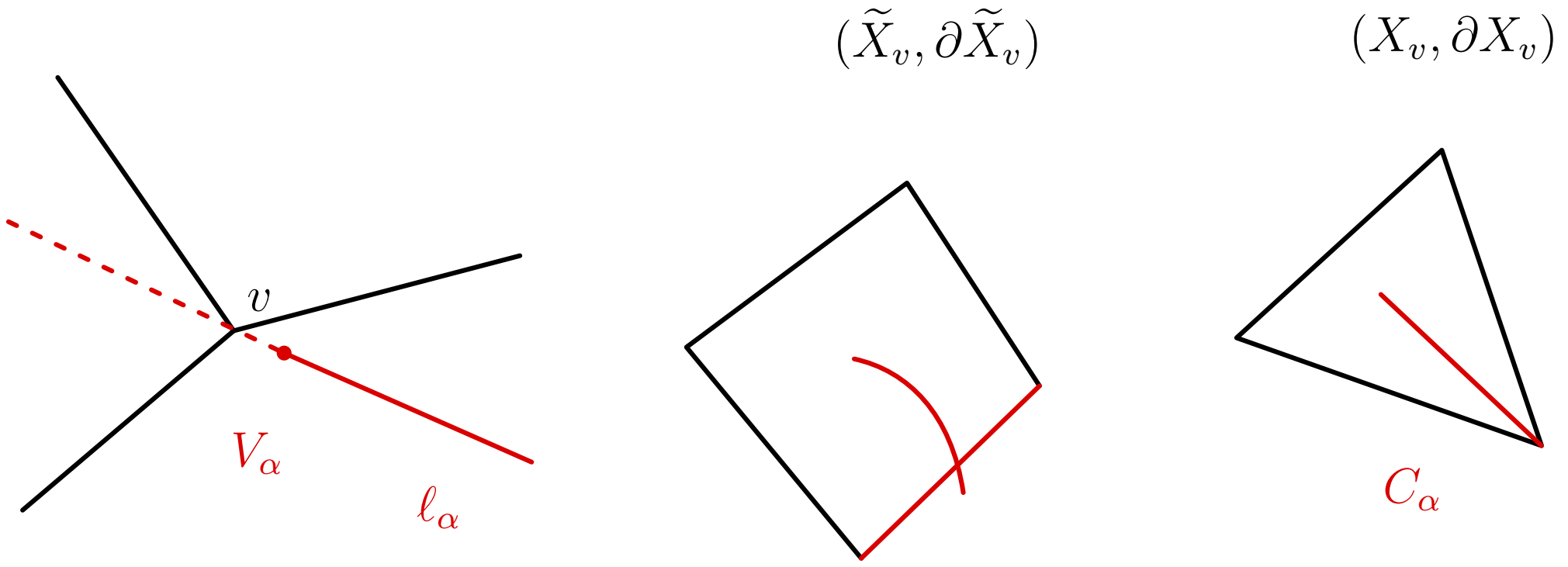

Let $I_e$ be an index set for the 7-branes.
As in Section \ref{sec:symington-E-lines}, we perturb infinitesimally the position of the focus--focus singularities $V_\alpha$ in the Symington polygons for every $\alpha \in I_e$.
Fix some $\alpha \in I_e$
and denote by $v$ the vertex of the Symington polygon at which the singularity $V_\alpha$ is originally located before perturbation. As also described in  Section \ref{sec:symington-E-lines}, there are two $E$-lines emanating from $V_\alpha$.
We denote by $\ell_\alpha$ the $E$-line starting from $V_\alpha$ and not passing through $v$, as illustrated in Figure \ref{fig_exuberant_curve}. 
As in the tropical correspondence theorem \cite{NS}, the $E$-line $\ell_\alpha$ determines an algebraic curve $\mathcal{L}_\alpha$ in the normal crossings surface $\widetilde{\mathcal X}_0$, whose tropicalization is $\ell_\alpha$. The irreducible components of $\mathcal{L}_\alpha$ correspond to the vertices of the triangulation traversed by $\ell_\alpha$ before perturbation, with adjacent components meeting at nodes corresponding to the segments joining these vertices. In particular, the initial vertex $v$ determines a distinguished irreducible component $C_\alpha$ of $\mathcal{L}_\alpha$, which we call the \emph{exuberant curve} associated with $\ell_\alpha$.

The exuberant curve $C_\alpha$ admits an explicit description in terms of a toric model of the log Calabi--Yau surface $(X_v,\partial X_v)$, that is, a presentation of $(X_v,\partial X_v)$ as a blow-down of a non-toric blow-up of a toric surface. As explained in \cite{GHK1}, a generic perturbation separating the integral-affine singularities at $v$ into distinct focus--focus singularities determines a choice of toric model for $(X_v,\partial X_v)$.
Let $\Sigma$ be the fan obtained from the primitive directions of the edges of $\tcP$ emanating from $v$ by adding rays in the monodromy-invariant directions of the perturbed focus--focus singularities. Denote by $\overline X_\Sigma$ the corresponding toric surface. For each singularity $V_\beta$ originally located at $v$, blow up a general point of the toric boundary component of $\overline X_\Sigma$ corresponding to its monodromy-invariant direction. These blow-ups are performed away from the toric fixed points. Let $\widetilde X_v$ be the resulting surface, with boundary $\partial\widetilde X_v$ given by the strict transform of the toric boundary of $\overline X_\Sigma$.
The surface $X_v$ is then obtained by contracting those irreducible components of $\partial\widetilde X_v$ corresponding to rays of $\Sigma$ that are not directions of edges of $\tcP$ emanating from $v$. Under this contraction $\widetilde{X}_v \rightarrow X_v$, the exceptional curve of the blow-up $\widetilde{X}_v \rightarrow \overline{X}_\Sigma$
associated with $V_\alpha$ maps to the exuberant curve $C_\alpha\subset X_v$, as illustrated in Figure~\ref{fig_exuberant_curve}.
In particular, $C_\alpha$ is a compact rational curve whose normalization is isomorphic 
to the projective line $\mathbb{P}^1$.
We denote by $A_e$ the
index set of these compact curves, with the natural bijection 
$A_e\simeq I_e$.


Finally, the \emph{exuberant divisor} $E_\alpha$ is the total space of a one-parameter smoothing of the curve $\mathcal{L}_\alpha$ in $\widetilde{\cX}_0$ with tropicalization the $E$-line $\ell_\alpha$. The existence of this smoothing follows as in the tropical correspondence theorem \cite{NS}. The exuberant divisor $E_\alpha$ is a non-compact divisor in $\widetilde{\mathcal{X}}$, whose intersection with the central fiber $\widetilde{\mathcal{X}}_0$ is the reducible curve $\mathcal{L}_\alpha$:
\be 
E_\alpha \cap \widetilde{\mathcal{X}}_0 = \mathcal{L}_\alpha \,.
\ee

The two types of divisors, components of the central fiber and exuberant divisors, may be summarized as follows: 
\be
\boxed{\begin{array}{ccccc}
\text{Web with pulled 7-branes}&&\text{Symington polygon}&&\text{divisor}\\[1mm]
\hline\\[-3mm]
\text{face}&\longleftrightarrow&\text{vertex}&\longleftrightarrow&D_i\ (i\in I_0),\\[2mm]
\text{7-brane and chosen ray}&\longleftrightarrow&\text{$E$-line }\ell_\alpha
&\longleftrightarrow&E_\alpha\ (\alpha\in I_e).
\end{array}}
\label{eq:ordinary-exuberant-dictionary}
\ee
Accordingly, the divisor index set used below is
\be
I=I_0\sqcup I_e=I_c\sqcup I_{0,\partial}\sqcup I_e\,,
\label{eq:divisorIndexSets}
\ee
and for every $i\in I$, we will denote by $D_i$ the corresponding divisor. In particular, for $i=\alpha \in I_e$, we will have $D_i = E_\alpha$.
The curve index set used below is \be
A=A_0\sqcup A_e
 =A_{0,c}\sqcup A_{0,\partial}\sqcup A_e\,,\qquad
A_c=A_{0,c}\sqcup A_e\,,
\label{eq:curveIndexSets}
\ee
where $A_0$ is the index set of double curves in the central fiber, $A_e$ is the index set of exuberant curves, $A_{0,c}$ is the index set of compact double curves, $A_{0, \partial}$ is the index set of non-compact double curves, and $A_c$ is the index set of compact curves (either double curves or exuberant). For every $a \in A$, we will denote by $C_a$ the corresponding curve.

\begin{remark} Every divisor on $\widetilde{\cX}$ defines by Poincar\'e duality a class in $H^2(\widetilde{\cX}, \ZZ)$.
    It follows from \cite[Lemma 3.11-Corollary 8.14]{AAB_KSBA} that the classes of the divisors $(D_i)_{i\in I_0}$ and $(E_\alpha)_{\alpha \in I_e}$ span $H^2(\widetilde{\cX}, \ZZ)$ and satisfy three independent linear relations, so that 
    \[ \mathrm{rank} \, H^2(\widetilde{\cX}, \ZZ) =  |I_0| + |I_e| - 3\,,\]
where:
\begin{enumerate}
\item $|I_0|$ is the cardinality of $I_0$, that is, the number of lattice points in the Symington polygon $\widetilde{P}$, 
\item $|I_e|$ is the cardinality of $I_e$, that is, the total number of 7-branes in the web description, or equivalently the total number of focus--focus singularities in the Symington polygon after generic perturbation. We also have 
\[ |I_e| = \sum_{v\in I_0} Q_{X_v}\,,\]
where $Q_{X_v}$ is the charge of the log Calabi--Yau surface $(X_v, \partial X_v)$ defined at the end of Section \ref{sec:LogCYFromSymington}.
\end{enumerate}

The divisors $(D_i)_{i\in I_0}$ always satisfy the relation 
\be \sum_{i \in I_0} D_i=0 \label{eq_sum}
\ee 
in $H^2(\widetilde{\cX},\ZZ)$. Indeed,  $\sum_{i\in I_0} D_i$ is the class of the central fiber $\widetilde{\cX}_0$, which is always a principal divisor since it is the zero-locus of the projection 
$\pi: \widetilde{\cX} \rightarrow \Delta$. The total number of independent linear relations among the divisors $(D_i)_{i \in I_0}$ in $H^2(\widetilde{\cX}, \ZZ)$ depends on the 7-brane configuration of the web underlying the GTP. If the web contains at least two 7-branes with distinct $[p,q]$ charges, \eqref{eq_sum} is the unique such relation. This number increases to two when all 7-branes have identical charges, and to three in the absence of 7-branes, corresponding to the toric case.

\end{remark}

\subsubsection{Example}

In this section, we illustrate the geometry of exuberant divisors in a particular example. Consider the GTP of Figure \ref{fig:BW7pull}, with $E$-lines represented in Figure \ref{fig:E-lines-first-example}.
We first consider the toric geometry, without 7-branes pulled in. 
The toric Calabi--Yau threefold can be visualized by thickening the web drawing in a third transverse direction, so that the faces of the web appear as surfaces in an ambient three-dimensional space. These faces give a representation of the toric divisors $D_i$ in the corresponding toric Calabi--Yau threefold.

\begin{equation}
\begin{split}
    \begin{tikzpicture}[scale=0.2]
        \draw[thick] (0,0)--(3,-6)--(12,-6)--(9,0)--(0,0);
        \draw[thick] (0,0)--(-6,3);
        \draw[thick] (3,-6)--(0,-9);
        \draw[thick] (12,-6)--(18,-9);
        \draw[thick] (9,0)--(12,3);
        \draw (-6,-2)--(-6,7) (-7,-2)--(-7,7);
        \draw (-6.5,7) ellipse (14pt and 8pt);
        \draw (-7,3) arc[start angle=-180, end angle=0, x radius=14pt, y radius=8pt];
        \draw[dash pattern=on 0pt off 1pt, line cap=round] (-7,3) arc[start angle=180, end angle=0, x radius=14pt, y radius=8pt];
        \draw (-7,-2) arc[start angle=-180, end angle=0, x radius=14pt, y radius=8pt];
        \draw[dash pattern=on 0pt off 1pt, line cap=round] (-7,-2) arc[start angle=180, end angle=0, x radius=14pt, y radius=8pt];
        \draw (0,-14)--(0,-5) (-1,-14)--(-1,-5);
        \draw (-.5,-5) ellipse (14pt and 8pt);
        \draw (-1,-9) arc[start angle=-180, end angle=0, x radius=14pt, y radius=8pt];
        \draw[dash pattern=on 0pt off 1pt, line cap=round] (-1,-9) arc[start angle=180, end angle=0, x radius=14pt, y radius=8pt];
        \draw (-1,-14) arc[start angle=-180, end angle=0, x radius=14pt, y radius=8pt];
        \draw[dash pattern=on 0pt off 1pt, line cap=round] (-1,-14) arc[start angle=180, end angle=0, x radius=14pt, y radius=8pt];
        \draw (18,-14)--(18,-5) (19,-14)--(19,-5);
        \draw (18.5,-5) ellipse (14pt and 8pt);
        \draw (18,-9) arc[start angle=-180, end angle=0, x radius=14pt, y radius=8pt];
        \draw[dash pattern=on 0pt off 1pt, line cap=round] (18,-9) arc[start angle=180, end angle=0, x radius=14pt, y radius=8pt];
        \draw (18,-14) arc[start angle=-180, end angle=0, x radius=14pt, y radius=8pt];
        \draw[dash pattern=on 0pt off 1pt, line cap=round] (18,-14) arc[start angle=180, end angle=0, x radius=14pt, y radius=8pt];
        \draw (12,-2)--(12,7) (13,-2)--(13,7);
        \draw (12.5,7) ellipse (14pt and 8pt);
        \draw (12,3) arc[start angle=-180, end angle=0, x radius=14pt, y radius=8pt];
        \draw[dash pattern=on 0pt off 1pt, line cap=round] (12,3) arc[start angle=180, end angle=0, x radius=14pt, y radius=8pt];
        \draw (12,-2) arc[start angle=-180, end angle=0, x radius=14pt, y radius=8pt];
        \draw[dash pattern=on 0pt off 1pt, line cap=round] (12,-2) arc[start angle=180, end angle=0, x radius=14pt, y radius=8pt];
        \fill[pattern={Lines[angle=45, distance=5pt, line width=0.3pt]}, pattern color=blue]
        (0,0)--(3,-6)--(12,-6)--(9,0)--(0,0);
        \node[circle,inner sep=0pt, draw=white,fill=white] at (6,-3) {\small\color{blue}$D_1$};
        \fill[pattern={Lines[angle=0, distance=5pt, line width=0.3pt]}, pattern color=blue]
        (0,0)--(9,0)--(12,3)--(-6,3)--(0,0);
        \node[circle,inner sep=0pt, draw=white,fill=white] at (4,1.5) {\small\color{blue}$D_2$};
        \fill[pattern={Lines[angle=-63.5, distance=5pt, line width=0.3pt]}, pattern color=blue]
        (3,-6)--(0,0)--(-6,3)--(-1.1,-6.8)--(-1.1,-4.5)--(0.1,-4.5)--(0.1,-8.9)--(3,-6);
        \node[circle,inner sep=0pt, draw=white,fill=white] at (-1.5,-2.5) {\small\color{blue}$D_3$};
        \fill[pattern={Lines[angle=0, distance=5pt, line width=0.3pt]}, pattern color=blue]
        (12,-6)--(3,-6)--(0,-9)--(18,-9)--(12,-6);
        \node[circle,inner sep=0pt, draw=white,fill=white] at (8,-7.5) {\small\color{blue}$D_4$};
        \fill[pattern={Lines[angle=-63.5, distance=5pt, line width=0.3pt]}, pattern color=blue]
        (9,0)--(12,-6)--(18,-9)--(12,3)--(9,0);
        \node[circle,inner sep=0pt, draw=white,fill=white] at (14,-5) {\small\color{blue}$D_5$};
    \end{tikzpicture}\;.
    \end{split}
\end{equation}

After the $[1,-1]$ 7-brane is pulled into a bounded face, the web
and its monodromy cut take the form

\begin{equation}
\begin{split}
    \begin{tikzpicture}[scale=0.2]
        \draw[thick] (0,0)--(3,-6)--(12,-6)--(9,0)--(0,0);
        \draw[thick] (0,0)--(-6,3);
        \draw[thick] (3,-6)--(0,-9);
        \draw[thick] (9,0)--(12,3);  
        \draw[red,dashed,thick] (6,-3)--(18,-9);
        \node[red] at (20,-9.5) { monodromy cut};
        \draw (-6,-2)--(-6,7) (-7,-2)--(-7,7);
        \draw (-6.5,7) ellipse (14pt and 8pt);
        \draw (-7,3) arc[start angle=-180, end angle=0, x radius=14pt, y radius=8pt];
        \draw[dash pattern=on 0pt off 1pt, line cap=round] (-7,3) arc[start angle=180, end angle=0, x radius=14pt, y radius=8pt];
        \draw (-7,-2) arc[start angle=-180, end angle=0, x radius=14pt, y radius=8pt];
        \draw[dash pattern=on 0pt off 1pt, line cap=round] (-7,-2) arc[start angle=180, end angle=0, x radius=14pt, y radius=8pt];
        \draw (0,-14)--(0,-5) (-1,-14)--(-1,-5);
        \draw (-.5,-5) ellipse (14pt and 8pt);
        \draw (-1,-9) arc[start angle=-180, end angle=0, x radius=14pt, y radius=8pt];
        \draw[dash pattern=on 0pt off 1pt, line cap=round] (-1,-9) arc[start angle=180, end angle=0, x radius=14pt, y radius=8pt];
        \draw (-1,-14) arc[start angle=-180, end angle=0, x radius=14pt, y radius=8pt];
        \draw[dash pattern=on 0pt off 1pt, line cap=round] (-1,-14) arc[start angle=180, end angle=0, x radius=14pt, y radius=8pt];
        \fill[white] (5.4,-3)--(6.6,-3)--(6.6,5.3)--(5.4,5.3)--(5.4,-3);
        \draw (5.5,-11)--(5.5,-6) (5.5,-3)--(5.5,5) (6.5,-11)--(6.5,-6) (6.5,-3)--(6.5,5);
        \draw[dash pattern=on 0pt off 1pt, line cap=round] (5.5,-6)--(5.5,-3) (6.5,-6)--(6.5,-3);
        \draw (6,5) ellipse (14pt and 8pt);
        \draw (5.5,-3) arc[start angle=-180, end angle=0, x radius=14pt, y radius=8pt];
        \draw[dash pattern=on 0pt off 1pt, line cap=round] (5.5,-3) arc[start angle=180, end angle=0, x radius=14pt, y radius=8pt];
        \draw (5.5,-11) arc[start angle=-180, end angle=0, x radius=14pt, y radius=8pt];
        \draw[dash pattern=on 0pt off 1pt, line cap=round] (5.5,-11) arc[start angle=180, end angle=0, x radius=14pt, y radius=8pt];
        \draw (12,-2)--(12,7) (13,-2)--(13,7);
        \draw (12.5,7) ellipse (14pt and 8pt);
        \draw (12,3) arc[start angle=-180, end angle=0, x radius=14pt, y radius=8pt];
        \draw[dash pattern=on 0pt off 1pt, line cap=round] (12,3) arc[start angle=180, end angle=0, x radius=14pt, y radius=8pt];
        \draw (12,-2) arc[start angle=-180, end angle=0, x radius=14pt, y radius=8pt];
        \draw[dash pattern=on 0pt off 1pt, line cap=round] (12,-2) arc[start angle=180, end angle=0, x radius=14pt, y radius=8pt];
    \end{tikzpicture}
       \end{split}
\end{equation}

The central-fiber divisors, now denoted by $\widetilde D_i$ in the
diagrams to distinguish them from the toric presentation, remain
associated with faces:

\begin{equation}
\begin{split}
    \begin{tikzpicture}[scale=0.3]
        \draw[thick] (0,0)--(3,-6)--(12,-6)--(9,0)--(0,0);
        \draw[thick] (0,0)--(-6,3);
        \draw[thick] (3,-6)--(0,-9);
        \draw[thick] (9,0)--(12,3);     
        \draw[red,dashed,thick] (6,-3)--(20,-10);
        \node[red] at (20,-10.5) {monodromy cut};
        \draw (-6,-2)--(-6,7) (-7,-2)--(-7,7);
        \draw (-6.5,7) ellipse (14pt and 8pt);
        \draw (-7,3) arc[start angle=-180, end angle=0, x radius=14pt, y radius=8pt];
        \draw[dash pattern=on 0pt off 1pt, line cap=round] (-7,3) arc[start angle=180, end angle=0, x radius=14pt, y radius=8pt];
        \draw (-7,-2) arc[start angle=-180, end angle=0, x radius=14pt, y radius=8pt];
        \draw[dash pattern=on 0pt off 1pt, line cap=round] (-7,-2) arc[start angle=180, end angle=0, x radius=14pt, y radius=8pt];
        \draw (0,-14)--(0,-5) (-1,-14)--(-1,-5);
        \draw (-.5,-5) ellipse (14pt and 8pt);
        \draw (-1,-9) arc[start angle=-180, end angle=0, x radius=14pt, y radius=8pt];
        \draw[dash pattern=on 0pt off 1pt, line cap=round] (-1,-9) arc[start angle=180, end angle=0, x radius=14pt, y radius=8pt];
        \draw (-1,-14) arc[start angle=-180, end angle=0, x radius=14pt, y radius=8pt];
        \draw[dash pattern=on 0pt off 1pt, line cap=round] (-1,-14) arc[start angle=180, end angle=0, x radius=14pt, y radius=8pt];
        \draw (5.5,-11)--(5.5,-6) (6.5,-11)--(6.5,-6);
        \draw[dash pattern=on 0pt off 1pt, line cap=round] (5.5,-6)--(5.5,-3) (6.5,-6)--(6.5,-3);
        \draw (5.5,-11) arc[start angle=-180, end angle=0, x radius=14pt, y radius=8pt];
        \draw[dash pattern=on 0pt off 1pt, line cap=round] (5.5,-11) arc[start angle=180, end angle=0, x radius=14pt, y radius=8pt];
        \draw (12,-2)--(12,7) (13,-2)--(13,7);
        \draw (12.5,7) ellipse (14pt and 8pt);
        \draw (12,3) arc[start angle=-180, end angle=0, x radius=14pt, y radius=8pt];
        \draw[dash pattern=on 0pt off 1pt, line cap=round] (12,3) arc[start angle=180, end angle=0, x radius=14pt, y radius=8pt];
        \draw (12,-2) arc[start angle=-180, end angle=0, x radius=14pt, y radius=8pt];
        \draw[dash pattern=on 0pt off 1pt, line cap=round] (12,-2) arc[start angle=180, end angle=0, x radius=14pt, y radius=8pt];
        \fill[pattern={Lines[angle=45, distance=5pt, line width=0.3pt]}, pattern color=blue]
        (0,0)--(3,-6)--(12,-6)--(9,0)--(0,0);
        \node[circle,inner sep=0pt, draw=white,fill=white] at (4,-2) {\small\color{blue}$\widetilde{D}_1$};
        \fill[pattern={Lines[angle=0, distance=5pt, line width=0.3pt]}, pattern color=blue]
        (0,0)--(9,0)--(12,3)--(-6,3)--(0,0);
        \node[circle,inner sep=0pt, draw=white,fill=white] at (3,1.5) {\small\color{blue}$\widetilde{D}_2$};
        \fill[pattern={Lines[angle=-63.5, distance=5pt, line width=0.3pt]}, pattern color=blue]
        (3,-6)--(0,0)--(-6,3)--(-1.1,-6.8)--(-1.1,-4.5)--(0.1,-4.5)--(0.1,-8.9)--(3,-6);
        \node[circle,inner sep=0pt, draw=white,fill=white] at (-1.5,-2.5) {\small\color{blue}$\widetilde{D}_3$};
        \fill[pattern={Lines[angle=0, distance=5pt, line width=0.3pt]}, pattern color=blue]
        (12,-6)--(3,-6)--(0,-9)--(18,-9)--(12,-6);
        \node[circle,inner sep=0pt, draw=white,fill=white] at (8,-7.5) {\small\color{blue}$\widetilde{D}_4'$};
        \fill[pattern={Lines[angle=-63.5, distance=5pt, line width=0.3pt]}, pattern color=blue]
        (9,0)--(12,-6)--(18,-9)--(12,3)--(9,0);
        \node[circle,inner sep=0pt, draw=white,fill=white] at (14,-5) {\small\color{blue}$\widetilde{D}_4''$};
        \fill[white] (5.4,-3)--(6.6,-3)--(6.6,5.3)--(5.4,5.3)--(5.4,-3);
        \draw (5.5,-3)--(5.5,5) (6.5,-3)--(6.5,5);
        \draw (6,5) ellipse (14pt and 8pt);
        \draw (5.5,-3) arc[start angle=-180, end angle=0, x radius=14pt, y radius=8pt];
        \draw[dash pattern=on 0pt off 1pt, line cap=round] (5.5,-3) arc[start angle=180, end angle=0, x radius=14pt, y radius=8pt];
    \end{tikzpicture}\;.
    \end{split}
\end{equation}

The cut identifies the pieces $\widetilde D_4'$ and
$\widetilde D_4''$ of the face marked $4$. Finally, we illustrate the exuberant divisors corresponding to the two $E$-lines: they are not represented by
faces of the web but by surfaces transverse to the face
presentation. The first is $\widetilde E_1$,

\begin{equation}
\begin{split}
    \begin{tikzpicture}[scale=0.2]
        \draw[thick] (0,0)--(3,-6)--(12,-6)--(9,0)--(0,0);
        \draw[thick] (0,0)--(-6,3);
        \draw[thick] (3,-6)--(0,-9);
        \draw[thick] (9,0)--(12,3);
        \draw (-6,-2)--(-6,7) (-7,-2)--(-7,7);
        \draw (-6.5,7) ellipse (14pt and 8pt);
        \draw (-7,3) arc[start angle=-180, end angle=0, x radius=14pt, y radius=8pt];
        \draw[dash pattern=on 0pt off 1pt, line cap=round] (-7,3) arc[start angle=180, end angle=0, x radius=14pt, y radius=8pt];
        \draw (-7,-2) arc[start angle=-180, end angle=0, x radius=14pt, y radius=8pt];
        \draw[dash pattern=on 0pt off 1pt, line cap=round] (-7,-2) arc[start angle=180, end angle=0, x radius=14pt, y radius=8pt];
        \draw (0,-14)--(0,-5) (-1,-14)--(-1,-5);
        \draw (-.5,-5) ellipse (14pt and 8pt);
        \draw (-1,-9) arc[start angle=-180, end angle=0, x radius=14pt, y radius=8pt];
        \draw[dash pattern=on 0pt off 1pt, line cap=round] (-1,-9) arc[start angle=180, end angle=0, x radius=14pt, y radius=8pt];
        \draw (-1,-14) arc[start angle=-180, end angle=0, x radius=14pt, y radius=8pt];
        \draw[dash pattern=on 0pt off 1pt, line cap=round] (-1,-14) arc[start angle=180, end angle=0, x radius=14pt, y radius=8pt];
        \fill[white] (5.4,-3)--(6.6,-3)--(6.6,5.3)--(5.4,5.3)--(5.4,-3);
        \draw (5.5,-11)--(5.5,-6) (5.5,-3)--(5.5,5) (6.5,-11)--(6.5,-6) (6.5,-3)--(6.5,5);
        \draw[dash pattern=on 0pt off 1pt, line cap=round] (5.5,-6)--(5.5,-3) (6.5,-6)--(6.5,-3);
        \draw (6,5) ellipse (14pt and 8pt);
        \draw (5.5,-3) arc[start angle=-180, end angle=0, x radius=14pt, y radius=8pt];
        \draw[dash pattern=on 0pt off 1pt, line cap=round] (5.5,-3) arc[start angle=180, end angle=0, x radius=14pt, y radius=8pt];
        \draw (5.5,-11) arc[start angle=-180, end angle=0, x radius=14pt, y radius=8pt];
        \draw[dash pattern=on 0pt off 1pt, line cap=round] (5.5,-11) arc[start angle=180, end angle=0, x radius=14pt, y radius=8pt];
        \draw (12,-5)--(12,11) (13,-5)--(13,11);
        \draw (12.5,11) ellipse (14pt and 8pt);
        \draw (12,3) arc[start angle=-180, end angle=0, x radius=14pt, y radius=8pt];
        \draw[dash pattern=on 0pt off 1pt, line cap=round] (12,3) arc[start angle=180, end angle=0, x radius=14pt, y radius=8pt];
        \draw (12,-5) arc[start angle=-180, end angle=0, x radius=14pt, y radius=8pt];
        \draw[dash pattern=on 0pt off 1pt, line cap=round] (12,-5) arc[start angle=180, end angle=0, x radius=14pt, y radius=8pt];
        \fill[pattern={Lines[angle=45, distance=5pt, line width=0.3pt]}, pattern color=red]
        (6.5,-10.5)--(12,-5)--(12,10.5)--(6.5,5)--(6.5,-10.5);
        \fill[pattern={Lines[angle=-45, distance=2pt, line width=1pt]}, pattern color=white]
        (6.6,-2.6)--(6.6,-5.9)--(11.9,-5.9)--(8.9,-0.2)--(6.6,-2.6);
        \node[circle,inner sep=0pt, draw=white,fill=white] at (9,3) {\small\color{red}$\widetilde{E}_1$};  
        \draw[red,dashed,thick] (6,-3)--(18,-9);
        \node[red] at (21,-9.5) {monodromy cut};
    \end{tikzpicture}\;,
\end{split}
\end{equation}

and the second is $\widetilde E_2$:

\begin{equation}
\begin{split}
    \begin{tikzpicture}[scale=0.2]
        \draw[thick] (0,0)--(3,-6)--(12,-6)--(9,0)--(0,0);
        \draw[thick] (0,0)--(-6,3);
        \draw[thick] (3,-6)--(0,-9);
        \draw[thick] (9,0)--(12,3);
        \draw[red,dashed,thick] (6,-3)--(18,-9);
        \node[red] at (21,-9.5) {monodromy cut};
        \draw (-6,-2)--(-6,7) (-7,-2)--(-7,7);
        \draw (-6.5,7) ellipse (14pt and 8pt);
        \draw (-7,3) arc[start angle=-180, end angle=0, x radius=14pt, y radius=8pt];
        \draw[dash pattern=on 0pt off 1pt, line cap=round] (-7,3) arc[start angle=180, end angle=0, x radius=14pt, y radius=8pt];
        \draw (-7,-2) arc[start angle=-180, end angle=0, x radius=14pt, y radius=8pt];
        \draw[dash pattern=on 0pt off 1pt, line cap=round] (-7,-2) arc[start angle=180, end angle=0, x radius=14pt, y radius=8pt];
        \draw (0,-17)--(0,-1) (-1,-17)--(-1,-1);
        \draw (-.5,-1) ellipse (14pt and 8pt);
        \draw (-1,-9) arc[start angle=-180, end angle=0, x radius=14pt, y radius=8pt];
        \draw[dash pattern=on 0pt off 1pt, line cap=round] (-1,-9) arc[start angle=180, end angle=0, x radius=14pt, y radius=8pt];
        \draw (-1,-17) arc[start angle=-180, end angle=0, x radius=14pt, y radius=8pt];
        \draw[dash pattern=on 0pt off 1pt, line cap=round] (-1,-17) arc[start angle=180, end angle=0, x radius=14pt, y radius=8pt];
        \fill[white] (5.4,-3)--(6.6,-3)--(6.6,5.3)--(5.4,5.3)--(5.4,-3);
        \draw (5.5,-11)--(5.5,-6) (5.5,-3)--(5.5,5) (6.5,-11)--(6.5,-6) (6.5,-3)--(6.5,5);
        \draw[dash pattern=on 0pt off 1pt, line cap=round] (5.5,-6)--(5.5,-3) (6.5,-6)--(6.5,-3);
        \draw (6,5) ellipse (14pt and 8pt);
        \draw (5.5,-3) arc[start angle=-180, end angle=0, x radius=14pt, y radius=8pt];
        \draw[dash pattern=on 0pt off 1pt, line cap=round] (5.5,-3) arc[start angle=180, end angle=0, x radius=14pt, y radius=8pt];
        \draw (5.5,-11) arc[start angle=-180, end angle=0, x radius=14pt, y radius=8pt];
        \draw[dash pattern=on 0pt off 1pt, line cap=round] (5.5,-11) arc[start angle=180, end angle=0, x radius=14pt, y radius=8pt];
        \draw (12,-2)--(12,7) (13,-2)--(13,7);
        \draw (12.5,7) ellipse (14pt and 8pt);
        \draw (12,3) arc[start angle=-180, end angle=0, x radius=14pt, y radius=8pt];
        \draw[dash pattern=on 0pt off 1pt, line cap=round] (12,3) arc[start angle=180, end angle=0, x radius=14pt, y radius=8pt];
        \draw (12,-2) arc[start angle=-180, end angle=0, x radius=14pt, y radius=8pt];
        \draw[dash pattern=on 0pt off 1pt, line cap=round] (12,-2) arc[start angle=180, end angle=0, x radius=14pt, y radius=8pt];
        \fill[pattern={Lines[angle=45, distance=5pt, line width=0.3pt]}, pattern color=red]
        (0,-17)--(5.5,-11.5)--(5.5,4.5)--(0,-1)--(0,-17);
        \fill[pattern={Lines[angle=-45, distance=2pt, line width=1pt]}, pattern color=white]
        (5.4,-2.9)--(3.1,-5.9)--(5.4,-5.9)--(5.4,-2.6);
        \node[circle,inner sep=0pt, draw=white,fill=white] at (3.5,-1.5) {\small\color{red}$\widetilde{E}_2$};
    \end{tikzpicture}\;.
\end{split}
\end{equation}

\section{Intersection Theory}
\label{sec:intersectionTheory}

\subsection{Intersection Products}
\label{sec:intersection-products}

In this section, we establish the intersection-theoretic framework
used to compute triple intersection numbers on the generalized toric 
Calabi--Yau threefold $\widetilde{\cX}$ constructed in Section~4.
Our ultimate goal is to determine the intrinsic triple intersection
numbers of divisors, namely those involving at least one compact
divisor. However, the reconstruction of these numbers will require
a larger collection of intersection products, including curve-valued
products of non-compact divisors and numerical intersections with
non-compact curves whenever these are well-defined.

We first recall the relevant intersection products on a non-compact
threefold. We then introduce the divisor--curve intersection
coefficients and characterize, in terms of the combinatorial data
of the Symington polygon and its $E$-lines, the pairs for which
numerical intersections are intrinsically defined.

\subsubsection{Intersection Products on non-compact
Calabi--Yau Threefolds}
\label{sec:noncompact-intersections}

Let $\widetilde{\cX}$ be the generalized toric Calabi--Yau
threefold associated with a GTP, as constructed in Section~4.
We are interested in the triple intersection form
\be
\kappa_{ijk}:=D_i\cdot D_j\cdot D_k,
\qquad
\{i,j,k\}\cap I_c\neq\varnothing,
\label{eq:intrinsic-triple-form}
\ee
where $I_c$ denotes the set of compact divisors.

On a non-compact threefold, it is important to distinguish
between curve-valued intersections of divisors and numerical
intersection products. In particular, while intersections of
non-compact divisors may be well-defined as curve classes,
their numerical triple intersections are not necessarily intrinsic.

Compact divisors define Poincar\'e dual classes in the cohomology $H^2_c(\widetilde{\cX},\ZZ)$ with compact support, whereas possibly non-compact divisors define Poincar\'e dual classes in the cohomology $H^2(\widetilde{\cX}, \ZZ)$. 
The intersection of two possibly non-compact divisors produces a possibly non-compact curve class, which defines a Poincar\'e dual class in $H^4(\widetilde{\cX}, \ZZ)$. Equivalently, this intersection product for possibly non-compact divisors is given by the cup-product
\be
H^2(\widetilde{\cX}, \ZZ)
\times
H^2(\widetilde{\cX}, \ZZ)
\longrightarrow
H^4(\widetilde{\cX}, \ZZ)\,.
\label{eq:BM-divisor-product}
\ee
Thus, neither divisor needs to be compact for the curve-valued
product $D_i\cdot D_j$ to be defined.

Now let $C\subset\widetilde{\cX}$ be a possibly non-compact
curve. Whenever the geometric intersection $D\cap C$ is compact,
the corresponding zero-dimensional intersection cycle has a
well-defined degree. In this case, we write
\be
D\cdot C\in\ZZ.
\label{eq:proper-divisor-curve-intersection}
\ee
If instead $D\cap C$ contains a non-compact component,
no intrinsic numerical intersection is assigned without
additional data at infinity.

Finally, if at least one of the three divisors is compact,
their triple intersection is intrinsically defined through
\be
H_c^2(\widetilde{\cX})
\times H^2(\widetilde{\cX})
\times H^2(\widetilde{\cX})
\longrightarrow
H_c^6(\widetilde{\cX})
\xrightarrow{\sim}\ZZ.
\label{eq:compact-triple-product}
\ee
In particular, if $D_i$ is compact, then
\be
D_i\cdot D_j\cdot D_k
=
\int_{\widetilde{\cX}}
\omega_i\wedge\omega_j\wedge\omega_k
\label{eq:intrinsic-triple-integral}
\ee
is well-defined for arbitrary divisors $D_j,D_k$.

These distinctions will play an essential role in what follows.
We will use both curve-valued divisor products and admissible
numerical divisor--curve intersections to obtain compatibility
relations, including relations involving non-compact cycles.
The final output, however, will consist only of the intrinsic
triple intersection numbers containing at least one compact divisor.
As reviewed in Section \ref{sec:CanSing5d}, these triple intersection numbers determine the prepotential of the 5d SCFT engineered by M-theory on $\widetilde{\cX}$, that is, the 5d SCFT associated with the GTP.

\subsubsection{Divisors, Curves, and Intersection Coefficients}
\label{sec:intersection-data}

We use the notation introduced in Section~\ref{sec:DivisorsCurves}.
Recall that the divisors are indexed by
\be
I=I_0\sqcup I_e,
\qquad
I_0=I_c\sqcup I_{0,\partial},
\label{eq:intersection-divisor-indices}
\ee
where $I_0$ indexes the components of the central fiber,
$I_c$ the compact components, $I_{0,\partial}$ the
non-compact components, and $I_e$ the exuberant divisors
associated with $E$-lines.

Similarly, the curves are indexed by
\be
A=A_0\sqcup A_e,
\qquad
A_0=A_{0,c}\sqcup A_{0,\partial},
\label{eq:intersection-curve-indices}
\ee
where $A_0$ indexes the double curves of the central fiber
and $A_e$ the exuberant curves.
We denote by $A_c\subset A$ the set of compact curves.

We describe the relevant intersection theory in terms of
two types of coefficients.

\medskip
\noindent
\textbf{Curve-valued divisor products.}
For two central-fiber divisors $D_i,D_j$, their intersection
is represented by the double curves of the central fiber.
We write
\be
D_i\cdot D_j
=
\sum_{a\in A_0}\alpha^a_{ij}C_a,
\qquad
(i,j)\in I_0\times I_0,
\label{eq:alpha-central-definition}
\ee
where $\alpha^a_{ij}\in\ZZ$ and
$\alpha^a_{ij}=\alpha^a_{ji}$.

Products of the form $E_\alpha \cdot D_j$ with $(\alpha, j) \in I_e \times I_0$ will
likewise admit an expansion into curve classes:
\be
E_\alpha\cdot D_j
=
\sum_{a\in A}\alpha^a_{\alpha j}C_a.
\label{eq:alpha-exuberant-definition}
\ee
We will not require a curve-class expansion of
$E_\alpha\cdot E_\beta$.

\medskip
\noindent
\textbf{Numerical divisor--curve intersections.}
For a divisor $D_i$ and a curve $C_a$, we introduce
the coefficients
\be
\beta_{ia}:=D_i\cdot C_a
\ee
whenever this numerical intersection is intrinsically defined.
Unlike the curve-valued products above, the coefficients
$\beta_{ia}$ are not defined for every divisor--curve pair.
Their domain of definition is determined by the compactness
of the corresponding geometric intersection, which we now
characterize combinatorially.

\subsubsection{Admissible Divisor--Curve Intersections}
\label{sec:admissible-intersections}

\begin{definition}[Admissible divisor--curve pairs]
\label{def:admissible-beta}
Let $\mathcal B\subset I\times A$ be the set of pairs
$(i,a)$ for which the geometric intersection
$D_i\cap C_a$ is compact.
For every $(i,a)\in\mathcal B$, we define
the intrinsic numerical intersection coefficient
\be
\beta_{ia}:=D_i\cdot C_a\in\ZZ.
\label{eq:beta-definition}
\ee
\end{definition}

Every compact curve can be paired with every divisor.
Consequently,
\be
I\times A_c\subset\mathcal B.
\label{eq:B-compact-curves}
\ee

The only non-trivial admissibility conditions arise for
non-compact boundary curves $C_a$, with
$a\in A_{0,\partial}$.
These conditions can be read directly from the Symington
polygon and its $E$-lines.

Let $e_a$ be the boundary edge corresponding to $C_a$,
and denote its relative interior by $e_a^\circ$.

First, consider a central-fiber divisor $D_i$, with
$i\in I_0$, corresponding to a vertex $v_i$ of the
Symington polygon. Then
\be
D_i\cap C_a\text{ is non-compact}
\quad\Longleftrightarrow\quad
v_i\in e_a.
\label{eq:B-central-geometric}
\ee
Indeed, when $v_i\in e_a$, the curve $C_a$ is contained
in $D_i$, and hence their intersection is non-compact.
Otherwise, the intersection consists of finitely many
triple points or is empty, and therefore defines an
intrinsic numerical intersection.

Next, consider an exuberant divisor $E_\alpha$, with
$\alpha\in I_e$, and let $\ell_\alpha$ be its associated
$E$-line. In this case,
\be
E_\alpha\cap C_a\text{ is non-compact}
\quad\Longleftrightarrow\quad
\ell_\alpha\cap e_a^\circ\neq\varnothing.
\label{eq:B-exuberant-geometric}
\ee
Thus, an intersection of the $E$-line with the relative
interior of a boundary edge records a non-compact common
locus of the corresponding exuberant divisor and boundary
curve. An intersection occurring only at an endpoint of
$e_a$ does not exclude the pair from $\mathcal B$.

It is convenient to combine these two conditions into
a single combinatorial criterion.
For every $a\in A_{0,\partial}$, define
\be
\epsilon_{ia}
=
\begin{cases}
1,
& i\in I_0\text{ and }v_i\in e_a,
\\[1mm]
1,
& i=\alpha\in I_e
\text{ and }\ell_\alpha\cap e_a^\circ\neq\varnothing,
\\[1mm]
0,
& \text{otherwise}.
\end{cases}
\label{eq:bad-incidence}
\ee

We obtain the following characterization of the
admissibility domain:
\be
\mathcal B
=
(I\times A_c)
\sqcup
\left\{
(i,a)\in I\times A_{0,\partial}:
\epsilon_{ia}=0
\right\}.
\label{eq:admissibility-domain}
\ee
Equivalently,
\be
(i,a)\notin\mathcal B
\quad\Longleftrightarrow\quad
a\in A_{0,\partial}
\text{ and }
\begin{cases}
v_i\in e_a,&i\in I_0,\\
\ell_i\cap e_a^\circ\neq\varnothing,&i\in I_e.
\end{cases}
\label{eq:B-complement}
\ee

In particular, the central-fiber and exuberant parts
of the admissibility domain are
\be
\mathcal B_0
=
(I_0\times A_0)
\setminus
\left\{
(i,a):
a\in A_{0,\partial},\ v_i\in e_a
\right\},
\label{eq:B-central}
\ee
and
\be
\mathcal B_e
=
(I_e\times A_c)
\sqcup
\left\{
(\alpha,a)\in I_e\times A_{0,\partial}:
\ell_\alpha\cap e_a^\circ=\varnothing
\right\}.
\label{eq:B-exuberant}
\ee

\subsection{Intersections among Central-Fiber Divisors}
\label{sec:central-fiber-intersections}

We now determine the intersection products involving the
components of the central fiber $\widetilde{\cX}_0$ from
the combinatorial data of the unimodular subdivision
$\tcP$ of the Symington polygon.

\subsubsection{Products of Central-Fiber Divisors}
\label{sec:central-divisor-products}

Let $D_i,D_j$ be central-fiber divisors corresponding
to vertices $v_i,v_j$ of $\tcP$.
Suppose first that $i\neq j$.
Their intersection is represented by the double curves
corresponding to edges joining $v_i$ and $v_j$.
Thus
\be
D_i\cdot D_j
=
\sum_{\substack{a\in A_0\\
C_a\subset D_i\cap D_j}}C_a.
\label{eq:distinct-central-divisors}
\ee
In terms of the coefficients introduced in
Section~\ref{sec:intersection-data}, this gives
\be
\alpha^a_{ij}
=
\begin{cases}
1,
& i\neq j\text{ and }e_a
\text{ joins }v_i\text{ and }v_j,
\\[1mm]
0,
& \text{otherwise},
\end{cases}
\label{eq:alpha-central-offdiag}
\ee
with contributions added if several edge representatives
are identified with the same curve class.

For the diagonal case $i=j$, note that the divisor
$\widetilde{\cX}_0 =\sum_{j \in I_0} D_j$ is a fiber of
$\widetilde{\cX}\rightarrow\Delta$ and so is linearly equivalent to zero.
Hence, 
\be
D_i|_{D_i}
=
-\sum_{\substack{j\in I_0\\j\neq i}}
D_j|_{D_i}.
\ee
On the other hand, adjunction in the Calabi--Yau
threefold gives
$D_i|_{D_i}=K_{D_i}$. 
Since $(D_i,\partial D_i)$ is log Calabi--Yau,
we have
$
D_i|_{D_i}
=
K_{D_i}
=
-\partial D_i$. 
Consequently,
\be
D_i^2
=
-\sum_{\substack{a\in A_0\\
C_a\subset\partial D_i}}C_a.
\label{eq:central-diagonal}
\ee
Thus
\be
\alpha^a_{ii}
=
\begin{cases}
-1,
& e_a\text{ is incident to }v_i,
\\[1mm]
0,
& \text{otherwise}.
\end{cases}
\label{eq:alpha-central-diagonal}
\ee

These formulas apply to both compact and non-compact
central-fiber divisors. In particular, the resulting
curve-valued products may contain non-compact boundary
curves $C_a$, with $a\in A_{0,\partial}$.

\subsubsection{Intersections with Double Curves}
\label{sec:central-divisor-curve-intersections}

We next compute the numerical intersection coefficients
\be
\beta_{ia}=D_i\cdot C_a
\ee
for central-fiber divisors $D_i$ and double curves
$C_a$, whenever $(i,a)\in\mathcal B_0$.

We distinguish intersections with the two divisors
containing a double curve from transverse intersections
with other central-fiber divisors.

\medskip
\noindent
\textbf{Intersections with the two adjacent divisors.}
Let $a\in A_{0,c}$ and let
\be
e_a=[v,w]
\label{eq:double-curve-edge}
\ee
be the corresponding interior edge of $\tcP$.
Denote its two adjacent unimodular triangles by
\be
(v,w,x),
\qquad
(v,w,y).
\label{eq:adjacent-triangles}
\ee
The edge $e_a$ corresponds to a compact double curve
\be
C_a\subset D_v\cap D_w.
\ee
We assume for simplicity that the two components are distinct and
meet transversely along $C_a$, as in the normal-crossings
central fiber of Section~\ref{sec:GlobalCY}.
The intersection numbers
\be
\beta_{va}=D_v\cdot C_a,
\qquad
\beta_{wa}=D_w\cdot C_a
\ee
are related to the self-intersections of $C_a$ in
the two surfaces.
Indeed, since $D_w|_{D_v}$ contains $C_a$ with
multiplicity one, and locally along $C_a$ the two
divisors meet transversely, we have
\be
D_w\cdot C_a=C_a^2|_{D_v},
\qquad
D_v\cdot C_a=C_a^2|_{D_w}.
\label{eq:double-curve-crossed-restrictions}
\ee

We first compute $C_a^2|_{D_v}$.
The primitive tangent vector
\be
u=w-v
\ee
determines the ray corresponding to $C_a$ in the
integral-affine fan at $v$.
The two adjacent rays are obtained from the directions
towards $x$ and $y$.

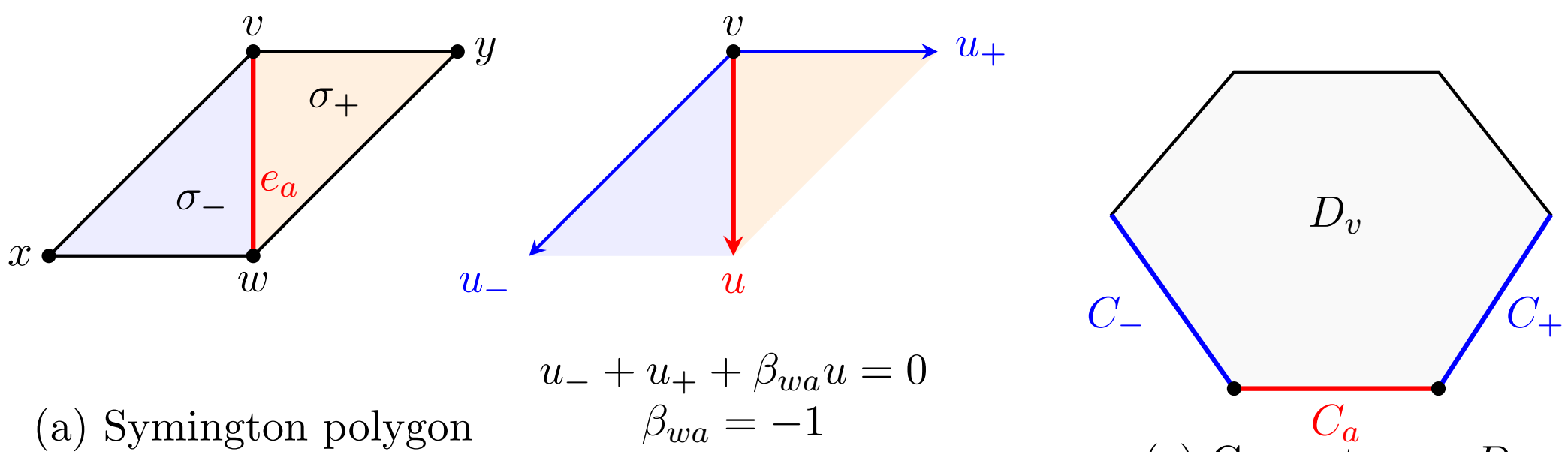
\begin{figure}[htbp]
\centering
\begin{tikzpicture}[
    >=stealth,
    scale=0.9,
    every node/.style={font=\small}
]

\begin{scope}[shift={(0,0)}]

    \coordinate (v) at (2,3);
    \coordinate (w) at (2,1);
    \coordinate (x) at (0,1);
    \coordinate (y) at (4,3);

    \fill[blue!7] (v)--(w)--(x)--cycle;
    \fill[orange!12] (v)--(w)--(y)--cycle;

    \draw[thick] (v)--(x)--(w)--(y)--cycle;

    \draw[very thick,red] (v)--(w);

    \foreach \P in {v,w,x,y}
        \fill (\P) circle (2pt);

    \node[above] at (v) {$v$};
    \node[below] at (w) {$w$};
    \node[left] at (x) {$x$};
    \node[right] at (y) {$y$};

    \node[red,right] at (1.9,1.7) {$e_a$};

    \node at (1.5,1.5) {$\sigma_-$};
    \node at (2.8,2.5) {$\sigma_+$};

    \node at (2,-0.7)
        {(a) Symington polygon};

\end{scope}

\begin{scope}[shift={(6.7,3)}]

    \coordinate (O) at (0,0);


    \coordinate (U)  at (0,-2);
    \coordinate (Um) at (-2,-2);
    \coordinate (Up) at (2,0);

    \fill[blue!7] (O)--(Um)--(U)--cycle;
    \fill[orange!12] (O)--(U)--(Up)--cycle;

    \draw[->,thick,blue]
    (O)--(-2,-2)
    node[below left] {$u_-$};

\draw[->,very thick,red]
    (O)--(0,-2)
    node[below] {$u$};

\draw[->,thick,blue]
    (O)--(2,0)
    node[right] {$u_+$};

    \fill (O) circle (2pt);
    \node[above] at (O) {$v$};

    \node[align=center] at (0,-3.15)
        {$u_-+u_++\beta_{wa}u=0$};

    \node at (0,-3.65)
        {$\beta_{wa}=-1$};

    \node at (0,-4.25)
        {(b) Fan at $v$};

\end{scope}

\begin{scope}[shift={(10.4,0.4)}]

    \coordinate (A) at (0,1);
    \coordinate (B) at (1.2,-0.7);
    \coordinate (C) at (3.2,-0.7);
    \coordinate (D) at (4.3,1);
    \coordinate (E) at (3.2,2.4);
    \coordinate (F) at (1.2,2.4);

    \fill[gray!5]
        (A)--(B)--(C)--(D)--(E)--(F)--cycle;

    \draw[thick]
        (A)--(B)--(C)--(D)--(E)--(F)--cycle;

    \draw[very thick,red] (B)--(C);

    \draw[very thick,blue] (A)--(B);
    \draw[very thick,blue] (C)--(D);

    \fill (B) circle (2pt);
    \fill (C) circle (2pt);

    \node[red,below] at (2.2,-0.7)
        {$C_a$};

    \node[blue,left] at (0.5,0.0)
        {$C_-$};

    \node[blue,right] at (3.7,0.0)
        {$C_+$};

    \node at (2.2,1.0)
        {$D_v$};


    \node at (2.2,-1.45)
        {(c) Geometry on $D_v$};

\end{scope}

\end{tikzpicture}

\caption{
(a) Two adjacent unimodular triangles sharing the
edge $e_a=[v,w]$, corresponding to the double curve
$C_a=D_v\cap D_w$.
(b) The primitive directions in the fan at $v$
satisfy
$u_-+u_++\beta_{wa}u=0$.
(c) The corresponding boundary curves on
$D_v$.
}
\label{fig:integral-affine-double-curve}
\end{figure}

These three rays must be expressed in a common
integral-affine chart.
If the two triangles are separated by an affine cut,
we parallel transport the tangent vectors across the
cut using the corresponding integral-affine transition.
Choose the chart containing $(v,w,x)$, and let $M\in SL(2,\ZZ)$ denote the linear part of the
transition from the chart containing $(v,w,y)$
to this chosen chart.
Thus, as illustrated in Figure \ref{fig:integral-affine-double-curve}, the three primitive rays at $v$ are
\be
u=w-v,
\qquad
u_-=x-v,
\qquad
u_+=M(y-v).
\label{eq:double-curve-fan-rays}
\ee
The notation $M(y-v)$ means that the tangent vector
$y-v$ is transported into the chosen chart.
If no cut is crossed, $M$ is the identity.

Since the two adjacent cones are unimodular,
the primitive rays satisfy a unique integral relation
\be
u_-+u_++n_vu=0,
\label{eq:double-curve-fan-relation}
\ee
where $n_v$ is the self-intersection of the
corresponding boundary curve in $D_v$:
\be
n_v=C_a^2|_{D_v}.
\ee
Using \eqref{eq:double-curve-crossed-restrictions},
we obtain
\be
n_v=\beta_{wa}.
\ee
Therefore the integral-affine self-intersection rule
at $v$ is
\be
(x-v)+M(y-v)+\beta_{wa}(w-v)=0.
\label{eq:double-curve-vector-rule-v}
\ee
This vector relation uniquely determines the integer
$\beta_{wa}$.

Similarly, at the other endpoint $w$, the primitive
ray corresponding to $C_a$ is $v-w$.
Using the same parallel-transport convention, the
three consecutive rays are
\be
v-w,
\qquad
x-w,
\qquad
M(y-w).
\ee
The corresponding fan relation gives
\be
(x-w)+M(y-w)+\beta_{va}(v-w)=0.
\label{eq:double-curve-vector-rule-w}
\ee

For computational purposes, one may extract the
integer coefficients by taking an ordinary Euclidean
inner product with the corresponding edge direction.
This gives
\be
\begin{aligned}
\beta_{va}
&=
-\frac{
\bigl((x-w)+M(y-w)\bigr)\cdot(v-w)
}{
|v-w|^2
},
\\[2mm]
\beta_{wa}
&=
-\frac{
\bigl((x-v)+M(y-v)\bigr)\cdot(w-v)
}{
|w-v|^2
}.
\end{aligned}
\label{eq:double-curve-self-intersections}
\ee

The Euclidean inner product is used only to extract
the integer coefficient from the vector relation.
The underlying rule is integral-affine and does not
depend on a choice of Euclidean metric.

When no affine cut is crossed, $M=1$, and the
formulas reduce to the standard toric
self-intersection rules.

\medskip
\noindent
\textbf{Transverse intersections with double curves.}

Now let $D_i$ be a central-fiber divisor whose
corresponding vertex $v_i$ is not an endpoint
of $e_a$.
The intersection of $D_i$ with $C_a$ occurs at
the triple points corresponding to triangles
containing both $e_a$ and $v_i$.
Thus, whenever $(i,a)\in\mathcal B_0$,
\be
\beta_{ia}
=
\#\left\{
\sigma\in\tcP^{[2]}:
e_a\subset\sigma,\quad
v_i\in\sigma\setminus e_a
\right\}.
\label{eq:beta-triangle-count}
\ee

This formula also applies to non-compact boundary
curves $C_a$, with $a\in A_{0,\partial}$, provided
$v_i\notin e_a$.
In that case the intersection is a finite collection
of triple points and therefore has a well-defined
degree.
If instead $v_i\in e_a$, the non-compact curve
$C_a$ is contained in $D_i$.
The pair $(i,a)$ is then inadmissible, in agreement
with Section~\ref{sec:admissible-intersections}.

\begin{remark}
For a compact smooth rational double curve
$C_a=D_v\cap D_w$, the Calabi--Yau condition
implies
\be
\beta_{va}+\beta_{wa}=-2.
\label{eq:CY-double-curve-relation}
\ee
Indeed, since $D_v$ and $D_w$ are distinct smooth
divisors meeting transversely along $C_a$,
the adjunction formula for the complete intersection
gives
\be
K_{C_a}
=
(K_{\widetilde{\cX}}+D_v+D_w)|_{C_a}.
\ee
Using $K_{\widetilde{\cX}}=0$, we obtain
\be
\deg K_{C_a}
=
D_v\cdot C_a+D_w\cdot C_a.
\ee
Since $C_a\simeq\PP^1$, this becomes
\be
\beta_{va}+\beta_{wa}=-2.
\ee
This relation provides an independent consistency
check of our calculations.
\end{remark}

\subsection{Intersections involving Exuberant Divisors}
\label{sec:exuberantIntersectionTheory}

We now turn to intersections involving the exuberant divisors
$E_\alpha$, $\alpha\in I_e$, associated with the $E$-lines
of the Symington polygon.
Unlike intersections among central-fiber divisors, which are
determined by the integral-affine triangulation, intersections
involving exuberant divisors require additional geometric data.
The primitive direction of an $E$-line determines the 
intersection numbers of its associated exuberant curve, while
its incidence with the triangulation constrains the remaining
divisor--curve intersections. These rules determine part of the
intersection data directly. The remaining coefficients will be
determined by the compatibility relations developed in
Section~\ref{sec:compatibility}.

\subsubsection{Intersection Numbers of Exuberant Curves}
\label{sec:local-exuberant-intersections}

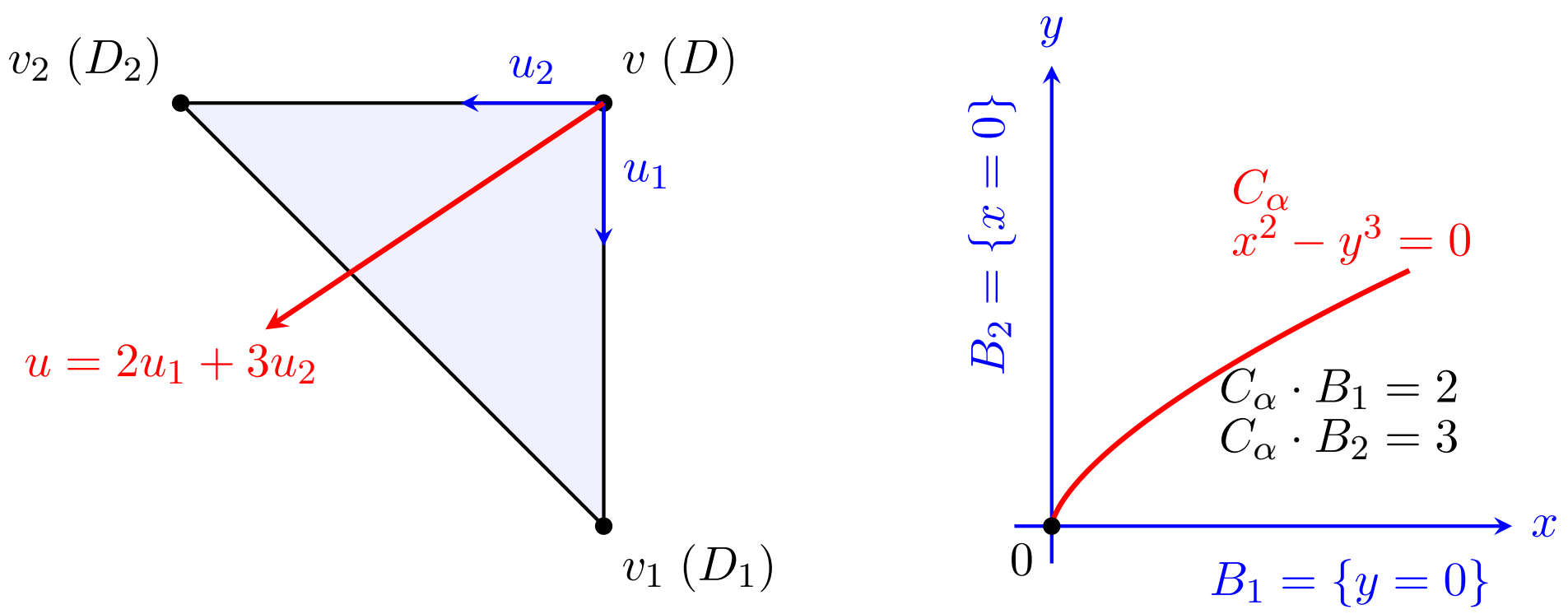
\begin{figure}[ht]
\centering
\begin{tikzpicture}[
    >=stealth,
    scale=1,
    every node/.style={font=\small}
]

\begin{scope}[shift={(0,0)}]

    \coordinate (v)  at (3.4,3.4);
    \coordinate (v1) at (3.4,0);
    \coordinate (v2) at (0,3.4);

    \fill[blue!6] (v)--(v1)--(v2)--cycle;
    \draw[thick] (v)--(v1)--(v2)--cycle;

    \fill (v) circle (2pt);
    \fill (v1) circle (2pt);
    \fill (v2) circle (2pt);

    \node[above right] at (v) {$v\;(D)$};
    \node[below right] at (v1) {$v_1\;(D_1)$};
    \node[above left] at (v2) {$v_2\;(D_2)$};

    \draw[->,thick,blue]
        (v)--(3.4,2.25)
        node[midway,right] {$u_1$};

    \draw[->,thick,blue]
        (v)--(2.25,3.4)
        node[midway,above] {$u_2$};

    \draw[very thick,red,->]
        (v)--(0.68,1.58);

    \node[red,anchor=east] at (1.25,1.3)
        {$u=2u_1+3u_2$};

    \node at (1.7,-1.05)
        {(a) Local Symington polygon};

\end{scope}

\begin{scope}[shift={(7.0,0)}]

    \draw[thick,blue,->]
        (-0.3,0)--(3.7,0)
        node[right] {$x$};

    \draw[thick,blue,->]
        (0,-0.3)--(0,3.7)
        node[above] {$y$};

    \node[blue,below] at (2.4,-0.12)
        {$B_1=\{y=0\}$};

    \node[blue,rotate=90] at (-0.48,2.35)
        {$B_2=\{x=0\}$};

    \draw[
        very thick,
        red,
        domain=0:1.4,
        samples=100,
        smooth,
        variable=\t
    ]
        plot ({1.05*\t*\t*\t},
              {1.05*\t*\t});

    \fill (0,0) circle (2pt);
    \node[below left] at (0,0) {$0$};

    \node[red,anchor=west] at (1.3,2.7)
        {$C_\alpha$};

    \node[red,anchor=west] at (1.3,2.3)
        {$x^2-y^3=0$};

    \node[anchor=west] at (1.2,1.10)
        {$C_\alpha\cdot B_1=2$};

    \node[anchor=west] at (1.2,0.7)
        {$C_\alpha\cdot B_2=3$};

    \node at (1.7,-1.05)
        {(b) Local geometry on $D$};

\end{scope}

\end{tikzpicture}

\caption{
Local geometry of an exuberant curve $C_\alpha$
at the initial vertex of its $E$-line.
The primitive direction $u=pu_1+qu_2$ lies
in a unimodular cone of the Symington polygon,
illustrated for $(p,q)=(2,3)$ (left).
The corresponding curve on $D$ has local
equation $x^p-y^q=0$, with boundary components
$B_1=D_1|_D=\{y=0\}$ and
$B_2=D_2|_D=\{x=0\}$ (right).
}
\label{fig:exuberant-local-geometry}
\end{figure}

Let $E_\alpha$, $\alpha\in I_e$, be an exuberant divisor
whose $E$-line $\ell_\alpha$ starts at the vertex $v$
corresponding to a compact central-fiber divisor $D$.
Let $u$ be the primitive direction of $\ell_\alpha$
at $v$. Choose a unimodular cone in the star of $v$,
with primitive generators $u_1,u_2$, containing $u$.
We may write
\be
u=pu_1+qu_2,
\qquad
p,q\in\ZZ_{\geq0},
\qquad
\gcd(p,q)=1.
\label{eq:E-direction}
\ee
Let $D_1,D_2$ be the central-fiber divisors corresponding
to the far endpoints of the edges in the directions
$u_1,u_2$, respectively. 
As reviewed in Section \ref{sec:ExuberantGeometry}, the 
intersection of $E_\alpha$
with $D$ is the compact exuberant curve
$C_\alpha:=D\cdot E_\alpha$. The corresponding local geometry is illustrated in 
Figure \ref{fig:exuberant-local-geometry}.

\begin{proposition}
\label{prop:local-E-line-intersections}
With the notation above,
\be
 D\cdot C_\alpha=-p-q,
\qquad
D_1\cdot C_\alpha=p,
\qquad
D_2\cdot C_\alpha=q,
\qquad
E_\alpha\cdot C_\alpha=pq-1  \, .
\label{eq:exuberant-local-data}
\ee
All other central-fiber divisors have zero intersection
with $C_\alpha$.
\end{proposition}

\begin{proof}
Near the point where $C_\alpha$ meets the boundary of $D$,
choose local coordinates $x,y$ such that the boundary components
corresponding to $u_1,u_2$ are given by $y=0$ and $x=0$,
respectively. Since $u=pu_1+qu_2$, the curve $C_\alpha$
has local equation
\[
x^p-y^q=0,
\]
and hence admits the parametrization $x=t^q$, $y=t^p$.
Its intersection multiplicities with the two boundary components
are therefore $p$ and $q$, respectively.
As these boundary components are the restrictions of $D_1$
and $D_2$ to $D$, we obtain
\[
D_1\cdot C_\alpha=p,\qquad D_2\cdot C_\alpha=q.
\]
No other central-fiber divisor distinct from $D$ meets $C_\alpha$.

To compute the intersection with $D$, recall
that $D$ is a log Calabi--Yau surface and
that its normal bundle in $\widetilde{\cX}$
is its canonical bundle. Thus
$D|_D=K_D=-\partial D$.
The curve $C_\alpha$ meets the boundary
$\partial D$ at a single point, with total
intersection multiplicity
\be
C_\alpha\cdot\partial D=p+q.
\label{eq:E-boundary-intersection}
\ee
Consequently,
\be
D\cdot C_\alpha
=
K_D\cdot C_\alpha
=
-(p+q).
\label{eq:E-D-intersection}
\ee

It remains to determine $E_\alpha \cdot C_\alpha$.
Suppose first that $p,q>0$.
The local geometric construction describes
$C_\alpha$ as the image of an exceptional
rational curve under the corresponding
toric blow-down. Its normalization is
therefore $\PP^1$, while its unique
singular point has local analytic equation
\be
f:= x^p-y^q=0.
\label{eq:E-cusp-local-equation}
\ee
Hence, the arithmetic genus of $C_\alpha$ equals the delta invariant $\delta$ of this singularity. Moreover, $\gcd(p,q)=1$ implies that the singularity is unibranch. Its Milnor number is
\be
\mu
=
\dim_\CC
\frac{\CC[x,y]}
{(\partial_x f,\partial_y f)}
=
(p-1)(q-1).
\ee
Since the singularity is unibranch,
the Milnor formula, \cite[Theorem 10.5]{Milnor1968}, gives
$2\delta=\mu$. 
Hence
\be
p_a(C_\alpha)
=
\delta
=
\frac{(p-1)(q-1)}{2}.
\label{eq:E-arithmetic-genus}
\ee
We now apply the adjunction formula for
an integral curve on the smooth surface $D$:
\be
2p_a(C_\alpha)-2
=
C_\alpha^2|_D
+
K_D\cdot C_\alpha.
\label{eq:E-curve-adjunction}
\ee
Using
\eqref{eq:E-arithmetic-genus}
and \eqref{eq:E-D-intersection}, we obtain
\be
\begin{aligned}
C_\alpha^2|_D
&=
2p_a(C_\alpha)-2
-K_D\cdot C_\alpha
\\
&=
(p-1)(q-1)-2+p+q
\\
&=pq-1.
\end{aligned}
\label{eq:E-self-intersection-derivation}
\ee
Since $C_\alpha=D\cap E_\alpha$, the
restriction of $E_\alpha$ to $D$ is
the divisor $C_\alpha$. Therefore
\be
E_\alpha\cdot C_\alpha
=
(E_\alpha|_D)\cdot C_\alpha
=
C_\alpha^2|_D
=
pq-1.
\label{eq:E-self-intersection-threefold}
\ee
Finally, suppose that one of $p,q$ vanishes.
Primitivity forces the other to be one.
Then 
$C_\alpha$ is an exceptional curve,
and hence
\be
C_\alpha^2|_D=-1=pq-1.
\ee
Thus the formula holds in all cases.
\end{proof}

\begin{remark}
The local intersection numbers satisfy
\be
C_\alpha\cdot(D+D_1+D_2)
=
-(p+q)+p+q
=
0.
\ee
This is consistent with the fact that
the central fiber $\widetilde{\cX}_0$ is linearly equivalent to zero, and therefore has zero intersection number with 
every compact curve.
\end{remark}

\subsubsection{Divisor Products
determined by $E$-lines}
\label{sec:exuberant-incidence}

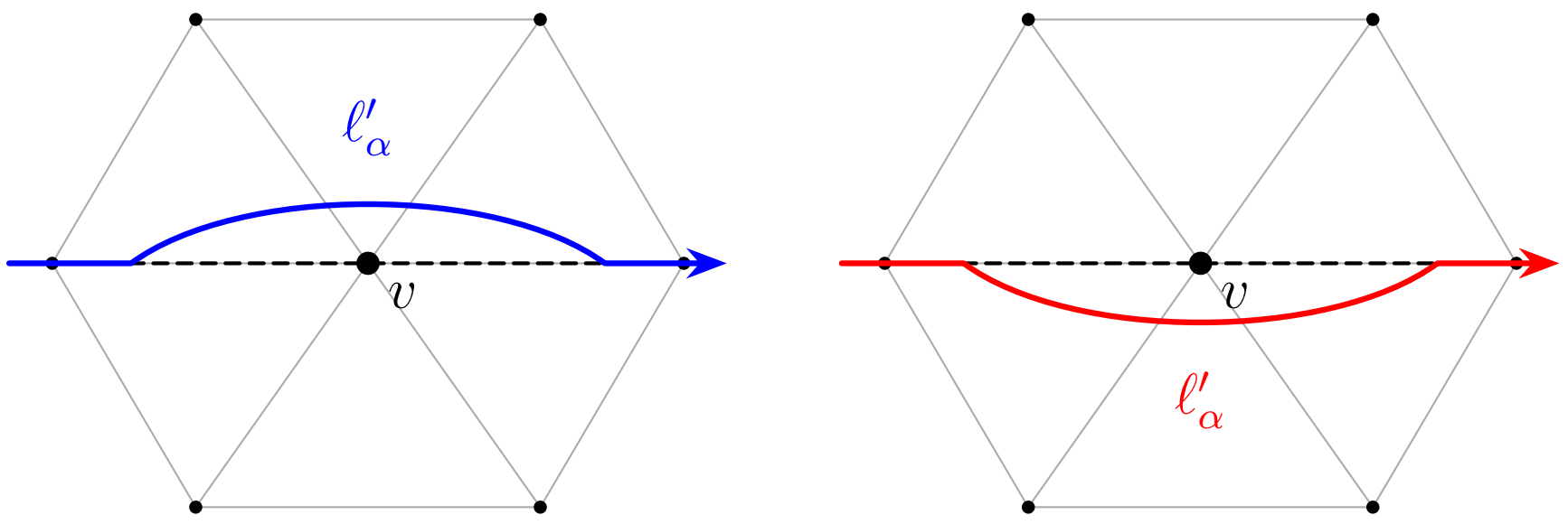
\begin{figure}[htbp]
\centering
\begin{tikzpicture}[
    scale=1.05,
    line cap=round,
    line join=round,
    >=Stealth
]

\foreach \s/\col in {0/blue,5.8/red} {

    \begin{scope}[xshift=\s cm]
        \coordinate (O) at (0,0);
        \coordinate (L) at (-2.2,0);
        \coordinate (R) at (2.2,0);
        \coordinate (TL) at (-1.2,1.7);
        \coordinate (TR) at (1.2,1.7);
        \coordinate (BL) at (-1.2,-1.7);
        \coordinate (BR) at (1.2,-1.7);

        \draw[gray!65,thin]
          (TL)--(TR)--(R)--(BR)--(BL)--(L)--cycle;
        \draw[gray!65,thin]
          (O)--(L) (O)--(R)
          (O)--(TL) (O)--(TR)
          (O)--(BL) (O)--(BR);

        \draw[black,densely dashed,thick]
          (-2.5,0)--(2.5,0);

        \foreach \p in {L,R,TL,TR,BL,BR}
            \fill[black] (\p) circle (1.3pt);

        \fill[black] (O) circle (2.3pt);
        \node[below right] at (O) {$v$};
    \end{scope}
}

\begin{scope}
    \draw[blue,very thick,->]
      (-2.5,0)
      --(-1.65,0)
      .. controls (-0.9,0.55) and (0.9,0.55) ..
      (1.65,0)
      --(2.5,0);

    \node[blue] at (0,0.95) {$\ell_\alpha'$};
    \node at (0,-2.25) {(a) Bypass above $v$};
\end{scope}

\begin{scope}[xshift=5.8cm]
    \draw[red,very thick,->]
      (-2.5,0)
      --(-1.65,0)
      .. controls (-0.9,-0.55) and (0.9,-0.55) ..
      (1.65,0)
      --(2.5,0);

    \node[red] at (0,-0.95) {$\ell_\alpha'$};
    \node at (0,-2.25) {(b) Bypass below $v$};
\end{scope}

\draw[black,densely dashed,thick]
    (2.3,-3.0)--(3.1,-3.0);
\node[right] at (3.2,-3.0)
    {$\ell_\alpha$ (original $E$-line)};

\end{tikzpicture}

\caption{
Two possible infinitesimal deformations of an
$E$-line $\ell_\alpha$ passing through a vertex
$v$ of the triangulation $\widetilde{\mathcal P}$.
The dashed line represents $\ell_\alpha$.
The deformed path $\ell_\alpha'$ avoids $v$
by passing on either side of it.
The two choices intersect the interiors of
different edges adjacent to $v$, and hence
determine different allowed supports for the
curve expansion of $E_\alpha\cdot D_j$.
}
\label{fig:E-line-vertex-bypass}
\end{figure}

Let $j \in I_0$. 
As introduced in
Section~\ref{sec:intersection-data},
we write
\be
E_\alpha\cdot D_j
=
\sum_{a\in A}
\alpha^a_{\alpha j}C_a.
\label{eq:E-D-expansion}
\ee
To specify a unique curve expansion of
$E_\alpha \cdot D_j$ in \eqref{eq:E-D-expansion}, we choose a small
perturbation $\ell'_\alpha$ of the $E$-line $\ell_\alpha$
that avoids all vertices of the triangulation except
its initial vertex. At each vertex traversed by
$\ell_\alpha$, this amounts to choosing one of the
two possible sides along which $\ell'_\alpha$ bypasses
the vertex, as illustrated in Figure \ref{fig:E-line-vertex-bypass}.
This choice determines the set of curve classes
allowed to appear in the expansion, as explained below.

First, denote by $v_j$ the vertex of $\tcP$ corresponding to $D_j$.
Define 
\be A_{\alpha j} \subset A 
\label{eq_A_alpha_j}
\ee
to be the set of $a \in A$ such that either:
\begin{enumerate}
    \item $a \in A_0$, the corresponding edge 
    $e_a$ of $\tcP$ is incident to $v_j$, and $\ell_\alpha'$ intersects the interior of $e_a$, or
     \item $a = \alpha$, or
    \item $a=\beta \in A_e$, the corresponding focus--focus singularity $V_\beta$ is originally located before perturbation at $v_j$, and $\ell_\alpha'$ passes between $V_\beta$ and $v_j$.
\end{enumerate}

\begin{lemma} \label{lem}
There exists a unique curve expansion of $E_\alpha \cdot D_j$ such that 
\be  \label{eq_zero}
\alpha^a_{\alpha j}=0\ee 
unless $a \in A_{\alpha j}$.
\end{lemma}

\begin{proof}
If $v_j$ does not lie on $\ell_\alpha$, then every irreducible component of the geometric intersection $E_\alpha \cap D_j$
is of the form $C_a$, where $a \in A_0$ is necessarily in $A_{\alpha j}$ (Case 1.).
If $v_j$ is the initial point of $\ell_\alpha$, then $E_\alpha \cap D_j=C_\alpha$, with $\alpha \in A_{\alpha j}$ (Case 2.).

The more interesting situation is when $v_j$ lies on $\ell_\alpha$ but is not the initial point. We focus on the 
case where $v_j$ is a smooth point of the integral-affine structure, so that the corresponding surface $X_{v_j}$ is toric. The case of a singular point can be treated similarly using a toric model, as in Section \ref{sec:ExuberantGeometry}.
Let $u_1,\ldots,u_N$ be the primitive directions of the edges outgoing from  $v_j$, spanning the fan of $X_{v_j}$ and corresponding to toric divisors
$B_1,\ldots,B_N$.
Write the outgoing and incoming directions of $\ell_\alpha$
in their respective unimodular cones as
\be
u=pu_k+qu_{k+1},
\qquad
-u=ru_l+su_{l+1}.
\ee

Then, arguing as in the beginning of the proof of Proposition \ref{prop:local-E-line-intersections}, the boundary intersection numbers of
the curve $C := E_\alpha \cap X_{v_j}$ are determined by these
decompositions:
\be
\begin{aligned}
B_k\cdot C&=p,
&
B_{k+1}\cdot C&=q,
\\
B_l\cdot C&=r,
&
B_{l+1}\cdot C&=s.
\end{aligned}
\ee
All other boundary intersections vanish, so these intersection numbers uniquely determine the class of $C$ in the toric surface $X_{v_j}$. Geometrically, $C$ is the closure of a general translate of a one-parameter subgroup of the torus $(\mathbb C^\star)^2\subset X_{v_j}$ with primitive direction $u$. The choice of how $\ell'_\alpha$ bypasses $v_j$ selects one of the two chains of toric boundary curves onto which $C$ can degenerate. This yields a unique expansion of $[C]$ in terms of the boundary curve classes belonging to $A_{\alpha j}$ (Case 3.).
 More precisely, orient the fan so
that the chosen chain lies counterclockwise from $u$ to $-u$.
Then
\begin{equation}
    [C] = \sum_{m=1}^N
    \max\{0,\det(u,u_m)\}\,[B_m].
\end{equation}
Indeed, the right-hand side is the effective toric boundary divisor obtained by taking the positive part of the principal divisor of a character whose restriction to the one-parameter subgroup with direction $u$ is trivial.

Uniqueness follows from the toric linear equivalences. Indeed, since the fan is complete, the rays in the complementary closed half-plane span the ambient two-dimensional vector space. Thus, any principal toric divisor supported on the selected chain must vanish.
\end{proof}


For $\alpha\in I_e$, we denote by $b(\alpha)\in I_c$ the index of the initial vertex $v_{b(\alpha)}$ of $\ell_\alpha$, so that $D_{b(\alpha)}$ is the compact divisor containing the exuberant curve $C_\alpha$.
At the initial vertex
$v_{b(\alpha)}$ of $\ell_\alpha$,
the intersection with the corresponding
compact divisor contains the exuberant
curve $C_\alpha$ with multiplicity one.
Thus
\be
\alpha^\alpha_{\alpha,b(\alpha)}=1.
\label{eq:E-base-alpha}
\ee
The $E$-line geometry also constrains
the numerical divisor--curve intersections
\be
\beta_{\alpha a}
=
E_\alpha\cdot C_a.
\ee
For compact curves, these intersections
are always admissible:
\be
(\alpha,a)\in\mathcal B,
\qquad
a\in A_c.
\ee

For a non-compact boundary curve
$C_a$, with $a\in A_{0,\partial}$,
the admissibility criterion of
Section~\ref{sec:admissible-intersections}
gives
\be
(\alpha,a)\in\mathcal B
\quad\Longleftrightarrow\quad
\ell_\alpha\cap e_a^\circ=\varnothing.
\label{eq:E-beta-admissible}
\ee
Thus the numerical intersection
$\beta_{\alpha a}$ is intrinsically
defined precisely when the corresponding
$E$-line does not meet the relative
interior of the boundary edge $e_a$.

Moreover, for every admissible pair $(\alpha,a) \in\mathcal B$, we have
\be
\beta_{\alpha a}=0 
\label{eq:E-beta-zero}
\ee
unless the exuberant divisor $E_\alpha$ and the curve $C_a$ have a non-empty geometric intersection, that is, unless either
\begin{enumerate}
    \item $a \in A_0$ and the corresponding edge $e_a$ of $\tcP$ is an edge of a triangle traversed by $\ell_\alpha$, or
    \item  $a = \beta \in A_e$, and $\ell_\alpha$ traverses a triangle with vertex $v_{b(\beta)}$ and containing the initial direction of $\ell_\beta$. 
\end{enumerate}

The intersection data involving exuberant
divisors are therefore partially determined
by three types of geometric information:

\begin{enumerate}

\item
The primitive direction of the $E$-line
at its initial vertex determines the
intersection numbers
\eqref{eq:exuberant-local-data}.

\item
The incidence of the $E$-line with
the triangulation determines the
possible support of the curve-valued
products \eqref{eq:E-D-expansion}.

\item
The admissibility and disjointness
conditions determine which numerical
intersections $\beta_{\alpha a}$
are defined and which of them vanish.

\end{enumerate}

These rules determine part of the
exuberant $\alpha$- and $\beta$-data
directly. However, some coefficients
allowed by the incidence conditions
are not fixed yet.
To determine the remaining coefficients,
we impose compatibility relations
arising from associativity of the
intersection product, which we discuss
in the following subsection.

\subsection{Compatibility Relations from Associativity}
\label{sec:compatibility}

The intersection rules derived in
Sections~\ref{sec:central-fiber-intersections}
and~\ref{sec:exuberantIntersectionTheory}
determine part of the intersection data directly
from the geometry of the Symington polygon and
its $E$-lines. 
However, some coefficients in the curve-valued
products $E_\alpha\cdot D_j$ and the admissible
numerical intersections $E_\alpha\cdot C_a$
are not fixed by these local rules.
We now explain how associativity of the
intersection product provides compatibility
relations among these coefficients.

Suppose that a curve-valued divisor product
admits an expansion
\be
D_j\cdot D_k
=
\sum_{a\in A}\alpha^a_{jk}C_a.
\label{eq:general-alpha-expansion}
\ee

We say that this expansion is
\emph{admissible against $D_i$} if 
one of the following holds:
\begin{enumerate}
    \item $(j,k) \in I_0 \times I_0$, so that the coefficients $\alpha^a_{jk}$ are known by \eqref{eq:alpha-central-offdiag}-\eqref{eq:alpha-central-diagonal}, and we require \be
\alpha^a_{jk}\neq0
\qquad\Longrightarrow\qquad
(i,a)\in\mathcal B\,.
\label{eq:pairing-admissibility}
\ee
    \item $(j,k) \in I_e \times I_0$, and we require 
    \be a \in A_{jk}  \qquad\Longrightarrow\qquad
(i,a)\in\mathcal B \,,
\ee
where $A_{jk} \subset A$ is defined in 
\eqref{eq_A_alpha_j}.
\end{enumerate}

Whenever this condition holds, we define
the corresponding numerical iterated
intersection by
\be
T_{i;jk}
:=
D_i\cdot(D_j\cdot D_k)
=
\sum_{a\in A}\alpha^a_{jk}\beta_{ia}.
\label{eq:T-definition}
\ee

The intersection product is associative
and commutative. Consequently, different
admissible presentations of the same
triple intersection must agree.
We therefore obtain the 
compatibility relation
\be
T_{i;jk}=T_{j;ik}
\label{eq:master-compatibility}
\ee
whenever both sides are defined.
If all three presentations are admissible,
then
\be
T_{i;jk}
=
T_{j;ik}
=
T_{k;ij}.
\label{eq:three-way-compatibility}
\ee

These equations provide constraints on
the intersection coefficients that are
not determined by the local geometric
rules.
An important feature is that the
compatibility relations are not restricted
to products containing a compact divisor.
Indeed, a non-compact boundary curve
$C_a$, with $a\in A_{0,\partial}$,
may appear in the expansion
\eqref{eq:general-alpha-expansion}
and still contribute to
\eqref{eq:T-definition}, provided
its intersection with $D_i$ is
admissible.

\subsection{Triple Intersection Numbers}
\label{sec:compact-first}

The compatibility relations of the preceding
subsection determine the intersection coefficients
needed to reconstruct the triple intersection
numbers of divisors.
Although these relations may involve non-compact
curves and non-compact divisors, the intrinsic
numerical triple intersections relevant for the
five-dimensional prepotential are precisely
those containing at least one compact divisor.
We now explain how these intrinsic triple
intersections are extracted from the
intersection coefficients.

Let $i\in I_c$ and let $D_j,D_k$ be arbitrary
divisors, with $j,k\in I$.
Since $D_i$ is compact, its intersection
with any divisor $D_k$ is a compact
curve class. Hence it admits an expansion
in compact curves:
\be
D_i\cdot D_k
=
\sum_{a\in A_c}\alpha^a_{ik}C_a.
\label{eq:compact-divisor-product}
\ee
Every compact curve can be paired
numerically with every divisor.
In particular,
\be
(j,a)\in\mathcal B,
\qquad
j\in I,\quad a\in A_c.
\ee
It follows that the intrinsic triple
intersection number is
\be
D_i\cdot D_j\cdot D_k
=
\sum_{a\in A_c}
\alpha^a_{ik}\beta_{ja},
\qquad
(i,j,k)\in I_c\times I\times I.
\label{eq:intersection-master-formula}
\ee

This is the basic formula for extracting
the intrinsic triple intersection form
from the intersection coefficients.

For $i\in I_c$, the intrinsic
triple intersection
\be
D_i\cdot E_\alpha\cdot E_\beta
\ee
can be computed by first expanding
the compact curve class
\be
D_i\cdot E_\beta
=
\sum_{a\in A_c}
\alpha^a_{i\beta}C_a
\ee
and then intersecting with $E_\alpha$:
\be
D_i\cdot E_\alpha\cdot E_\beta
=
\sum_{a\in A_c}
\alpha^a_{i\beta}\beta_{\alpha a}.
\label{eq:two-exuberant-triple}
\ee

Thus the curve-valued products
involving at least one central-fiber
divisor, together with the admissible
divisor--curve intersection coefficients,
suffice to determine all intrinsic
triple intersection numbers.

\begin{remark}
For every compact component $D_i$ of $\widetilde{\cX}_0$, the triple self-intersection number $D_i^3$ is given by
\be
D_i^3
=
12-Q_i-\chi_{\rm top}(\partial D_i),
\label{eq:compact-triple-self-general}
\ee
where
$Q_i
=
\chi_{\rm top}(D_i\setminus\partial D_i)$ is the charge of $D_i$.
If $\partial D_i$ is a cycle of
$r_i$ rational curves, then
\be
D_i^3=12-Q_i-r_i.
\label{eq:compact-triple-self-cycle}
\ee
Indeed, since $\widetilde{\cX}$ is
Calabi--Yau, adjunction gives
$D_i|_{D_i}=K_{D_i}$. 
Since $(D_i,\partial D_i)$
is log Calabi--Yau,
we have $K_{D_i}=-\partial D_i$. 
Therefore
$D_i^3
=
K_{D_i}^2$. 
The surface $D_i$ is rational,
so Noether's formula gives
$K_{D_i}^2
=
12-\chi_{\rm top}(D_i)$. 
By additivity of the topological
Euler characteristic, 
$\chi_{\rm top}(D_i)
=
\chi_{\rm top}(D_i\setminus\partial D_i)
+
\chi_{\rm top}(\partial D_i)$.
Using the definition of $Q_i$,
we obtain
$\chi_{\rm top}(D_i)
=
Q_i+\chi_{\rm top}(\partial D_i)$, which proves the formula \eqref{eq:compact-triple-self-general} for $D_i^3$.

This formula
will not be needed as input to the
reconstruction algorithm.
Instead, it provides an independent
consistency check on the intersection
coefficients obtained from the
combinatorial rules.
\end{remark}



\section{Algorithm for Triple Intersection Numbers}
\label{sec:intersectionAlgorithm}

In this section, we show that the intersection rules developed in
Section~\ref{sec:intersectionTheory} allow us to reconstruct
the triple intersection numbers of the Calabi--Yau threefold
associated with a GTP directly from
its combinatorial data.

We fix a GTP, or equivalently the associated
Symington polygon, together with a choice of perturbation of the integral-affine singularities, which induces a choice of exuberant divisors.
The goal is to determine the triple intersection numbers
\be
\kappa_{ijk}:=D_i\cdot D_j\cdot D_k,
\qquad
\{i,j,k\}\cap I_c\neq\varnothing  
\label{eq:algorithm-output}
\ee
between divisors on the generalized toric Calabi--Yau threefold $\widetilde{\cX}$ associated to the GTP as in Section \ref{sec:CYGTP}. As reviewed in Section \ref{sec:CanSing5d}, these triple intersection numbers determine the prepotential of the 5d SCFT associated with the GTP.

In Section~\ref{sec:algorithm-reconstruction}, we formulate an explicit combinatorial algorithm. This algorithm involves solving a system of compatibility equations, expressing the associativity of the intersection product.
In Section~\ref{sec:algorithm-existence-uniqueness}, we establish that this system always admits a unique solution, and therefore that the algorithm computes the desired triple intersection numbers.

\subsection{Reconstruction Problem and Algorithm}
\label{sec:algorithm-reconstruction}

We formulate the reconstruction of triple intersection numbers
as a finite system of equations for the intersection coefficients
introduced in Section~\ref{sec:intersectionTheory}.
The combinatorial data determine some of these coefficients
directly and restrict the support of the others.
The remaining coefficients are determined by imposing the
compatibility relations of Section~\ref{sec:compatibility}.
We first specify the input data and the unknowns of the
reconstruction, and then give the algorithm.

\subsubsection{Combinatorial Input}
\label{sec:algorithm-input}

We fix a GTP $(P, \cP)$, or equivalently the associated
Symington polygon $(\widetilde P,\tcP)$, together with a choice of generic infinitesimal perturbation of the integral-affine singularities, which induces a choice of $E$-lines as in Section \ref{sec:symington-E-lines} and of exuberant divisors as in Section \ref{sec:ExuberantGeometry}.
We record the index sets
\be
I_0=I_c\sqcup I_{0,\partial},
\qquad
A_0=A_{0,c}\sqcup A_{0,\partial},
\label{eq:algorithm-central-indices}
\ee
for vertices and edges of $\tcP$,
together with the vertex--edge--triangle incidence
relations.

For each $E$-line $\ell_\alpha$, we record its
initial vertex $v_{b(\alpha)}$, corresponding to
a compact divisor $D_{b(\alpha)}$, and the associated
exuberant divisor $E_\alpha$ and compact curve
$C_\alpha
=
D_{b(\alpha)}\cdot E_\alpha$.
As in Section \ref{sec:exuberant-incidence}, we introduce a small perturbation $\ell_\alpha'$ of $\ell_\alpha$, containing the information on how to bypass vertices lying on $\ell_\alpha$.
We also record the primitive direction of
$\ell_\alpha$ at its initial vertex, the
unimodular cone containing that direction,
and the sequence of edges and triangles met
by $\ell_\alpha'$.
Whenever an $E$-line crosses an affine cut,
its direction is transported using the
corresponding integral-affine transition map.

Together, these data determine the index sets
\be
I=I_0\sqcup I_e,
\qquad
A=A_0\sqcup A_e,
\qquad
A_c=A_{0,c}\sqcup A_e.
\label{eq:algorithm-index-sets}
\ee

The reconstruction involves two families of
intersection coefficients.
The first consists of the coefficients
$\alpha^a_{ij}$ in the curve class expansions
\be
D_i\cdot D_j
=
\sum_{a\in A}\alpha^a_{ij}C_a\,,
\label{eq:algorithm-alpha-definition}
\ee
where $(i,j)\in (I \times I) \setminus (I_e \times I_e)$, as in Section~\ref{sec:intersection-data}.

The second family consists of the numerical
divisor--curve intersections
\be
\beta_{ia}=D_i\cdot C_a\,,
\qquad
(i,a)\in\mathcal B,
\label{eq:algorithm-beta-definition}
\ee
where $\mathcal B\subset I\times A$ is the
admissibility domain defined in
Section~\ref{sec:admissible-intersections}.

The required output is the collection of
triple intersection numbers
\be
\kappa_{ijk}
=
D_i\cdot D_j\cdot D_k,
\qquad
\{i,j,k\}\cap I_c\neq\varnothing.
\label{eq:algorithm-output2}
\ee

\subsubsection{Reconstruction Algorithm}
\label{sec:algorithm-procedure}

Starting from the combinatorial data specified
above, we reconstruct the intersection coefficients
and the triple intersection form in five steps.

\begin{enumerate}[label=\textbf{Step \arabic*.}]

\item
\textbf{Central-fiber intersection data.}

For every $i,j\in I_0$, determine the
curve-valued product
\be
D_i\cdot D_j
=
\sum_{a\in A_0}\alpha^a_{ij}C_a
\label{eq:algorithm-central-product}
\ee
using the incidence rules of
Section~\ref{sec:central-divisor-products}:
\be
\alpha^a_{ij}
=
\begin{cases}
1,
& i\neq j\text{ and }e_a
  \text{ joins }v_i,v_j,
\\[1mm]
-1,
& i=j\text{ and }e_a
  \text{ is incident to }v_i,
\\[1mm]
0,
& \text{otherwise}.
\end{cases}
\label{eq:algorithm-central-alpha}
\ee

Next, compute the numerical intersections
of central-fiber divisors with double curves.
For an interior edge $e_a=[v,w]$,
$a\in A_{0,c}$, determine
$\beta_{va}$ and $\beta_{wa}$ from
the integral-affine fan relations of
Section~\ref{sec:central-divisor-curve-intersections}.

For a vertex $v_i$ which is not an endpoint
of $e_a$, the intersection is determined
by the number of incident triangles:
\be
\beta_{ia}
=
\#\left\{
\sigma\in\tcP^{[2]}:
e_a\subset\sigma,\quad
v_i\in\sigma\setminus e_a
\right\}.
\label{eq:algorithm-beta-transverse}
\ee

In particular, all central-fiber
intersection coefficients are determined
directly from the combinatorics of the unimodular triangulation and of the
integral-affine structure on the Symington polygon.

\item
\textbf{Admissibility conditions.}

Construct the set $\mathcal B$ of admissible
divisor--curve pairs, that is, for which a numerical
intersection is defined.

Every compact curve is admissible against
every divisor:
\be
I\times A_c\subset\mathcal B.
\ee

For each non-compact boundary curve
$C_a$, $a\in A_{0,\partial}$,
exclude the pair $(i,a)$ if
\be
v_i\in e_a,
\qquad i\in I_0,
\label{eq:algorithm-boundary-exclusion}
\ee
or if
\be
\ell_\alpha\cap e_a^\circ
\neq\varnothing,
\qquad i=\alpha\in I_e.
\label{eq:algorithm-E-exclusion}
\ee

Thus
\be
\mathcal B
=
(I\times A_c)
\sqcup
\left\{
(i,a)\in I\times A_{0,\partial}:
\epsilon_{ia}=0
\right\},
\label{eq:algorithm-B}
\ee
where $\epsilon_{ia}$ is the incidence
function defined in
\eqref{eq:bad-incidence}.

Only coefficients $\beta_{ia}$ with
$(i,a)\in\mathcal B$ are considered.

\item
\textbf{Exuberant intersection data.}

For each $E$-line $\ell_\alpha$,
let $D=D_{b(\alpha)}$ be the compact
divisor corresponding to its initial
vertex.
Write the primitive direction of
$\ell_\alpha$ in a unimodular cone as
\be
u=pu_1+qu_2,
\qquad
p,q\in\ZZ_{\geq0},
\qquad
\gcd(p,q)=1.
\label{eq:algorithm-E-direction}
\ee

Let $D_1,D_2$ be the central-fiber
divisors corresponding to the two vertices adjacent to $v_{b(\alpha)}$ in the directions $u_1$ and $u_2$ respectively.
The local intersection rule
\eqref{eq:exuberant-local-data}
gives
\be
\begin{aligned}
D\cdot C_\alpha&=-p-q,
\\
D_1\cdot C_\alpha&=p,
\\
D_2\cdot C_\alpha&=q,
\\
E_\alpha\cdot C_\alpha&=pq-1.
\end{aligned}
\label{eq:algorithm-E-local}
\ee
The intersections of $C_\alpha$
with all other central-fiber
divisors vanish.

For each central-fiber divisor $D_j$, $j \in I_0$,
consider the curve-valued product
\be
E_\alpha\cdot D_j
=
\sum_{a\in A}
\alpha^a_{\alpha j}C_a.
\label{eq:algorithm-E-support}
\ee

Set to zero all coefficients with $a \notin A_{\alpha j}$, where $A_{\alpha j} \subset A$ is defined in \eqref{eq_A_alpha_j},
and set
\be
\alpha^\alpha_{\alpha,b(\alpha)}=1.
\ee

The remaining coefficients allowed
by the incidence conditions are
retained as unknown integers.

Similarly, consider the numerical
intersections
\be
\beta_{\alpha a}
=
E_\alpha\cdot C_a,
\qquad
(\alpha,a)\in\mathcal B.
\ee
Insert the values fixed by the local
$E$-line geometry (Proposition \ref{prop:local-E-line-intersections}), and set to zero
all coefficients excluded by the incidence conditions described immediately after \eqref{eq:E-beta-zero}.
All remaining admissible coefficients
are retained as unknown integers.

At this stage, every coefficient
entering the reconstruction is either
known, forced to vanish, or represented
by an integer unknown.

\item
\textbf{Compatibility equations.}

We now impose the compatibility relations
of Section~\ref{sec:compatibility}.

For each available curve-valued product
\be
D_j\cdot D_k
=
\sum_{a\in A}\alpha^a_{jk}C_a,
\label{eq:algorithm-available-product}
\ee
and each divisor $D_i$, determine
whether the expansion is admissible
against $D_i$, namely whether
one of the following holds:
\begin{enumerate}
    \item $(j,k) \in I_0 \times I_0$, so that the coefficients $\alpha^a_{jk}$ are known by \eqref{eq:alpha-central-offdiag}-\eqref{eq:alpha-central-diagonal}, and we require \be
\alpha^a_{jk}\neq0
\qquad\Longrightarrow\qquad
(i,a)\in\mathcal B.
\label{eq:algorithm-admissibility-test}
\ee
    \item $(j,k) \in I_e \times I_0$, and we require 
    \be a \in A_{jk}  \qquad\Longrightarrow\qquad
(i,a)\in\mathcal B \,,
\ee
where $A_{jk} \subset A$ is defined in 
\eqref{eq_A_alpha_j}.
\end{enumerate}

Whenever this condition holds, form
\be
T_{i;jk}
=
\sum_{a\in A}
\alpha^a_{jk}\beta_{ia}.
\label{eq:algorithm-T}
\ee

For every pair of admissible
presentations of the same triple
intersection, impose
\be
T_{i;jk}=T_{j;ik}.
\label{eq:algorithm-compatibility}
\ee

These equations form a
compatibility system of quadratic equations.

Solve this system for the unknown
intersection coefficients. 
The existence and uniqueness of the solution are established by Theorem \ref{thm:compatibility-unique}
in Section~\ref{sec:algorithm-existence-uniqueness}.

\item
\textbf{Triple intersection numbers.}

Once the compatibility system has
been solved, compute the triple
intersection numbers containing
at least one compact divisor.

Let $i\in I_c$ and $j,k\in I$.
Since $D_i$ is compact, the
curve-valued product $D_i\cdot D_k$
is supported on compact curves:
\be
D_i\cdot D_k
=
\sum_{a\in A_c}
\alpha^a_{ik}C_a.
\label{eq:algorithm-compact-product}
\ee

Every compact curve can be paired
with $D_j$, so the required
triple intersection is
\be
\kappa_{ijk}
=
\sum_{a\in A_c}
\alpha^a_{ik}\beta_{ja},
\qquad
i\in I_c,\quad j,k\in I.
\label{eq:algorithm-kappa}
\ee

\end{enumerate}

\subsection{Existence and Uniqueness of the Reconstruction}
\label{sec:algorithm-existence-uniqueness}

In this section, we prove the existence and uniqueness of the solution to the compatibility system used in the algorithm in
Section~\ref{sec:algorithm-reconstruction}.

\begin{theorem}
\label{thm:compatibility-unique}
Let $(P, \cP)$ be a GTP. Fix a generic perturbation of the focus--focus singularities of the associated Symington polygon, as in Section \ref{sec:symington-E-lines}, and small deformations of the $E$-lines bypassing non-initial vertices, as in Section \ref{sec:exuberant-incidence}.
Then, the compatibility system of
Section~\ref{sec:algorithm-reconstruction} has a unique solution. 
\end{theorem}

\begin{proof}

The existence is guaranteed by the existence of a corresponding Calabi--Yau threefold $\widetilde{\cX}$, established in Section~\ref{sec:CYGTP}. The geometric intersection coefficients provide a solution of the compatibility system by 
the intersection theory arguments presented in Section \ref{sec:intersectionTheory}.
Therefore, we only need to establish the uniqueness of a solution.
The key observation is that the initially undetermined intersection coefficients can be determined recursively by applying the compatibility relations along each $E$-line, propagating from its initial vertex through successive crossings of edges and vertices.

Fix an exuberant divisor $E=E_\alpha$
and orient its $E$-line
$\ell_\alpha$ from its initial
vertex towards its terminal point.
We want to determine the intersection coefficients 
$\alpha^a_{\alpha j}$ in the expansions
\be
E_\alpha \cdot D_j = \sum_{a \in A} \alpha^a_{\alpha j} C_a
\ee
for $j \in I_0$ corresponding to a vertex of $\tcP$ on $\ell_\alpha$.
At the initial vertex $v_{b(\alpha)}$, we have $E_\alpha \cdot D_{v_{b(\alpha)}} = C_\alpha$.
Next, we explain how to uniquely propagate these curve expansions along the $E$-line $\ell_\alpha$, through intersection with edges and vertices of $\widetilde{\cP}$.

We first consider an intersection with an interior edge of $\widetilde{\cP}$. The case of a boundary edge can be treated similarly. Denote the interior edge by $e=[v,w]$, 
separating two unimodular triangles
\be
\sigma_-=(v,w,x),
\qquad
\sigma_+=(v,w,y),
\label{eq:propagation-triangles}
\ee
where $\sigma_-$ lies before the crossing
and $\sigma_+$ lies after it.
Let $C_e=D_v\cap D_w$ 
be the corresponding compact double curve. Assume that the $E$-line $\ell_\alpha$ goes out of the triangle $\sigma_+$ through the edge $[w,y]$.

We first recall the relevant central-fiber
intersection numbers. We express the two
triangles in a common integral-affine
chart, using parallel transport across
any intervening affine cut.
After an integral-affine change of
coordinates, we may take
\be
v=(0,0),
\qquad
w=(1,0),
\qquad
x=(0,1),
\qquad
y=(m,-1),
\label{eq:propagation-local-coordinates}
\ee
for some $m\in\ZZ$.
The integral-affine fan relations of
Section~\ref{sec:central-divisor-curve-intersections}
give
\be
D_v\cdot C_e=m-2,
\qquad
D_w\cdot C_e=-m.
\label{eq:propagation-self-intersections}
\ee
The two remaining central-fiber divisors
intersect $C_e$ transversely at the
triple points corresponding to
$\sigma_-$ and $\sigma_+$.
Consequently,
$D_x\cdot C_e
=
D_y\cdot C_e
=
1$.

By the assumption that $\ell_\alpha$ goes out of $\sigma_+$ through the edge $[w,y]$, we have 
\be
E_\alpha  \cdot D_v
=
\Gamma_{\rm old}+n_eC_e,
\ee
where $\Gamma_{\rm old}$ is determined by the
preceding data and $n_e$ is the only new
coefficient.
Then, the compatibility relation
\be
D_x \cdot (E_\alpha \cdot D_v)
=
(E_\alpha \cdot  D_x) \cdot D_v
\ee
uniquely determines $n_e$ by
\be
n_e
=
(E_\alpha \cdot D_x) \cdot D_v
-
D_x \cdot \Gamma_{\rm old}\,,
\ee
where the right-hand side is already determined.

Finally, we consider an intersection of the $E$-line $\ell_\alpha$ with a vertex $v$ of $\widetilde{\cP}$. 
Then, the uniqueness of the curve expansion follows as in the proof of Lemma \ref{lem}.
\end{proof}

\section{Examples of Calabi--Yau Threefolds from GTPs}
\label{sec:examples}

In this section, we work out explicitly a few examples, following the algorithm of Section \ref{sec:intersectionAlgorithm}. The first example is the one already discussed in Section \ref{sec_webs57}, which we work out in full detail. We have chosen this example because it displays most of the subtleties of the problem. The second example is one for which a gauge theory phase is known, which allows us to compare our results with expectations from the gauge theoretic prepotential.  

\subsection{A fully worked out Example}
\label{sec:ex1}

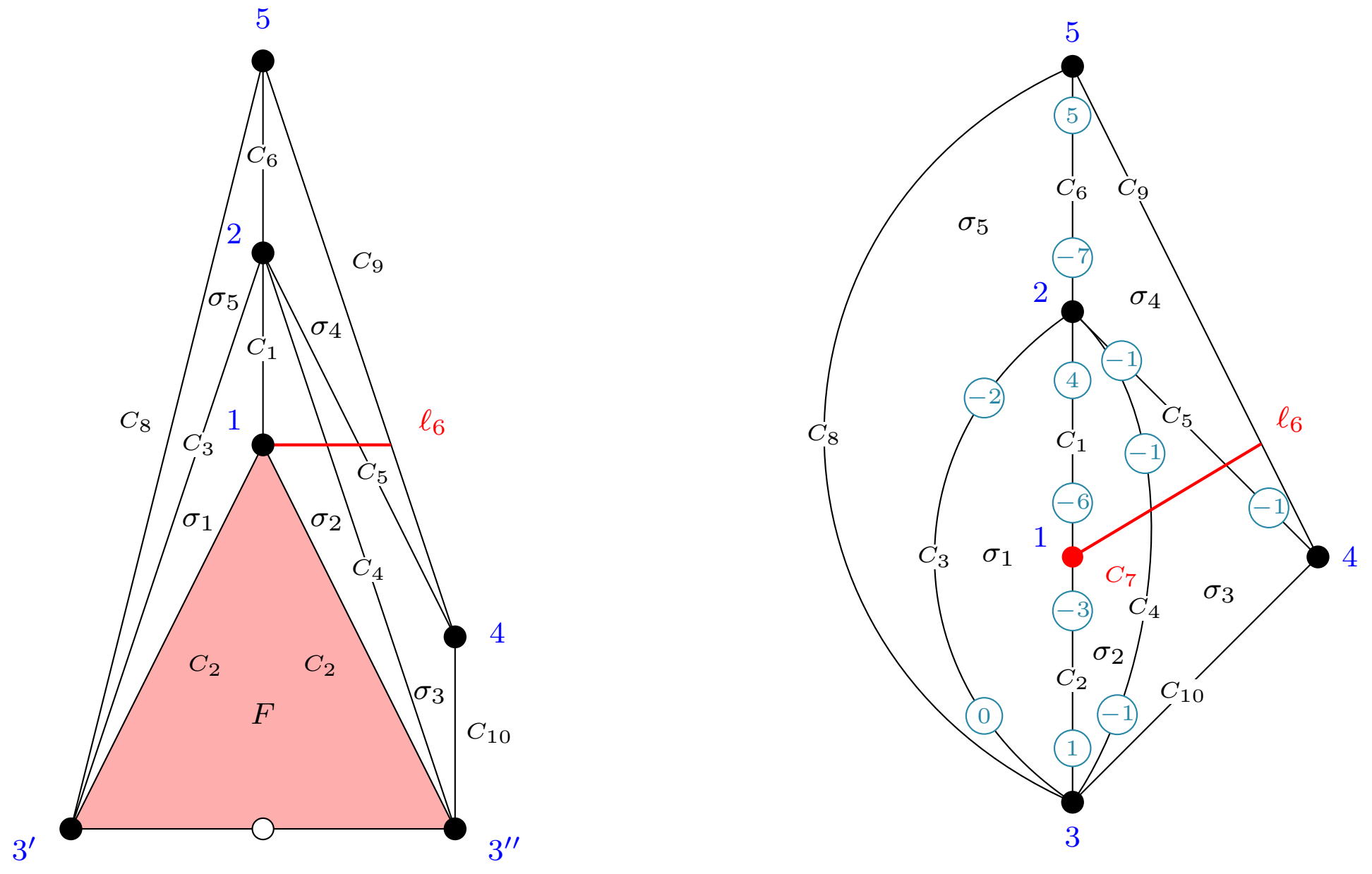
\begin{figure}
    \centering
    \begin{tikzpicture}
        \node at (0,0) {\scalebox{1}{%
        \begin{tikzpicture}[scale=1.8]
            \coordinate (v1)   at ( 0, 0);
            \coordinate (v2)   at ( 0, 1);
            \coordinate (v5)   at ( 0, 2);
            \coordinate (v4)   at ( 1,-1);
            \coordinate (v3p)  at (-1,-2);
            \coordinate (v3pp) at ( 1,-2);
            \coordinate (w)    at ( 0,-2);
            \coordinate (x9)   at (2/3,0);
            \fill[red,opacity=.32] (v1)--(v3p)--(v3pp)--cycle;
            \draw
                (v5)--(v4)--(v3pp)--(v3p)--cycle
                (v1)--(v2)--(v5)
                (v2)--(v3p) (v2)--(v3pp) (v2)--(v4)
                (v1)--(v3p) (v1)--(v3pp);
            \draw[red,thick] (v1)--(x9);
            \begin{scope}[every node/.style={fill=white,inner sep=.6pt,font=\tiny}]
                \node at (0,.5)       {$C_1$};
                \node at (-1/3,0)     {$C_3$};
                \node at (.55,-.65)   {$C_4$};
                \node at (.575,-.15)  {$C_5$};
                \node at (0,1.5)      {$C_6$};
            \end{scope}
            \begin{scope}[every node/.style={inner sep=.6pt,font=\tiny}]
                \node at (-.3,-1.15)  {$C_2$};
                \node at (.3,-1.15)   {$C_2$};
                \node at (-.66,.12)   {$C_8$};
                \node at (.55,.95)    {$C_9$};
                \node at (1.18,-1.5)  {$C_{10}$};
            \end{scope}
            \begin{scope}[every node/.style={inner sep=.6pt,font=\scriptsize}]
                \node at (-1/3,-.4)   {$\sigma_1$};
                \node at (1/3,-.4)    {$\sigma_2$};
                \node at (.87,-1.3)   {$\sigma_3$};
                \node at (1/3,.6)     {$\sigma_4$};
                \node at (-.2,.75)    {$\sigma_5$};
                \node at (0,-1.4)     {$F$};
            \end{scope}
            \node[bd] at (v1)   {};
            \node[bd] at (v2)   {};
            \node[bd] at (v5)   {};
            \node[bd] at (v4)   {};
            \node[bd] at (v3p)  {};
            \node[bd] at (v3pp) {};
            \node[wd] at (w)    {};
            \node[blue] at ($(v1)+(-.15,.13)$)
                {\scriptsize $1$};
            \node[blue] at ($(v2)+(-.15,.10)$)
                {\scriptsize $2$};
            \node[blue] at ($(v3p)+(-.24,-.10)$)
                {\scriptsize $3'$};
            \node[blue] at ($(v3pp)+(.26,-.10)$)
                {\scriptsize $3''$};
            \node[blue] at ($(v4)+(.22,.02)$)
                {\scriptsize $4$};
            \node[blue] at ($(v5)+(0,.22)$)
                {\scriptsize $5$};
            \node[red] at ($(x9)+(.22,.12)$)
                {\scriptsize \color{red}$\ell_6$};
        \end{tikzpicture}}};
        \node at (7.4,0) {\scalebox{1}{%
        \begin{tikzpicture}[scale=2.3,
            beta/.style={circle,draw=cyan!60!black,text=cyan!60!black,fill=white,
                inner sep=0pt,minimum size=3.4mm,font=\tiny},
            clab/.style={fill=white,inner sep=.6pt,font=\tiny}]
            \coordinate (s1) at (0,1);
            \coordinate (s2) at (0,2);
            \coordinate (s3) at (0,0);
            \coordinate (s4) at (1,1);
            \coordinate (s5) at (0,3);
            \coordinate (s9) at (10/13,19/13);
            \draw (s5)--(s2)--(s1)--(s3)
                  (s2)--(s4)--(s3) (s5)--(s4);
            \draw (s2) .. controls (-.75,1.5) and (-.75,.5) .. (s3);
            \draw (s2) .. controls (.5,1.55) and (.35,.5) .. (s3);
            \draw (s5) .. controls (-1.35,2.4) and (-1.35,.6) .. (s3);
            \draw[red,thick] (s1)--(s9);
            \path (s5)--(s2) node[beta,pos=.2] {$5$} node[clab,pos=.5] {$C_6$} node[beta,pos=.78] {$-7$};
            \path (s2)--(s1) node[beta,pos=.28] {$4$} node[clab,pos=.53] {$C_1$} node[beta,pos=.78] {$-6$};
            \path (s1)--(s3) node[beta,pos=.22] {$-3$} node[clab,pos=.5] {$C_2$} node[beta,pos=.78] {$1$};
            \path (s2) .. controls (-.75,1.5) and (-.75,.5) .. (s3)
                node[beta,pos=.2] {$-2$} node[clab,pos=.5] {$C_3$} node[beta,pos=.8] {$0$};
            \path (s2) .. controls (.5,1.55) and (.35,.5) .. (s3)
                node[beta,pos=.32] {$-1$} node[clab,pos=.6] {$C_4$} node[beta,pos=.8] {$-1$};
            \path (s2)--(s4) node[beta,pos=.2] {$-1$} node[clab,pos=.43] {$C_5$} node[beta,pos=.8] {$-1$};
            \path (s5)--(s4) node[clab,pos=.25] {$C_9$};
            \begin{scope}[every node/.style={inner sep=.6pt,font=\scriptsize}]
                \node at (-.3,1)    {$\sigma_1$};
                \node at (.15,.6)   {$\sigma_2$};
                \node at (.6,.85)   {$\sigma_3$};
                \node at (.3,2.05)  {$\sigma_4$};
                \node at (-.4,2.35) {$\sigma_5$};
            \end{scope}
            \path (s4)--(s3) node[clab,pos=.55] {$C_{10}$};
            \path (s5) .. controls (-1.35,2.4) and (-1.35,.6) .. (s3) node[clab,pos=.5] {$C_8$};
            \node[bd] at (s2) {};
            \node[bd] at (s3) {};
            \node[bd] at (s4) {};
            \node[bd] at (s5) {};
            \node[circle,fill=red,inner sep=0pt,minimum size=2mm] at (s1) {};
            \node[blue] at ($(s1)+(-.13,.08)$) {\scriptsize $1$};
            \node[blue] at ($(s2)+(-.13,.08)$) {\scriptsize $2$};
            \node[blue] at ($(s3)+(0,-.14)$)   {\scriptsize $3$};
            \node[blue] at ($(s4)+(.13,0)$)    {\scriptsize $4$};
            \node[blue] at ($(s5)+(0,.14)$)    {\scriptsize $5$};
            \node[red] at ($(s1)+(.2,-.08)$)   {\tiny $C_7$};
            \node[red] at ($(s9)+(.12,.1)$)    {\scriptsize $\ell_6$};
        \end{tikzpicture}}};
    \end{tikzpicture} 
    \caption{Left: the GTP $P$, with the generalized face $F$ shaded, the white dot on the decorated boundary segment $[v_{3'},v_{3''}]$, and the $E$-line $\ell_6$ of the exuberant divisor $E_6$ in red. Each edge is labelled by its double curve $C_a$. The two edges labelled $C_2$ are glued into a single edge of $\widetilde P$.  In both panels, the triangles are labelled $\sigma_1,\dots,\sigma_5$. Right: the Symington polygon $\widetilde P$ with its triangulation $\tcP$, drawn with curved edges. The vertex labelled $i$ is $v_i$, dual to the component $D_i$ of $\widetilde{\cX}_0$, and $v_3$ is the common image of $v_{3'}$ and $v_{3''}$. The focus--focus singularity at $v_1$ is marked in red. Each edge is labelled by its double curve $C_a$, and the circled number next to vertex $i$ on the edge $C_a$ is $\beta_{ia}$. The $E$-line $\ell_6$ starts at $v_1$, where the exuberant curve $C_7=D_1\cdot E_6$ lies, crosses $C_4$ and $C_5$, and ends on $C_9$.} 
    \label{fig:ex1}
\end{figure}

Consider the GTP of Section \ref{sec_webs57}, see Figure~\ref{fig:ex1}. 
The GTP $P$, with its triangulation $\mathcal{P}$, has 6 vertices, conveniently labelled by $\{1,2,3',3'',4,5\}$, which after identification correspond to the 5 vertices, labelled by $\{1,2,3,4,5\}$, of the Symington polygon $\widetilde P$, drawn with a monodromy cut on the right of Figure~\ref{fig:GTP-affine-cut}.
The identification is the affine map fixing $v_1$ with linear part
\be
M=\begin{pmatrix}1&-1\\0&1\end{pmatrix}=M_{[0,1]}^T\,,
\qquad
M\begin{pmatrix}-1\\-2\end{pmatrix}=\begin{pmatrix}1\\-2\end{pmatrix}\,,
\qquad
M\begin{pmatrix}1\\0\end{pmatrix}=\begin{pmatrix}1\\0\end{pmatrix}\,.
\label{eq:ex1-M}
\ee
The Symington polygon thus has 5 vertices and 5 triangles. The vertex $v_1$ is the focus--focus singularity, with monodromy $M$ and monodromy-invariant direction $(1,0)$.  
We denote by $p_\sigma$ the triple point of $\widetilde{\cX}_0$ dual to a triangle $\sigma$.

We label divisors and curves as in Figure~\ref{fig:ex1}.
The divisor index sets are
\be
I_0=\{1,2,3,4,5\}\,,\qquad
I_c=\{1,2\}\,,\qquad
I_{0,\partial}=\{3,4,5\}\,,\qquad
I_e=\{6\}\,,\qquad
I=\{1,\dots,6\}\,,
\label{eq:ex1-I}
\ee
with $D_i=X_{v_i}$ for $i\in I_0$, and $E_6$ the exuberant divisor of the single 7-brane. Of the two monodromy-invariant rays $\pm(1,0)$ at $v_1$, we take $\ell_6$ to start in the direction $u=(1,0)$, as in Figure~\ref{fig:ex1}.
Its initial vertex is $v_{b(6)}=v_1$, so $b(6)=1$. 
The curve index sets are
\be
A_{0,c}=\{1,\dots,6\}\,,\quad
A_{0,\partial}=\{8,9,10\}\,,\quad
A_e=\{7\}\,,\quad
A_c=\{1,\dots,7\}\,,\quad
A=\{1,\dots,10\}\,,
\label{eq:ex1-A}
\ee
where $C_7=D_1\cdot E_6$ is the exuberant curve, so that the bijection $A_e\simeq I_e$ of Section~\ref{sec:ExuberantGeometry} reads $7\leftrightarrow6$.

\textbf{Step 1: central-fiber intersection data.}
The incidence rule \eqref{eq:algorithm-central-alpha} gives the curve-valued products
\be
\ba
D_1^2&=-C_1-C_2\,, & D_1D_2&=C_1\,, & D_1D_3&=C_2\,,\\
D_2^2&=-C_1-C_3-C_4-C_5-C_6\,, & D_2D_3&=C_3+C_4\,, & D_2D_4&=C_5\,,\\
D_3^2&=-C_2-C_3-C_4-C_8-C_{10}\,, & D_2D_5&=C_6\,, & D_3D_4&=C_{10}\,,\\
D_4^2&=-C_5-C_9-C_{10}\,, & D_3D_5&=C_8\,, & D_4D_5&=C_9\,,\\
D_5^2&=-C_6-C_8-C_9\,, & D_1D_4&=0\,, & D_1D_5&=0\,.
\ea
\label{eq:ex1-alpha-central}
\ee
In particular, note that $D_2\cdot D_3$ has the two components $C_3$ and $C_4$! This feature is specific to generalized toric geometry. 

To compute the coefficients $\beta_{va}$ and $\beta_{wa}$ for $e_a=[v,w]$, we use the fan relations \eqref{eq:double-curve-vector-rule-v} and \eqref{eq:double-curve-vector-rule-w}:  
\be
\setlength{\arraycolsep}{3pt}
\begin{array}{c|c|c|c|c|c}
a & v & w & \text{relation at } v & \text{relation at } w & (\beta_{va},\beta_{wa})\\ \hline
1 & v_1 & v_2 & (-1,-2)+(1,-2)+4\,(0,1)=0 & (-1,-3)+(1,-3)-6\,(0,-1)=0 & (-6,4)\\
2 & v_1 & v_{3''} & (0,1)+M(0,1)+1\,(1,-2)=0 & (-1,3)+M(1,3)-3\,(-1,2)=0 & (-3,1)\\
3 & v_2 & v_{3'} & (0,-1)+(0,1)+0\,(-1,-3)=0 & (1,2)+(1,4)-2\,(1,3)=0 & (-2,0)\\
4 & v_2 & v_{3''} & (0,-1)+(1,-2)-1\,(1,-3)=0 & (-1,2)+(0,1)-1\,(-1,3)=0 & (-1,-1)\\
5 & v_2 & v_4 & (1,-3)+(0,1)-1\,(1,-2)=0 & (0,-1)+(-1,3)-1\,(-1,2)=0 & (-1,-1)\\
6 & v_2 & v_5 & (1,-2)+(-1,-3)+5\,(0,1)=0 & (1,-3)+(-1,-4)-7\,(0,-1)=0 & (-7,5)
\end{array}
\label{eq:ex1-fan-relations}
\ee
In every row $\beta_{va}+\beta_{wa}=-2$, as required by \eqref{eq:CY-double-curve-relation}. This gives the circled numbers in Figure~\ref{fig:ex1}.

The rows $a=1,2$ determine the surface $D_1$.
The rays of $\overline X_{v_1}$ satisfy $(-1,-2)+(1,-2)+4\,(0,1)=0$, so $\overline X_{v_1}\simeq\PP(1,1,4)$, and the cut cone $\langle(-1,-2),(1,-2)\rangle$ is the Wahl singularity $\CC^2/(\ZZ/4\ZZ)$ with weights $(1,1)$, that is $n=2$.
Its smoothing is $D_1\simeq\PP^2$, with $\partial D_1=C_1+C_2$.
The curve $C_1$ of the ray $(0,1)$ is a conic, $(C_1|_{D_1})^2=\beta_{21}=4$.
The two boundary curves of the rays $(\mp1,-2)$ have self-intersection $\tfrac14$ and meet with intersection number $\tfrac14$ at the singular point. The smoothing merges them into the line $C_2$, with $(C_2|_{D_1})^2=\tfrac14+\tfrac14+2\cdot\tfrac14=1=\beta_{32}$.

\begin{figure}[t]
\centering
\begin{tikzpicture}[x=0.8cm,y=0.8cm,
    lat/.style={gray!60},
    lab/.style={font=\scriptsize,inner sep=1.5pt}]
    \begin{scope}
        \foreach \x in {-1,0,1} \foreach \y in {-2,-1,0,1} \fill[lat] (\x,\y) circle (1pt);
        \fill[red,opacity=.32] (0,0)--(-1,-2)--(1,-2)--cycle;
        \draw (0,0)--(0,1.6) (0,0)--(-1.15,-2.3) (0,0)--(1.15,-2.3);
        \fill (0,1) circle (1.5pt) (-1,-2) circle (1.5pt) (1,-2) circle (1.5pt);
        \node[circle,fill=red,inner sep=0pt,minimum size=2mm] at (0,0) {};
        \node[blue,font=\scriptsize] at (-.25,.2) {$1$};
        \node[lab,above] at (0,1.6) {$C_1\ [4]$};
        \node[lab,below left] at (-1.15,-2.3) {$C_2'\ [\tfrac14]$};
        \node[lab,below right] at (1.15,-2.3) {$C_2''\ [\tfrac14]$};
        \node[font=\tiny] at (0,-1.4) {$\tfrac14(1,1)$};
        \node at (0,2.6) {$\overline X_{v_1}\simeq\PP(1,1,4)$};
    \end{scope}
    \draw[->] (2.4,-0.4)--(4,-0.4) node[midway,above,font=\scriptsize] {smoothing};
    \begin{scope}[shift={(6.3,-0.4)}]
        \draw (0,0) ellipse (1.3 and 0.85);
        \draw (-1.8,-0.4)--(1.8,-0.4);
        \draw[red,thick] (0.578,-1.098)--(1.716,0.298);
        \fill (-1.147,-0.4) circle (1.5pt) (1.147,-0.4) circle (1.5pt);
        \node[lab,above] at (0,0.85) {$C_1\ [4]$};
        \node[lab,above] at (0,-0.4) {$C_2\ [1]$};
        \node[lab,red,above right] at (1.716,0.298) {$C_7\ [1]$};
        \node[lab,below left] at (-1.147,-0.4) {$p_{\sigma_1}$};
        \node[lab,below right] at (1.147,-0.4) {$p_{\sigma_2}$};
        \node at (0,3.0) {$X_{v_1} = D_1\simeq\PP^2$};
    \end{scope}
\end{tikzpicture}
\caption{ The surface $D_1$ of Example 1. Each ray is labelled by its curve, with the self-intersection of that curve in the surface in brackets. (a) The star of $v_1$ in $\cP$ spans the fan of $\overline X_{v_1}\simeq\PP(1,1,4)$, with $(-1,-2)+(1,-2)+4\,(0,1)=0$. The shaded cut cone is the cone over $F$ and gives the Wahl singularity $\frac14(1,1)$. The boundary curves $C_2'$ and $C_2''$ of the rays towards $v_{3'}$ and $v_{3''}$ are merged into $C_2$ by the smoothing. (b) $D_1\simeq\PP^2$, with $\partial D_1=C_1+C_2$ a conic and a line meeting at the triple points $p_{\sigma_1}$ and $p_{\sigma_2}$. The exuberant curve $C_7$ of Step 3 is the tangent line to $C_1$ at $p_{\sigma_2}$. }
\label{fig:ex1-surfaces}
\end{figure}
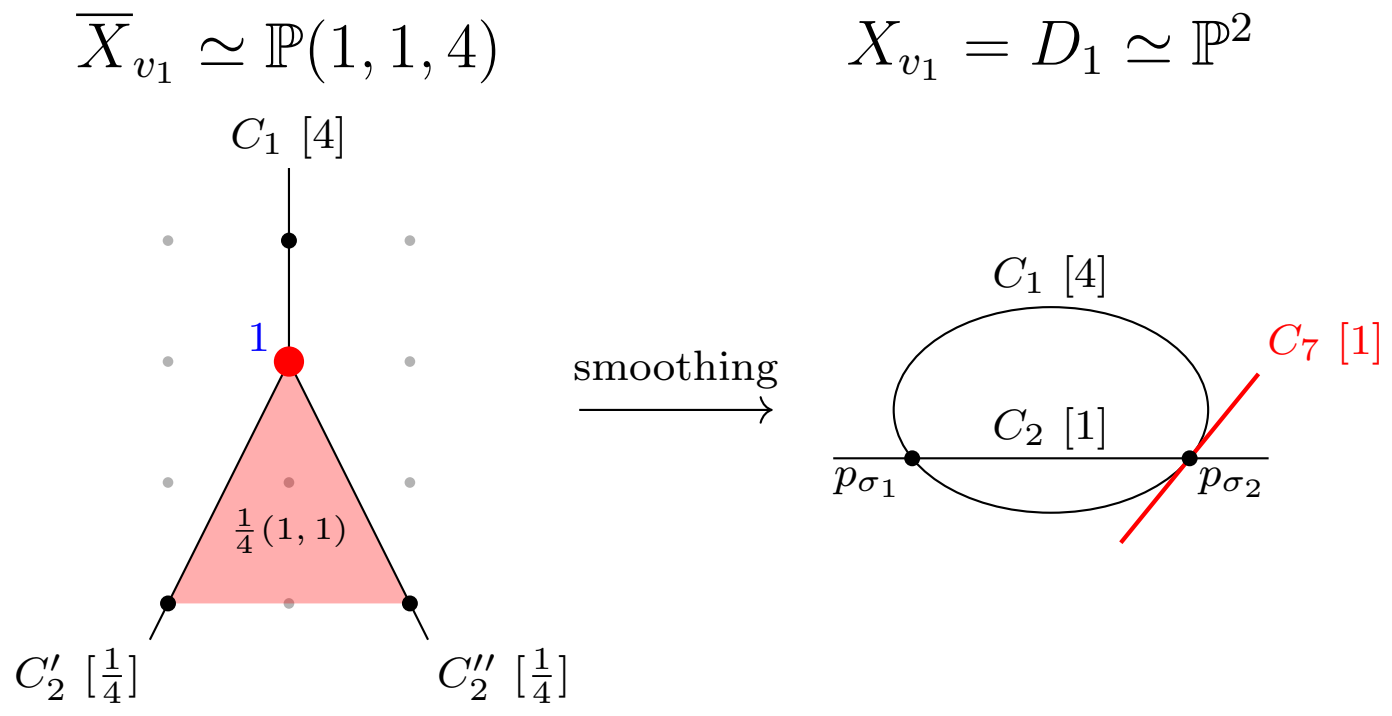

For transverse intersections, \eqref{eq:algorithm-beta-transverse} gives the non-zero values
\be
\ba
&\beta_{31}=2\,,\quad
\beta_{22}=2\,,\quad
\beta_{13}=\beta_{53}=1\,,\quad
\beta_{14}=\beta_{44}=1\,,\quad
\beta_{35}=\beta_{55}=1\,,\quad
\beta_{36}=\beta_{46}=1\,,\\
&\beta_{28}=\beta_{29}=\beta_{2,10}=1\,.
\ea
\label{eq:ex1-beta-transverse}
\ee
Here note in particular that $\beta_{31}=2$ and $\beta_{22}=2$ because the two triangles containing $e_1$, respectively $e_2$, have the same third vertex $v_3$, respectively $v_2$, in $\tcP$. Again, this is specific to generalized toric geometry. All other transverse coefficients vanish.

\medskip
\noindent
\textbf{Step 2: admissibility conditions.}
All pairs $(i,a)$ with $a\in A_c$ are admissible. 
For the boundary curves, \eqref{eq:algorithm-boundary-exclusion} excludes the pairs with $v_i\in e_a$, namely $(3,8)$, $(5,8)$, $(4,9)$, $(5,9)$, $(3,10)$ and $(4,10)$.
The $E$-line $\ell_6$ meets the relative interior of $e_9$ and does not meet $e_8$ or $e_{10}$, so \eqref{eq:algorithm-E-exclusion} excludes $(6,9)$ only.
Thus
\be
\mathcal B=(I\times A)\setminus\big\{(3,8),\,(5,8),\,(4,9),\,(5,9),\,(6,9),\,(3,10),\,(4,10)\big\}\,.
\label{eq:ex1-B}
\ee
Collecting the results, with $\times$ marking the pairs excluded in \eqref{eq:ex1-B}, one finds the $\beta$-matrix 
\be
\big(\beta_{ia}\big)_{i\in I_0,\,a\in A_0}=
\begin{array}{c|cccccc|ccc}
 & C_1 & C_2 & C_3 & C_4 & C_5 & C_6 & C_8 & C_9 & C_{10}\\ \hline
D_1 & -6 & -3 & 1 & 1 & 0 & 0 & 0 & 0 & 0\\
D_2 & 4 & 2 & -2 & -1 & -1 & -7 & 1 & 1 & 1\\
D_3 & 2 & 1 & 0 & -1 & 1 & 1 & \times & 0 & \times\\
D_4 & 0 & 0 & 0 & 1 & -1 & 1 & 0 & \times & \times\\
D_5 & 0 & 0 & 1 & 0 & 1 & 5 & \times & \times & 0
\end{array}
\label{eq:ex1-beta-central}
\ee
Each column indexed by a compact curve sums to zero, as it must since $\widetilde{\cX}_0=\sum_{i\in I_0}D_i$ is linearly equivalent to zero.

\medskip
\noindent
\textbf{Step 3: exuberant intersection data.}
The direction $u=(1,0)$ of $\ell_6$ lies in the unimodular cone at $v_1$, generated by $u_1=v_{3''}-v_1=(1,-2)$ and $u_2=v_2-v_1=(0,1)$:
\be
u=(1,0)=1\cdot(1,-2)+2\cdot(0,1)\,,
\qquad
(p,q)=(1,2)\,.
\label{eq:ex1-pq}
\ee
The local rule \eqref{eq:algorithm-E-local} gives
\be
\beta_{17}=-p-q=-3\,,\quad
\beta_{37}=p=1\,,\quad
\beta_{27}=q=2\,,\quad
\beta_{67}=pq-1=1\,,\quad
\beta_{47}=\beta_{57}=0\,.
\label{eq:ex1-local-E}
\ee
In $D_1\simeq\PP^2$, this says $C_7\cdot C_1=2$, $C_7\cdot C_2=1$ and $(C_7|_{D_1})^2=1$: the exuberant curve $C_7$ is a line, meeting $\partial D_1$ at the single point $p_{\sigma_2}\in C_1\cap C_2$ with multiplicity $p+q=3$, i.e.\ the tangent line to the conic $C_1$ at $p_{\sigma_2}$, see Figure \ref{fig:ex1-surfaces}. 

Next, the incidence of $\ell_6$ restricts the curve-valued products $E_6\cdot D_j$.
By \eqref{eq:E-base-alpha}, $\alpha^7_{61}=1$.
Besides $C_7$, the only curves that can appear are the double curves $C_4$, $C_5$, $C_9$ of the edges met by $\ell_6$, in the products with the divisors attached to their endpoints.
Hence
\be
\ba
E_6\cdot D_1&=C_7\,, &
E_6\cdot D_2&=\alpha^4_{62}C_4+\alpha^5_{62}C_5\,, &
E_6\cdot D_3&=\alpha^4_{63}C_4\,,\\
E_6\cdot D_4&=\alpha^5_{64}C_5+\alpha^9_{64}C_9\,, &
E_6\cdot D_5&=\alpha^9_{65}C_9\,, & &
\ea
\label{eq:ex1-alpha-E-ansatz}
\ee
with six unknown integers $\alpha^4_{62}$, $\alpha^5_{62}$, $\alpha^4_{63}$, $\alpha^5_{64}$, $\alpha^9_{64}$, $\alpha^9_{65}$.

For the numerical intersections $\beta_{6a}=E_6\cdot C_a$, $(6,a)\in\mathcal B$, the value $\beta_{67}=1$ is fixed by \eqref{eq:ex1-local-E}.
The curve $E_6\cap\widetilde{\cX}_0$ meets the double locus of $\widetilde{\cX}_0$ only along $C_4\cup C_5\cup C_9$, which contains the triple points $p_{\sigma_2},p_{\sigma_3},p_{\sigma_4}$ of the triangles traversed by $\ell_6$.
Since two double curves meet only at the triple points of common triangles, a double curve $C_a$ can meet $E_6$ only if $e_a$ is an edge of $\sigma_2$, $\sigma_3$ or $\sigma_4$.
The edges $e_3$ and $e_8$ belong only to $\sigma_1$ and $\sigma_5$. Hence $C_3$ and $C_8$ are disjoint from $E_6$, and by \eqref{eq:E-beta-zero}
\be
\beta_{63}=\beta_{68}=0\,.
\label{eq:ex1-beta-E-zero}
\ee
The six remaining admissible coefficients $\beta_{61}$, $\beta_{62}$, $\beta_{64}$, $\beta_{65}$, $\beta_{66}$, $\beta_{6,10}$ are unknown integers (recall that $\beta_{69}$ is not defined).
Altogether, the compatibility system has twelve unknowns.

\medskip
\noindent
\textbf{Step 4: compatibility equations.}
For $i,j,k\in I_0$, all admissible presentations $T_{i;jk}$ follow from \eqref{eq:ex1-alpha-central} and \eqref{eq:ex1-beta-central}, and the relations $T_{i;jk}=T_{j;ik}$ hold identically.
For instance, $T_{1;22}=T_{2;12}=4$, $T_{3;12}=T_{1;23}=T_{2;13}=2$, $T_{2;33}=T_{3;23}=-1$ and $T_{5;22}=T_{2;25}=-7$.

The non-trivial relations involve $E_6$.
For $j,k\in I_0$, the triple product $E_6\cdot D_j\cdot D_k$ has the presentations $T_{6;jk}$, $T_{j;6k}$ and $T_{k;6j}$ of \eqref{eq:algorithm-T}, whenever these are admissible in the sense of \eqref{eq:algorithm-admissibility-test}.
Using \eqref{eq:ex1-alpha-central}, \eqref{eq:ex1-beta-central}, \eqref{eq:ex1-local-E}, \eqref{eq:ex1-alpha-E-ansatz} and \eqref{eq:ex1-beta-E-zero}, we find
\be
\ba
(6,1,1):&\quad T_{6;11}=-\beta_{61}-\beta_{62}\,,\quad
T_{1;61}=\beta_{17}=-3\,,\\
(6,1,2):&\quad T_{6;12}=\beta_{61}\,,\quad
T_{1;62}=\alpha^4_{62}\beta_{14}+\alpha^5_{62}\beta_{15}=\alpha^4_{62}\,,\quad
T_{2;61}=\beta_{27}=2\,,\\
(6,1,3):&\quad T_{6;13}=\beta_{62}\,,\quad
T_{1;63}=\alpha^4_{63}\beta_{14}=\alpha^4_{63}\,,\quad
T_{3;61}=\beta_{37}=1\,,\\
(6,1,4):&\quad T_{6;14}=0\,,\quad
T_{1;64}=\alpha^5_{64}\beta_{15}+\alpha^9_{64}\beta_{19}=0\,,\quad
T_{4;61}=\beta_{47}=0\,,\\
(6,1,5):&\quad T_{6;15}=0\,,\quad
T_{1;65}=\alpha^9_{65}\beta_{19}=0\,,\quad
T_{5;61}=\beta_{57}=0\,,\\
(6,2,2):&\quad T_{6;22}=-\beta_{61}-\beta_{63}-\beta_{64}-\beta_{65}-\beta_{66}\,,\\
&\quad T_{2;62}=\alpha^4_{62}\beta_{24}+\alpha^5_{62}\beta_{25}=-\alpha^4_{62}-\alpha^5_{62}\,,\\
(6,2,3):&\quad T_{6;23}=\beta_{63}+\beta_{64}\,,\quad
T_{2;63}=\alpha^4_{63}\beta_{24}=-\alpha^4_{63}\,,\\
&\quad T_{3;62}=\alpha^4_{62}\beta_{34}+\alpha^5_{62}\beta_{35}=-\alpha^4_{62}+\alpha^5_{62}\,,\\
(6,2,4):&\quad T_{6;24}=\beta_{65}\,,\quad
T_{2;64}=\alpha^5_{64}\beta_{25}+\alpha^9_{64}\beta_{29}=-\alpha^5_{64}+\alpha^9_{64}\,,\\
&\quad T_{4;62}=\alpha^4_{62}\beta_{44}+\alpha^5_{62}\beta_{45}=\alpha^4_{62}-\alpha^5_{62}\,,\\
(6,2,5):&\quad T_{6;25}=\beta_{66}\,,\quad
T_{2;65}=\alpha^9_{65}\beta_{29}=\alpha^9_{65}\,,\quad
T_{5;62}=\alpha^4_{62}\beta_{54}+\alpha^5_{62}\beta_{55}=\alpha^5_{62}\,,\\
(6,3,3):&\quad T_{6;33}=-\beta_{62}-\beta_{63}-\beta_{64}-\beta_{68}-\beta_{6,10}\,,\quad
T_{3;63}=\alpha^4_{63}\beta_{34}=-\alpha^4_{63}\,,\\
(6,3,4):&\quad T_{6;34}=\beta_{6,10}\,,\quad
T_{3;64}=\alpha^5_{64}\beta_{35}+\alpha^9_{64}\beta_{39}=\alpha^5_{64}\,,\quad
T_{4;63}=\alpha^4_{63}\beta_{44}=\alpha^4_{63}\,,\\
(6,3,5):&\quad T_{6;35}=\beta_{68}\,,\quad
T_{3;65}=\alpha^9_{65}\beta_{39}=0\,,\quad
T_{5;63}=\alpha^4_{63}\beta_{54}=0\,.
\ea
\label{eq:ex1-T}
\ee
The triples $(6,4,4)$, $(6,4,5)$ and $(6,5,5)$ give no relation: $T_{6;44}$, $T_{6;45}$, $T_{6;55}$ are not admissible because $D_4^2$, $D_4D_5$, $D_5^2$ contain $C_9$ and $(6,9)\notin\mathcal B$, and $T_{4;64}$, $T_{4;65}$, $T_{5;64}$, $T_{5;65}$ are not admissible because $E_6D_4$, $E_6D_5$ may contain $C_9$ and $(4,9),(5,9)\notin\mathcal B$.
The triples $(6,6,k)$ give no relation either, since $E_6\cdot E_6$ is not expanded and $T_{6;6k}$ is the only presentation.

We impose that all the presentations of each triple in \eqref{eq:ex1-T} agree, and solve in the order in which $\ell_6$ traverses $\tcP$, as in the propagation argument of Section~\ref{sec:algorithm-existence-uniqueness}.
\begin{itemize}
\item
At the initial vertex, $E_6\cdot D_1=C_7$.
The relations $(6,1,2)$ and $(6,1,3)$, that is $D_1\cdot(E_6\cdot D_k)=(E_6\cdot D_1)\cdot D_k$ for $k=2,3$, fix the coefficients of the curve $C_4$ of the first crossed edge:
\be
\alpha^4_{62}=\beta_{61}=2\,,\qquad
\alpha^4_{63}=\beta_{62}=1\,.
\ee
The relation $(6,1,1)$ is then satisfied, $-2-1=-3$.
\item
Crossing $e_5$, the relations $(6,2,3)$ and $(6,3,4)$ give
\be
\beta_{64}=-\alpha^4_{63}=-1\,,\qquad
\alpha^5_{62}=\alpha^4_{62}-\alpha^4_{63}=1\,,\qquad
\alpha^5_{64}=\beta_{6,10}=\alpha^4_{63}=1\,.
\ee
\item
At the boundary edge $e_9$, the relations $(6,2,4)$ and $(6,2,5)$ give
\be
\beta_{65}=\alpha^4_{62}-\alpha^5_{62}=1\,,\qquad
\alpha^9_{64}=\beta_{65}+\alpha^5_{64}=2\,,\qquad
\alpha^9_{65}=\beta_{66}=\alpha^5_{62}=1\,.
\ee
\item
The remaining relations are now checks: $(6,2,2)$ reads $-(2-1+1+1)=-(2+1)$, $(6,3,3)$ reads $-(1-1+1)=-1$, and $(6,1,4)$, $(6,1,5)$, $(6,3,5)$ read $0=0=0$.
The last one, $T_{6;35}=\beta_{68}=0$, independently confirms \eqref{eq:ex1-beta-E-zero}.
\end{itemize}
The compatibility system therefore has the unique solution
\be
\ba
E_6\cdot D_1&=C_7\,, & E_6\cdot D_2&=2C_4+C_5\,, & E_6\cdot D_3&=C_4\,,\\
E_6\cdot D_4&=C_5+2C_9\,, & E_6\cdot D_5&=C_9\,, & &
\ea
\label{eq:ex1-alpha-E}
\ee
and the completed matrix of divisor--curve intersections is
\be
\big(\beta_{ia}\big)_{i\in I,\,a\in A}=
\begin{array}{c|ccccccc|ccc}
 & C_1 & C_2 & C_3 & C_4 & C_5 & C_6 & C_7 & C_8 & C_9 & C_{10}\\ \hline
D_1 & -6 & -3 & 1 & 1 & 0 & 0 & -3 & 0 & 0 & 0\\
D_2 & 4 & 2 & -2 & -1 & -1 & -7 & 2 & 1 & 1 & 1\\
D_3 & 2 & 1 & 0 & -1 & 1 & 1 & 1 & \times & 0 & \times\\
D_4 & 0 & 0 & 0 & 1 & -1 & 1 & 0 & 0 & \times & \times\\
D_5 & 0 & 0 & 1 & 0 & 1 & 5 & 0 & \times & \times & 0\\ \hline
E_6 & 2 & 1 & 0 & -1 & 1 & 1 & 1 & 0 & \times & 1
\end{array}
\label{eq:ex1-beta-full}
\ee

The solution passes several checks.
First, $\beta_{64}=-1<0$ forces $C_4\subset E_6$, in agreement with the incidence of $\ell_6$.
Second, $\beta_{61}=2\beta_{62}$ and, more generally, the columns of $C_1$, $C_2$, $C_7$ in \eqref{eq:ex1-beta-full} satisfy $\beta_{i1}=2\beta_{i2}=2\beta_{i7}$, as they must since $C_1$ is a conic and $C_2$, $C_7$ are lines in $D_1\simeq\PP^2$.
Third, the restriction $E_6|_{D_2}=2C_4+C_5$ reproduces, by intersections in $D_2$, the values $(\beta_{61},\beta_{63},\beta_{64},\beta_{65},\beta_{66})=(2,0,-1,1,1)$.
Finally, $\sum_{j\in I_0}E_6\cdot D_j=C_7+3C_4+2C_5+3C_9$ has vanishing intersection with both compact divisors, $-3+3=0$ and $2-3-2+3=0$, in agreement with $E_6\cdot\widetilde{\cX}_0=0$.

\medskip
\noindent
\textbf{Step 5: triple intersection numbers.}
For $i\in I_c$, \eqref{eq:algorithm-kappa} uses the compact expansions of $D_i\cdot D_k$, which are read off from \eqref{eq:ex1-alpha-central} and \eqref{eq:ex1-alpha-E}.
For every $j\in I$, with $D_j$ standing for $E_6$ when $j=6$, they give
\be
\ba
D_1D_jD_1&=-\beta_{j1}-\beta_{j2}\,, & D_1D_jD_2&=\beta_{j1}\,, & D_1D_jD_3&=\beta_{j2}\,,\\
D_1D_jD_4&=D_1D_jD_5=0\,, & D_1D_jE_6&=\beta_{j7}\,, & &\\
D_2D_jD_1&=\beta_{j1}\,, & D_2D_jD_2&=-\beta_{j1}-\beta_{j3}-\beta_{j4}-\beta_{j5}-\beta_{j6}\,, & D_2D_jD_3&=\beta_{j3}+\beta_{j4}\,,\\
D_2D_jD_4&=\beta_{j5}\,, & D_2D_jD_5&=\beta_{j6}\,, & D_2D_jE_6&=2\beta_{j4}+\beta_{j5}\,.
\ea
\label{eq:ex1-kappa-formulas}
\ee
Inserting \eqref{eq:ex1-beta-full}, we obtain
\be
\begin{array}{c|cccccc}
D_1 & D_1 & D_2 & D_3 & D_4 & D_5 & E_6\\ \hline
D_1 & 9 & -6 & -3 & 0 & 0 & -3\\
D_2 & * & 4 & 2 & 0 & 0 & 2\\
D_3 & * & * & 1 & 0 & 0 & 1\\
D_4 & * & * & * & 0 & 0 & 0\\
D_5 & * & * & * & * & 0 & 0\\
E_6 & * & * & * & * & * & 1
\end{array}
\qquad
\begin{array}{c|cccccc}
D_2 & D_1 & D_2 & D_3 & D_4 & D_5 & E_6\\ \hline
D_1 & -6 & 4 & 2 & 0 & 0 & 2\\
D_2 & * & 7 & -3 & -1 & -7 & -3\\
D_3 & * & * & -1 & 1 & 1 & -1\\
D_4 & * & * & * & -1 & 1 & 1\\
D_5 & * & * & * & * & 5 & 1\\
E_6 & * & * & * & * & * & -1
\end{array}
\label{eq:ex1-kappa}
\ee
These are all the intrinsic triple intersection numbers \eqref{eq:algorithm-output2}.
The result passes the following checks.
\begin{itemize}
\item
Every column sums to zero over $D_1,\dots,D_5$, since $\widetilde{\cX}_0\sim0$.
\item
$D_1^3=9=12-Q_1-r_1$ with charge $Q_1=1$ and $r_1=2$ boundary curves, as in \eqref{eq:compact-triple-self-cycle}, and $D_1^3=K_{\PP^2}^2$.
Similarly, $D_2^3=7=12-5$, as in \eqref{eq:toric-triple-self-intersection}, and $D_2^3=K^2_{\mathrm{Bl}_1\mathbb F_6}$.
\item
The row of $E_6$ in \eqref{eq:ex1-beta-full} coincides with the row of $D_3$ on all compact curves, so $E_6$ and $D_3$ have the same triple intersections with the compact divisors, as seen in the tables.
Writing $\equiv$ for equality of intersection numbers with all compact curves, $E_6\equiv D_3$.
Together with $\sum_{i\in I_0}D_i\sim0$ and the relation $D_2-2D_3-D_4+2D_5\equiv0$, which holds by \eqref{eq:ex1-beta-full} and corresponds to the $M$-invariant linear function $(x,y)\mapsto y$ on $\widetilde P$, the divisor classes modulo $\equiv$ span a space of rank $3$, corresponding to $r=2$ Coulomb parameters and one mass parameter.
\end{itemize}

\subsection{$SU(4)_k + 1\mathbf{AS} + 7\mathbf{F}$}
\label{sec:su4AS7F}
 
\begin{figure}
\centering
\begin{tikzpicture}
    \node at (0,0) {%
    \begin{tikzpicture}[scale=1.2]
        \coordinate (v1)   at (-1,4);
        \coordinate (v2)   at (-1,5);
        \coordinate (v3)   at (-2,4);
        \coordinate (v4p)  at (-3,3);
        \coordinate (v4pp) at ( 0,0);
        \foreach \j in {5,...,11} {\coordinate (v\j) at (0,\j-4);}
        \coordinate (v12)  at (-1,6);
        \coordinate (v13)  at (-2,5);
        \coordinate (v14)  at (-3,4);
        \coordinate (w1)   at (-2,2);
        \coordinate (w2)   at (-1,1);
        \fill[red,opacity=.32] (v1)--(v4p)--(v4pp)--cycle;
        \draw (v11)--(v12)--(v13)--(v14)--(v4p)--(v4pp)--cycle;
        \foreach \j in {4pp,5,6,7,8,9,2,3,4p} {\draw (v1)--(v\j);}
        \foreach \j in {9,10,11,12,13,3} {\draw (v2)--(v\j);}
        \foreach \j in {13,14,4p} {\draw (v3)--(v\j);}
        \draw[red,thick] (v1)--(v7);
        \begin{scope}[every node/.style={inner sep=.6pt,font=\tiny}]
            \node at (-1.95,3.22) {$C_{1,4}$};
            \node at (-.8,1.9)    {$C_{1,4}$};
        \end{scope}
        \node[font=\scriptsize] at (-1.33,2.4) {$F$};
        \foreach \p in {v1,v2,v3,v4p,v4pp,v5,v6,v7,v8,v9,v10,v11,v12,v13,v14}
            {\node[bd] at (\p) {};}
        \node[wd] at (w1) {};
        \node[wd] at (w2) {};
        \node[blue] at ($(v1)+(-.26,.22)$)   {\scriptsize $1$};
        \node[blue] at ($(v2)+(-.24,.20)$)   {\scriptsize $2$};
        \node[blue] at ($(v3)+(-.26,.18)$)   {\scriptsize $3$};
        \node[blue] at ($(v4p)+(-.30,-.08)$) {\scriptsize $4'$};
        \node[blue] at ($(v4pp)+(.30,-.10)$) {\scriptsize $4''$};
        \node[blue] at ($(v5)+(.24,0)$)      {\scriptsize $5$};
        \node[blue] at ($(v6)+(.24,0)$)      {\scriptsize $6$};
        \node[blue] at ($(v7)+(.24,0)$)      {\scriptsize $7$};
        \node[blue] at ($(v8)+(.24,0)$)      {\scriptsize $8$};
        \node[blue] at ($(v9)+(.24,0)$)      {\scriptsize $9$};
        \node[blue] at ($(v10)+(.27,0)$)     {\scriptsize $10$};
        \node[blue] at ($(v11)+(.27,.10)$)   {\scriptsize $11$};
        \node[blue] at ($(v12)+(-.10,.26)$)  {\scriptsize $12$};
        \node[blue] at ($(v13)+(-.14,.26)$)  {\scriptsize $13$};
        \node[blue] at ($(v14)+(-.30,.12)$)  {\scriptsize $14$};
        \node[red] at (-.27,3.68)            {\scriptsize $\ell_{15}$};
    \end{tikzpicture}};
    \node at (7.3,0) {%
    \begin{tikzpicture}[scale=2,
        beta/.style={circle,draw=cyan!60!black,text=cyan!60!black,fill=white,
            inner sep=0pt,minimum size=3.4mm,font=\tiny}]
        \coordinate (s1)  at (-1,4);
        \coordinate (s2)  at (-1,5);
        \coordinate (s3)  at (-2,4);
        \coordinate (s4)  at (-3,3);
        \coordinate (s5)  at (-2,3);
        \coordinate (s6)  at (-1,3);
        \foreach \j in {7,...,11} {\coordinate (s\j) at (0,\j-4);}
        \coordinate (s12) at (-1,6);
        \coordinate (s13) at (-2,5);
        \coordinate (s14) at (-3,4);
        \draw (s4)--(s7)--(s11)--(s14)--cycle;
        \foreach \j in {4,5,6,7,8,9,2,3} {\draw (s1)--(s\j);}
        \foreach \j in {9,10,11,12,13,3} {\draw (s2)--(s\j);}
        \foreach \j in {13,14,4} {\draw (s3)--(s\j);}
        \draw[red,thick] (s1)--(s7);
        \path (s1)--(s4)  node[beta,pos=.25] {$-1$} node[beta,pos=.75] {$-1$};
        \path (s1)--(s5)  node[beta,pos=.21] {$0$}  node[beta,pos=.79] {$-2$};
        \path (s1)--(s6)  node[beta,pos=.3]  {$0$}  node[beta,pos=.7]  {$-2$};
        \path (s1)--(s7)  node[beta,pos=.21] {$0$}  node[beta,pos=.79] {$-2$};
        \path (s1)--(s8)  node[beta,pos=.3]  {$0$}  node[beta,pos=.7]  {$-2$};
        \path (s1)--(s9)  node[beta,pos=.21] {$-1$} node[beta,pos=.79] {$-1$};
        \path (s1)--(s2)  node[beta,pos=.3]  {$-1$} node[beta,pos=.7]  {$-1$};
        \path (s1)--(s3)  node[beta,pos=.3]  {$0$}  node[beta,pos=.7]  {$-2$};
        \path (s2)--(s9)  node[beta,pos=.3]  {$-1$} node[beta,pos=.7]  {$-1$};
        \path (s2)--(s10) node[beta,pos=.21] {$0$}  node[beta,pos=.79] {$-2$};
        \path (s2)--(s11) node[beta,pos=.27] {$-1$} node[beta,pos=.84] {$-1$};
        \path (s2)--(s12) node[beta,pos=.3]  {$0$}  node[beta,pos=.7]  {$-2$};
        \path (s2)--(s13) node[beta,pos=.3]  {$-1$} node[beta,pos=.7]  {$-1$};
        \path (s2)--(s3)  node[beta,pos=.21] {$-1$} node[beta,pos=.79] {$-1$};
        \path (s3)--(s13) node[beta,pos=.3]  {$-1$} node[beta,pos=.7]  {$-1$};
        \path (s3)--(s14) node[beta,pos=.3]  {$-1$} node[beta,pos=.7]  {$-1$};
        \path (s3)--(s4)  node[beta,pos=.21] {$-2$} node[beta,pos=.79] {$0$};
        \foreach \p in {s2,s3,s4,s5,s6,s7,s8,s9,s10,s11,s12,s13,s14}
            {\node[bd] at (\p) {};}
        \node[circle,fill=red,inner sep=0pt,minimum size=2mm] at (s1) {};
        \node[blue] at ($(s1)+(-.13,.13)$)  {\scriptsize $1$};
        \node[blue] at ($(s2)+(-.13,.13)$)  {\scriptsize $2$};
        \node[blue] at ($(s3)+(-.13,.13)$)  {\scriptsize $3$};
        \node[blue] at ($(s4)+(-.13,-.11)$) {\scriptsize $4$};
        \node[blue] at ($(s5)+(0,-.14)$)    {\scriptsize $5$};
        \node[blue] at ($(s6)+(0,-.14)$)    {\scriptsize $6$};
        \node[blue] at ($(s7)+(.12,-.11)$)  {\scriptsize $7$};
        \node[blue] at ($(s8)+(.13,0)$)     {\scriptsize $8$};
        \node[blue] at ($(s9)+(.13,0)$)     {\scriptsize $9$};
        \node[blue] at ($(s10)+(.15,0)$)    {\scriptsize $10$};
        \node[blue] at ($(s11)+(.12,.11)$)  {\scriptsize $11$};
        \node[blue] at ($(s12)+(-.09,.13)$) {\scriptsize $12$};
        \node[blue] at ($(s13)+(-.11,.13)$) {\scriptsize $13$};
        \node[blue] at ($(s14)+(-.16,.05)$) {\scriptsize $14$};
        \node[red] at ($(s1)+(.46,-.19)$)   {\tiny $C_E$};
        \node[red] at ($(s1)+(.36,-.68)$)   {\scriptsize $\ell_{15}$};
    \end{tikzpicture}};
\end{tikzpicture}
\caption{Left: the GTP $P$ for $SU(4)_{-1/2}+1\mathbf{AS}+7\mathbf{F}$, with the generalized face $F$ shaded, with the $E$-line $\ell_{15}$ of the exuberant divisor $E_{15}$ in red. The two edges labelled $C_{1,4}$ are glued into a single edge of $\widetilde P$. Right: the Symington polygon $\widetilde P$ with its triangulation $\tcP$. The monodromy cut runs along $[v_1,v_7]$. In both panels, the vertex labelled $i$ is $v_i$, dual to the component $D_i$ of $\widetilde{\cX}_0$, and the edge $[v_i,v_j]$ carries the double curve $C_{i,j}$. The circled number next to vertex $i$ on a compact edge $[v_i,v_j]$ is $\beta_{ia}=D_i\cdot C_a$ for $C_a=C_{i,j}$. }
\label{fig:nf7}
\end{figure}
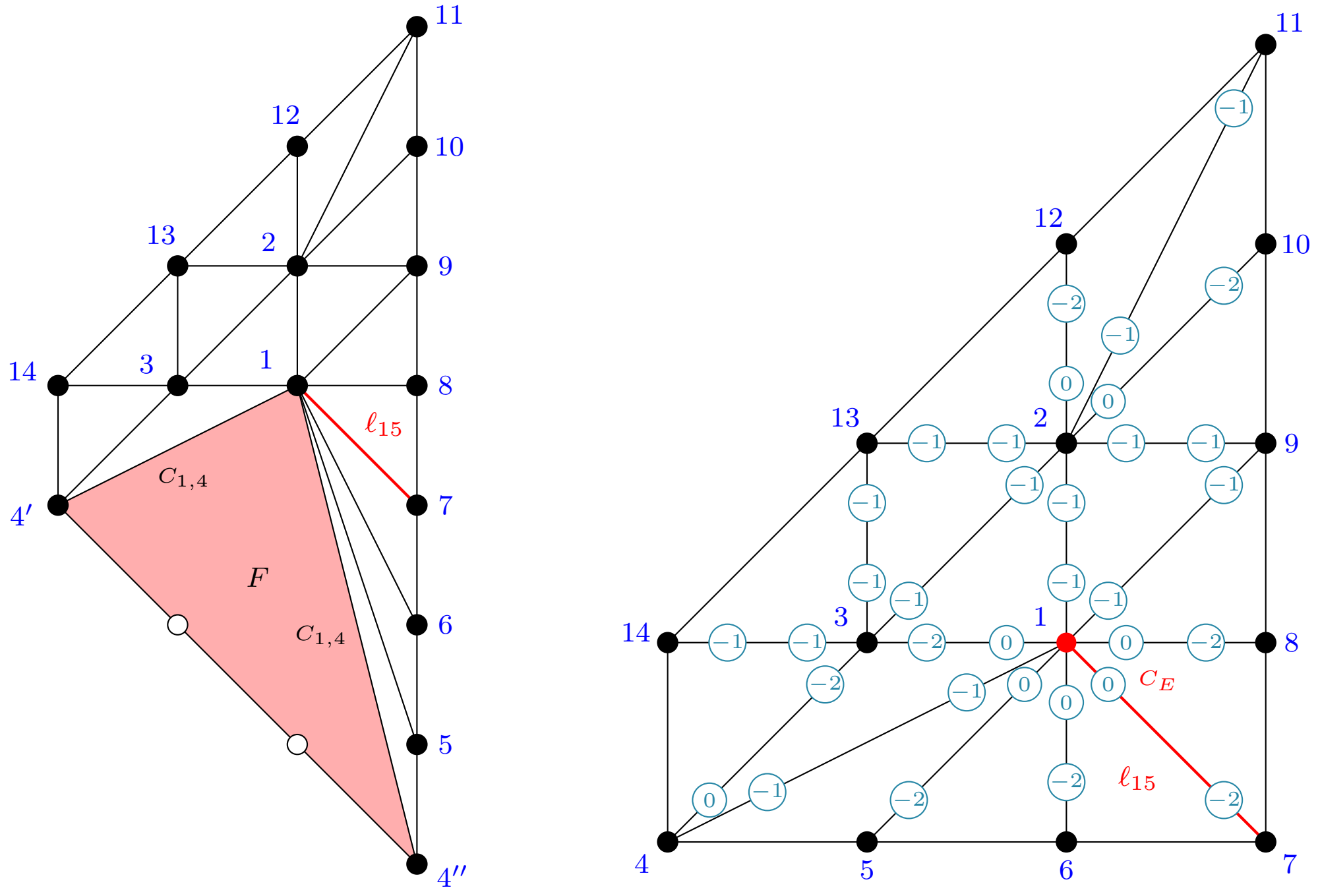

We close with a rank three example, the GTP for $SU(4)_k + 1\mathbf{AS} + 7\mathbf{F}$ with $|k| = \frac{1}{2}$, taken from Table 6 of \cite{VanBeest:2020kxw}, see Figure \ref{fig:nf7}. There are three compact divisors, and an additional 12 non-compact divisors, one of them being exuberant: 
\be
I_c=\{1,2,3\}\,,\qquad
I_{0,\partial}=\{4,\ldots,14\}\,,\qquad
I_e=\{15\}\,.
\label{nf7 I}
\ee
In this case no two vertices of $\tcP$ are joined by more than one edge (as opposed to the previous example), and we write $C_{i,j}$ for the double curve of the edge $[v_i,v_j]$.
There are $17$ compact double curves, $C_{1,j}$ ($2\leq j\leq9$), $C_{2,j}$ ($j=3$ and $9\leq j\leq13$) and $C_{3,j}$ ($j=4,13,14$), $11$ non-compact double curves along $\partial\widetilde P$, and the exuberant curve $C_E=D_1\cdot E_{15}$. We now proceed through the algorithm, giving only a few essential steps.

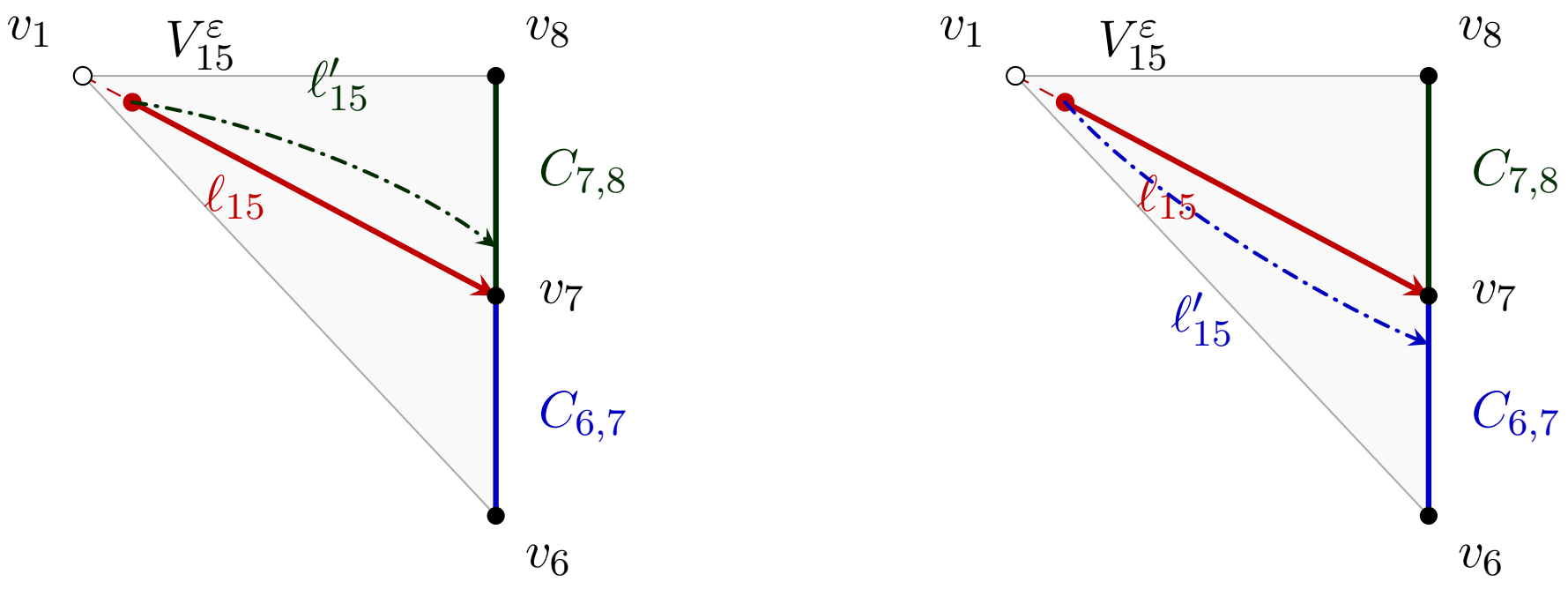
\begin{figure}
\centering
{\color{black}
\begin{tikzpicture}[font=\small,line cap=round,line join=round,>=stealth]
  \foreach \dx in {0,7} {
    \begin{scope}[xshift=\dx cm,x=3.1cm,y=1.65cm]
      \fill[gray!5] (0,0)--(1,-2)--(1,0)--cycle;
      \draw[gray!65] (0,0)--(1,-2) (0,0)--(1,0);
      \draw[gray!65,dashed] (0,0)--(1,-1);
      \draw[blue!75!black,very thick] (1,-2)--(1,-1);
      \draw[darkgreen!45!black,very thick] (1,-1)--(1,0);
      \draw[red!75!black,very thick,->] (.12,-.12)--(1,-1);
      \draw[red!75!black,dashed] (0,0)--(.12,-.12);
      \filldraw[fill=white,draw=black] (0,0) circle (1.9pt);
      \fill[red!75!black] (.12,-.12) circle (2pt);
      \fill (1,0) circle (1.9pt) (1,-1) circle (1.9pt) (1,-2) circle (1.9pt);
      \node[above left=3pt] at (0,0) {$v_1$};
      \node[above right=4pt] at (.12,-.12) {$V_{15}^\varepsilon$};
      \node[above right=3pt] at (1,0) {$v_8$};
      \node[right=5pt] at (1,-1) {$v_7$};
      \node[below right=3pt] at (1,-2) {$v_6$};
      \node[darkgreen!45!black,right=5pt] at (1,-.45) {$C_{7,8}$};
      \node[blue!75!black,right=5pt] at (1,-1.55) {$C_{6,7}$};
      \node[red!75!black,left=5pt] at (.55,-.55) {$\ell_{15}$};
    \end{scope}
  }
  \begin{scope}[x=3.1cm,y=1.65cm]
    \draw[darkgreen!45!black,dash dot,thick,->]
      (.12,-.12) .. controls (.4,-.19) and (.8,-.48) .. (1,-.78);
    \node[darkgreen!45!black,above=4pt] at (.62,-.33) {$\ell_{15}'$};
  \end{scope}
  \begin{scope}[xshift=7cm,x=3.1cm,y=1.65cm]
    \draw[blue!75!black,dash dot,thick,->]
      (.12,-.12) .. controls (.32,-.56) and (.78,-1.05) .. (1,-1.22);
    \node[blue!75!black,below left=3pt] at (.6,-.85) {$\ell_{15}'$};
  \end{scope}
  \node[align=center] at (1.65,-4.15)
    {(a) Auxiliary endpoint on $[v_7,v_8]$};
  \node[align=center] at (8.65,-4.15)
    {(b) Auxiliary endpoint on $[v_6,v_7]$};
\end{tikzpicture}}
\caption{Local enlargement near the edge $[v_1,v_7]$, showing the small deformations of the $\ell_{15}$ line needed so that the unicity part of Theorem~\ref{thm:compatibility-unique} applies. The dash-dotted paths illustrate two possible auxiliary support choices, selecting $C_{7,8}$ or $C_{6,7}$. Hence in the first case $(x,y)=(0,1)$ and in the second case $(x,y)=(1,0)$.  }
\label{fig:SU4-boundary-endpoint}
\end{figure}

\medskip
\noindent
\textbf{Steps 1--4.}
$D_1$ is the smoothing of the toric surface $\overline X_{v_1}$ with the nine rays towards $v_{4''},v_5,\ldots,v_9,v_2,v_3,v_{4'}$, which merges the curves of the first and last rays into $C_{1,4}$.
The direction $(1,-1)$ of $\ell_{15}$ is the ray towards $v_7$, so $(p,q)=(1,0)$ and \eqref{eq:algorithm-E-local} gives
\be
D_1\cdot C_E=-1\,,\qquad
D_7\cdot C_E=1\,,\qquad
E_{15}\cdot C_E=-1\,:
\ee
$C_E$ is a $(-1)$-curve in $D_1$ meeting $\partial D_1$ once, on $C_{1,7}$.
The compatibility relations, for instance $T_{15;17}=T_{1;15\,7}=T_{7;15\,1}$, then give\footnote{Here the support of $E_{15}\cdot D_7$ is restricted by \eqref{eq_zero}: with $\ell'_{15}$ displaced off the edge $e_{1,7}$ that $\ell_{15}$ runs along, it does not meet the interior of $e_{1,7}$, so $C_{1,7}$ does not occur in $E_{15}\cdot D_7$.
Without this restriction, a term $z\,C_{1,7}$ is allowed, and the compatibility relations leave $z$ free as well.
Indeed, $C_{1,7}$ is the compact section of $D_7\simeq\PP^1\times\CC$, with $C_{1,7}^2|_{D_7}=0$, so $D_i\cdot C_{1,7}=0$ for $i\in I_c$.
The only admissible presentation containing $z$ is $T_{15;15\,7}=z$, a presentation of $E_{15}^2\cdot D_7$, which has no compact factor and no second presentation.
Like the split of $x+y$, $z$ drops out of every triple intersection number with a compact divisor.}
\be
E_{15}\cdot D_7=x\,C_{6,7}+y\,C_{7,8}\,,\quad x+y=1\,,\qquad
E_{15}\cdot C_{1,7}=1\,,
\ee
with $E_{15}\cdot D_j=0$ for the other $j\in I_0$ and $E_{15}\cdot C_a=0$ for all other double curves.
Only $x+y$ is fixed, not $x$ and $y$ individually.  In order to apply Theorem~\ref{thm:compatibility-unique} which guarantees that all intersections are uniquely determined, one must fix a small deformation of the $E$-line, as shown in Figure \ref{fig:SU4-boundary-endpoint}. However, no triple intersection number depends on the individual values of $x$ and $y$, so in this particular situation the deformation of the line is not needed if one is just interested in triple intersections. 

 \medskip
\noindent
\textbf{Step 5.}
The non-vanishing triple intersection numbers with at least one compact divisor are
\be
\begin{array}{c|cccccccccc}
D_1 & D_1 & D_2 & D_3 & D_4 & D_5 & D_6 & D_7 & D_8 & D_9 & E_{15}\\ \hline
D_1 & 3 & -1 & 0 & -1 & 0 & 0 & 0 & 0 & -1 & -1\\
D_2 & * & -1 & 1 & 0 & 0 & 0 & 0 & 0 & 1 & 0\\
D_3 & * & * & -2 & 1 & 0 & 0 & 0 & 0 & 0 & 0\\
D_4 & * & * & * & -1 & 1 & 0 & 0 & 0 & 0 & 0\\
D_5 & * & * & * & * & -2 & 1 & 0 & 0 & 0 & 0\\
D_6 & * & * & * & * & * & -2 & 1 & 0 & 0 & 0\\
D_7 & * & * & * & * & * & * & -2 & 1 & 0 & 1\\
D_8 & * & * & * & * & * & * & * & -2 & 1 & 0\\
D_9 & * & * & * & * & * & * & * & * & -1 & 0\\
E_{15} & * & * & * & * & * & * & * & * & * & -1
\end{array}
\label{nf7 table D1}
\ee
\be
\begin{array}{c|cccccccc}
D_2 & D_1 & D_2 & D_3 & D_9 & D_{10} & D_{11} & D_{12} & D_{13}\\ \hline
D_1 & -1 & -1 & 1 & 1 & 0 & 0 & 0 & 0\\
D_2 & * & 5 & -1 & -1 & 0 & -1 & 0 & -1\\
D_3 & * & * & -1 & 0 & 0 & 0 & 0 & 1\\
D_9 & * & * & * & -1 & 1 & 0 & 0 & 0\\
D_{10} & * & * & * & * & -2 & 1 & 0 & 0\\
D_{11} & * & * & * & * & * & -1 & 1 & 0\\
D_{12} & * & * & * & * & * & * & -2 & 1\\
D_{13} & * & * & * & * & * & * & * & -1
\end{array}
\qquad
\begin{array}{c|cccccc}
D_3 & D_1 & D_2 & D_3 & D_4 & D_{13} & D_{14}\\ \hline
D_1 & 0 & 1 & -2 & 1 & 0 & 0\\
D_2 & * & -1 & -1 & 0 & 1 & 0\\
D_3 & * & * & 7 & -2 & -1 & -1\\
D_4 & * & * & * & 0 & 0 & 1\\
D_{13} & * & * & * & * & -1 & 1\\
D_{14} & * & * & * & * & * & -1
\end{array}
\label{nf7 table D2 D3}
\ee

\begin{figure}[!htbp]
\centering
{\color{black}
\begin{tikzpicture}[x=.9cm,y=.9cm,
    vol/.style={font=\scriptsize,inner sep=1pt},
    face/.style={blue,font=\scriptsize}]
    \coordinate (T1)  at (-6,-3);
    \coordinate (T2)  at (-3,-2);
    \coordinate (T3)  at (-1,-1);
    \coordinate (T4)  at (0,0);
    \coordinate (T5)  at (0,1);
    \coordinate (T6)  at (-3,4);
    \coordinate (T7)  at (-8,4);
    \coordinate (T8)  at (-8,-2);
    \coordinate (T9)  at (-3,6);
    \coordinate (T10) at (-4,7);
    \coordinate (T11) at (-6,8);
    \coordinate (T12) at (-9,8);
    \coordinate (T13) at (-9,5);
    \coordinate (T14) at (-12,5);
    \coordinate (T15) at (-12,2);
    \coordinate (B)   at (-22/3,-10/3); 
    \coordinate (S7)  at (-2.35,1.65);  
    \coordinate (CE)  at (-.35,-.35);   
    \coordinate (R)   at (2.3,0);
    \draw[red,dashed,thick] (S7)--(-8.5,-4.5);
    \draw[red,thick] (S7)--(CE)--(1.75,-.35);
    \draw (T1)--(T2)--(T3)--(T4)--(T5)--(T6)--(T7)--(T8)--(B)--(T1)
          (T6)--(T9)--(T10)--(T11)--(T12)--(T13)--(T7)
          (T13)--(T14)--(T15)--(T8);
    \foreach \t in {T1,T2,T3,T4,T5,T9,T10} {\draw (\t)--(\t -| R);}
    \foreach \t in {T11,T12,T14} {\draw (\t)--++(-1.5,1.5);}
    \draw (T15)--++(-1.5,0);
    \node[seven,fill=white] at (S7) {};
    \node[vol] at (-3.6,-2.62)  {$m_7-m_6$};
    \node[vol] at (-1.15,-1.62) {$m_6-m_5$};
    \node[vol] at (.7,-.72)     {$m_5-m_4$};
    \node[vol] at (.75,.5)      {$m_4-m_3$};
    \node[vol] at (-.55,2.75)   {$a_2+m_3$};
    \node[vol] at (-1.9,5)      {$-(a_2+m_2)$};
    \node[vol] at (-2.6,6.55)   {$m_2-m_1$};
    \node[vol] at (-4.3,7.62)   {$a_1+m_1$};
    \node[vol] at (-5.5,4.27)   {$h_0+a_1-a_3+m_1+m_2-\bar m$};
    \node[vol,rotate=90] at (-7.72,1) {$h_0+a_2+a_3-a_4+\bar m$};
    \node[vol] at (-6.35,-3.72) {$-(a_3+m_7)$};
    \node[vol] at (-6.85,4.8)   {$a_1+a_3+m_{\mathbf{AS}}$};
    \node[vol] at (-7.5,6.5)    {$-(a_2+a_3+m_{\mathbf{AS}})$};
    \node[vol] at (-7.62,8.27)  {$h_0-\bar m+m_{\mathbf{AS}}$};
    \node[vol] at (-10.42,5.3)  {$-(a_1+a_4+m_{\mathbf{AS}})$};
    \node[vol,rotate=90] at (-11.72,3.45) {$h_0-a_4+m_{\mathbf{AS}}+\bar m$};
    \node[vol,rotate=-45] at (-10.25,-.25) {$a_3-a_4$};
    \node[red,font=\scriptsize] at (-1.05,.95) {$C_E$};
    \node[face] at (-5.5,.5)    {$1$};
    \node[face] at (-4.8,5.6)   {$2$};
    \node[face] at (-10.2,2.8)  {$3$};
    \node[face] at (-10.5,-2.8) {$4$};
    \node[face] at (2.05,-2.5)  {$5$};
    \node[face] at (2.05,-1.5)  {$6$};
    \node[face] at (2.05,-.5)   {$7$};
    \node[face] at (2.05,.5)    {$8$};
    \node[face] at (2.05,3.5)   {$9$};
    \node[face] at (2.05,6.5)   {$10$};
    \node[face] at (0,8.2)      {$11$};
    \node[face] at (-8.3,9.1)   {$12$};
    \node[face] at (-11,6.3)    {$13$};
    \node[face] at (-13,3.5)    {$14$};
\end{tikzpicture}}
\caption{The 5-brane web dual to the GTP of Figure~\ref{fig:nf7}. It is labelled by $\mathrm{vol}(C_{i,j})$, taken from \eqref{nf7 dictionary}, with $\bar m=\frac12\sum_fm_f$. The $[1,1]$ 7-brane lies in face 1, and its monodromy cut (dashed) points in the direction $(-1,-1)$ of the three 5-branes that ended on it before it was pulled in. The red path from the 7-brane is dual to the $E$-line $\ell_{15}$. Its first segment, the $(1,1)$ string ending on $C_{1,7}$, is the exuberant curve $C_E$, with $\mathrm{vol}(C_E)=h_0+a_1+a_2+m_1+m_2+m_3+m_4-\bar m$. The path then continues into face 7, dual to the endpoint $v_7$ of $\ell_{15}$.}
\label{fig:nf7-web}
\end{figure}
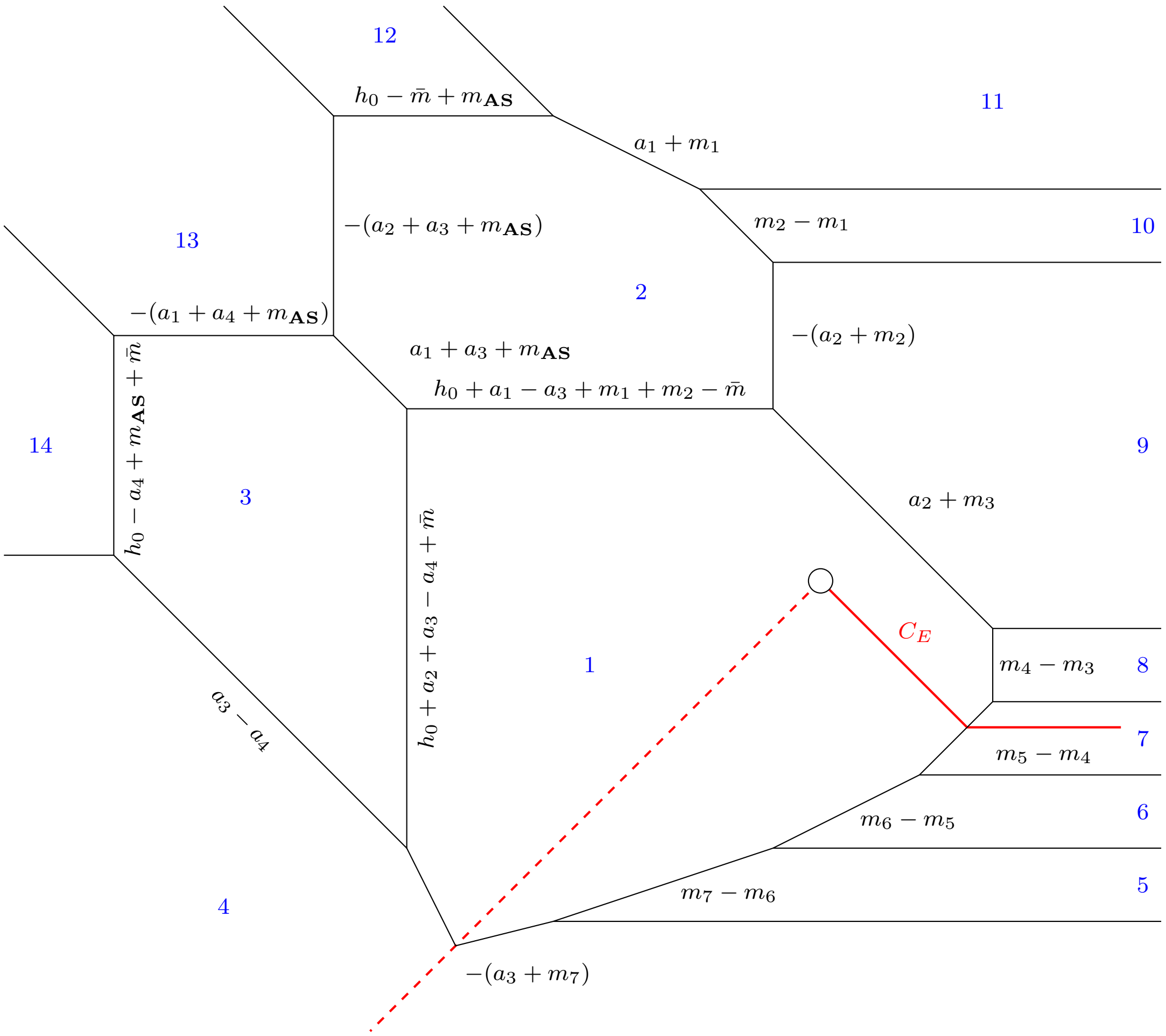

\paragraph{Match with the gauge theory.}
We now show that the triple intersection numbers above reproduce exactly the one-loop prepotential \eqref{eq:IMS-prepotential} of $SU(4)_{-1/2}+1\mathbf{AS}+7\mathbf F$. In particular, we show that one can read the Chern-Simons level $k=-\frac{1}{2}$ from geometry.\footnote{Of course, the overall sign of $k$ is a matter of convention and can be changed by charge conjugation. In the conventions used here though, we find $k=-\frac{1}{2}$. We insist on this to make the point that geometry does fix all the gauge theory parameters through the triple intersections. } 

The three compact surfaces are ruled, with fiber classes
\be
f_1=C_{2,9}+C_{2,10}+C_{2,11}\subset D_2\,,\qquad
f_2=C_{1,4}+C_{1,5}+\cdots+C_{1,9}\subset D_1\,,\qquad
f_3=C_{3,4}\subset D_3\,,
\ee
and the tables give $\big((D_2,D_1,D_3)\cdot f_i\big)=-\mathcal C_{A_3}$, minus the Cartan matrix of $SU(4)$.
Hence the fibers are the W-bosons, and $(D_2,D_1,D_3)$ correspond to the simple roots $(\alpha_1,\alpha_2,\alpha_3)$.
With $S=\sum_At^AD_A$ and $\mathrm{vol}(C)=-S\cdot C$, we define the Coulomb parameters by $\mathrm{vol}(f_i)=\sum_j(\mathcal C_{A_3})_{ij}\psi_j$ and set $(a_1,a_2,a_3,a_4)=(\psi_1,\psi_2-\psi_1,\psi_3-\psi_2,-\psi_3)$.

The hypermultiplets sit in the degenerate fibers.
The volumes of $C_{2,11}$, $C_{2,10}+C_{2,11}$, $C_{1,9}+\cdots+C_{1,9-j}$ ($0\leq j\leq4$) and $C_{2,3}$ define the masses $m_1,\ldots,m_7$ and $m_{\mathbf{AS}}$, the gauge-neutral curve $C_{2,12}$ defines the inverse coupling $h_0$, and the other entries of
\be
\begin{array}{c|c|c||c|c}
\text{curve} & \text{weight} & \mathrm{vol}(C) & \text{curve} & \mathrm{vol}(C)\\ \hline
C_{2,11} & e_1 & a_1+m_1 & C_{2,10} & m_2-m_1\\
C_{2,9} & -e_2 & -(a_2+m_2) & C_{1,j}\ (5\leq j\leq8) & m_{12-j}-m_{11-j}\\
C_{1,9} & e_2 & a_2+m_3 & C_{2,12} & h_0-\tfrac12\sum_fm_f+m_{\mathbf{AS}}\\
C_{1,4} & -e_3 & -(a_3+m_7) & & \\
C_{2,3} & e_1+e_3 & a_1+a_3+m_{\mathbf{AS}} & & \\
C_{2,13} & -(e_2+e_3) & -(a_2+a_3+m_{\mathbf{AS}}) & & \\
C_{3,13} & -(e_1+e_4) & -(a_1+a_4+m_{\mathbf{AS}}) & &
\end{array}
\label{nf7 dictionary}
\ee
are then consequences of the tables. See Figure \ref{fig:nf7-web}. 
Thus the flavors $f=1,2$ live in the fiber $(-1,-2,-1)$ of $D_2$, the flavors $f=3,\ldots,7$ in the fiber $(-1,-2,-2,-2,-2,-1)$ of $D_1$, and the antisymmetric in the fibers through the common curve $C_{2,3}$ of $D_2$ and $D_3$.
Positivity of the volumes selects the chamber
\be
a_1+m_{1,2}>0>a_2+m_{1,2}\,,\qquad
a_2+m_{3,\ldots,7}>0>a_3+m_{3,\ldots,7}\,,\qquad
s_{ij}(a_i+a_j+m_{\mathbf{AS}})>0\,,
\label{nf7 chamber}
\ee
with $s_{12}=s_{13}=1$ and $s_{ij}=-1$ otherwise. In this chamber, \eqref{eq:IMS-prepotential} reads
\be
\ba
\mathcal F_{\rm IMS}={}&\frac{h_0}2\sum_ia_i^2+\frac k6\sum_ia_i^3+\frac16\sum_{i<j}(a_i-a_j)^3
-\frac1{12}\sum_{f=1}^2\Big((a_1+m_f)^3-\sum_{i=2}^4(a_i+m_f)^3\Big)\\
&-\frac1{12}\sum_{f=3}^7\Big(\sum_{i=1}^2(a_i+m_f)^3-\sum_{i=3}^4(a_i+m_f)^3\Big)
-\frac1{12}\sum_{i<j}s_{ij}(a_i+a_j+m_{\mathbf{AS}})^3\,.
\ea
\label{nf7 IMS}
\ee

On the geometric side, the $\psi^3$ terms of $\mathcal F=\frac16S^3$ come from the entries of \eqref{nf7 table D1}--\eqref{nf7 table D2 D3} with $i,j,k\in I_c$:
\be
6\,\mathcal F\big|_{\psi^3}=5\psi_1^3+3\psi_2^3+7\psi_3^3-3\psi_1^2\psi_2-3\psi_1\psi_2^2-3\psi_1^2\psi_3-3\psi_1\psi_3^2-6\psi_2\psi_3^2+6\psi_1\psi_2\psi_3\,.
\label{nf7 cubic}
\ee
This coincides with the $\psi^3$ part of \eqref{nf7 IMS} if and only if $k=-\tfrac12$, so the geometry determines the Chern--Simons level. The agreement is complete. Expressing all $t^A$ through $(\psi_i,m_f,m_{\mathbf{AS}},h_0)$ via \eqref{nf7 dictionary}, $\frac16S^3$ coincides with \eqref{nf7 IMS} in every term of order $\psi^3$, $\psi^2m$ and $\psi m^2$, where $m$ is any of the nine mass parameters.
It differs only by $\psi$-independent terms, which involve non-compact divisors only and are not intrinsic.
Every intrinsic number in the tables, including those of $E_{15}$, enters this comparison.

\subsection*{Acknowledgements}

We thank Cyril Closset, Andres Collinucci, Quentin Lamouret, and Marcus Sperling for discussions at various stages of this long journey. 
We thank the organizers of String Math at ICTP, where the start of this collaboration across the physics-math divide was enabled.
PB acknowledges the support of
a Sloan Research Fellowship from the Alfred P. Sloan Foundation.
The work of SSN is  supported by the EPSRC Open Fellowship EP/X01276X/1 (Schafer-Nameki).
For the purpose of open
access, the authors have applied a CC BY public copyright licence to any author accepted
manuscript arising from this submission.

Anthropic's Claude Opus 5.5 and OpenAI ChatGPT-6.1 Sol and Astra were used in the final stages of the project to improve the writing, draw figures, perform the gauge-parameter match in Section \ref{sec:su4AS7F} and polish the Mathematica code. The authors take full responsibility for the content of the paper.

\appendix

\section{Description of the Mathematica Code}
\label{app:code}

\begin{figure}
    \centering
    \includegraphics[width=0.3\linewidth]{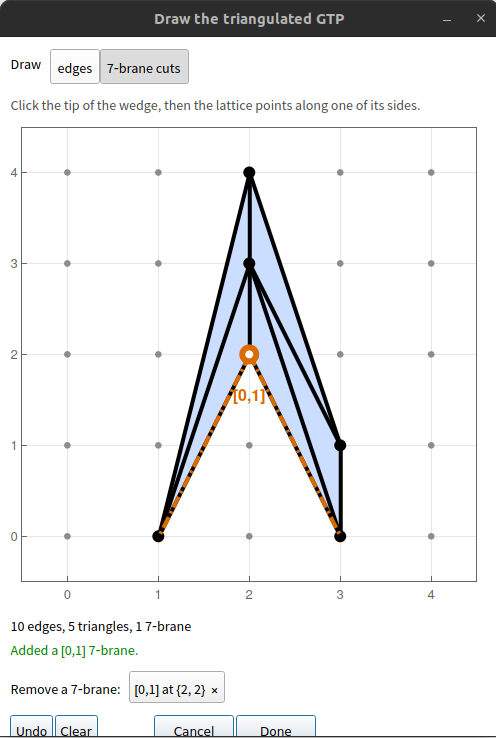}
    \caption{User-friendly input method for GTPs. We illustrate with the example of Section \ref{sec:ex1}. }
    \label{fig:code1}
\end{figure}

\begin{figure}
    \centering
    \includegraphics[width=0.7\linewidth]{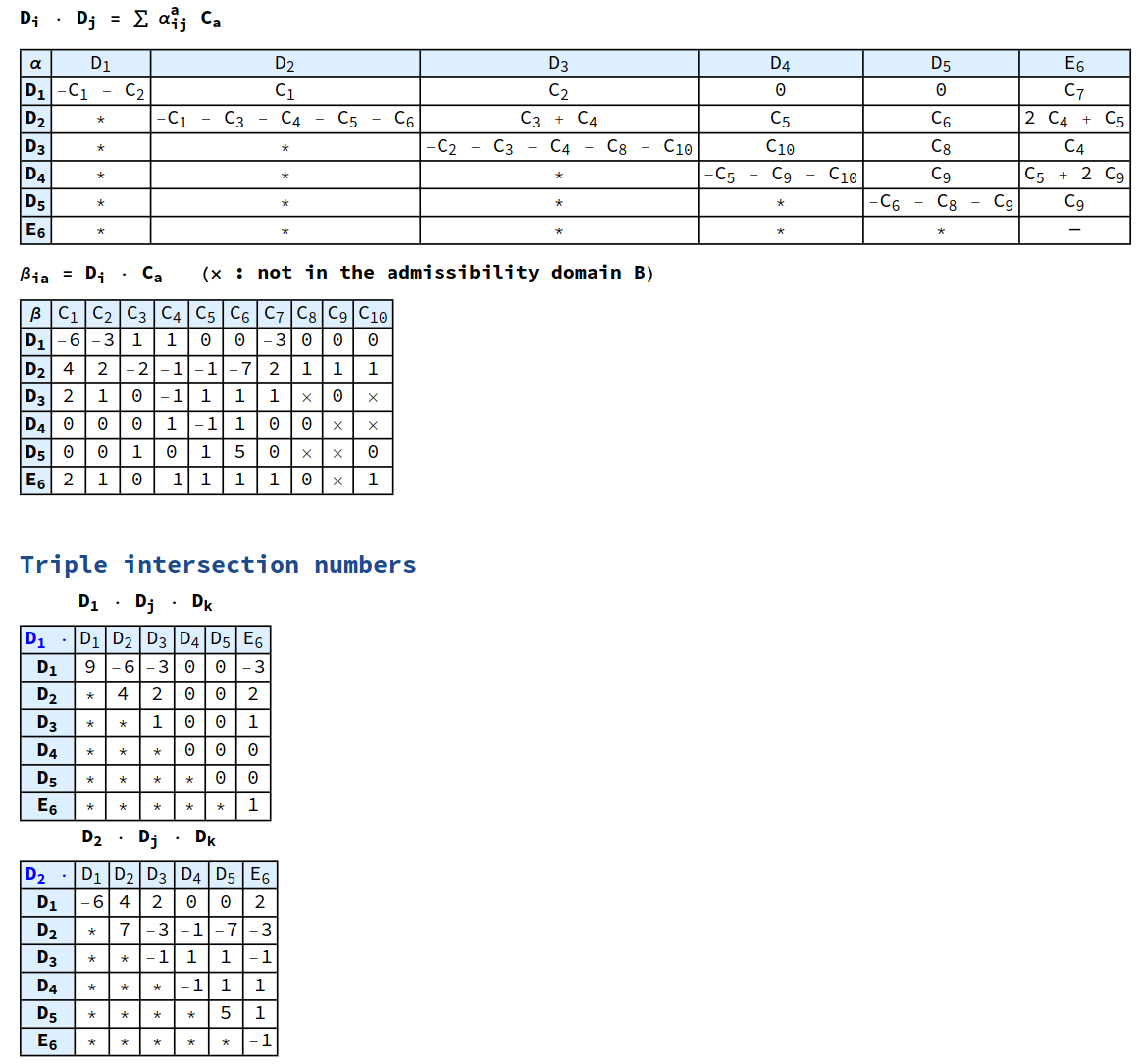}
    \caption{Output of the code on the example of Section \ref{sec:ex1}. }
    \label{fig:code2}
\end{figure}

For the reader's convenience, we provide an implementation of the algorithm of Section \ref{sec:intersectionAlgorithm} in a Mathematica notebook, available at 
\begin{center}
    \href{https://github.com/AntoineBrgt/GTP-Intersections}{https://github.com/AntoineBrgt/GTP-Intersections}
\end{center}
The code follows closely the steps outlined there. The input can be given in two ways. 
\begin{itemize}
    \item The first one uses text: one gives the list of edges of the GTP, followed by a list of pulled-in 7-branes. Each 7-brane is specified by its charge and the cut that gives rise to the Symington polygon. For instance, if one wants to check the computations of the example of Section \ref{sec:ex1}, one can write 
\begin{verbatim} 
gtpEx1 = {{{0, 0}, {0, 1}}, {{0, 0}, {1, -2}}, {{0, 1}, {-1, -2}}, 
{{0, 1}, {1, -2}}, {{0, 1}, {1, -1}}, {{0, 1}, {0, 2}}, 
{{0, 0}, {-1, -2}}, {{-1, -2}, {0, 2}}, {{1, -1}, {0, 2}}, 
{{1, -2}, {1, -1}}};
braneEx1 = {{{0, 1}, {{0, 0}, {1, -2}}}};
gtpRun[gtpEx1, braneEx1]
\end{verbatim}
Here, \texttt{gtpEx1} is the list of edges of the GTP, written using coordinates for the end points of each edge. And \texttt{braneEx1} is the list of 7-branes. Here there is a single 7-brane, so the list contains a single element, which is a pair: the first entry  \texttt{\{0, 1\}} is the charge of the brane, the second one \texttt{\{\{0, 0\}, \{1, -2\}\}} is the path followed by (one side of) the cut which defines the Symington polygon. 
\item While very explicit, the previous input method can be painful. The alternative is to just draw the GTP and the monodromy cuts in a pop-up window as shown in Figure \ref{fig:code1}. 
\end{itemize}
In either case, the output gives the intersection tables, as shown in Figure \ref{fig:code2}. 

\clearpage
 
\bibliographystyle{ytphys}
\bibliography{GeometryBib}

\end{document}